\documentclass[a4paper,12pt]{article}

\usepackage{amssymb,amsmath,empheq,epsfig,graphicx}

\usepackage{color}

\newcommand{\amp}{M}                     
\newcommand{\ampRid}{\mathfrak{M}}       
\newcommand{\bk}[1]{\langle #1 \rangle}  
\newcommand{\bs}{\boldsymbol}
\newcommand{\bt}{{\boldsymbol{b}}}
\newcommand{\C}{\mathbb{C}}              
\newcommand{\cA}{\bs{A}}
\newcommand{\cB}{\bs{B}}
\newcommand{\dif}{\mathrm{d}}            
\newcommand{\eik}{\mathrm{eik}}          
\newcommand{\e}{\varepsilon}             
\newcommand{\el}{{\mathrm{el}}}          
\newcommand{\esp}[1]{\mathrm{e}^{#1}}    
\newcommand{\GW}{\mathrm{GW}}            
\renewcommand{\Im}{\mathrm{Im\,}}        
\newcommand{\jw}{\mathrm{JW}}            
\newcommand{\ket}[1]{|#1\rangle}         
\newcommand{\lp}{l_P}                    
\newcommand{\ls}{l_s}                    
\newcommand{\match}{{\mathrm{matched}}}  
\newcommand{\ord}[1]{\mathcal{O}\left(#1\right)}
\newcommand{\pol}{\epsilon}              
\newcommand{\Qt}{{\boldsymbol{Q}}}
\newcommand{\qt}{{\boldsymbol{q}}}
\newcommand{\pt}{{\boldsymbol{p}}}
\newcommand{\R}{\mathbb{R}}              
\renewcommand{\Re}{\mathrm{Re}\,}        
\newcommand{\regge}{\mathrm{Regge}}
\newcommand{\res}{\mathrm{res}}
\newcommand{\sgn}{\epsilon}              
\newcommand{\soft}{\mathrm{soft}}
\newcommand{\spf}{\mathcal{F}}           
\newcommand{\tfa}{{\cal M}}              
\newcommand{\Tht}{{\boldsymbol{\Theta}}}
\newcommand{\tht}{{\boldsymbol{\theta}}}
\newcommand{\tts}{\tilde{\theta}_s}
\newcommand{\ui}{\mathrm{i}}             
\newcommand{\vp}{\vec{p}}                
\newcommand{\vq}{\vec{q}}                
\newcommand{\xt}{{\boldsymbol{x}}}
\newcommand{\Z}{\mathbb{Z}}              
\newcommand{\zt}{{\boldsymbol{z}}}

\numberwithin{equation}{section}

\newcommand{\preprint}{\raisebox{15mm}[0pt][0pt]{%
  \makebox[0pt][l]{\hspace{0.69\textwidth}\normalsize{CERN-PH-TH-2015-272}}}%
}

\title{\preprint%
{\bf Unified limiting form of graviton radiation at extreme energies}}

\author{
   Marcello~Ciafaloni
   \footnote{Email: ciafaloni@fi.infn.it}
   \\[1ex]
   {\sl\small Dipartimento di Fisica, Universit\`a di Firenze}\\
   {\sl\small Via Sansone 1, 50019 Sesto Fiorentino, Italy}\\[5mm]
   Dimitri~Colferai
   \footnote{Email: colferai@fi.infn.it}
   \ and 
   Francesco~Coradeschi
   \footnote{Email: coradeschi@fi.infn.it}
   \\
   {\sl\small Dipartimento di Fisica, Universit\`a di Firenze and INFN, Sezione di Firenze}\\
   {\sl\small Via Sansone 1, 50019 Sesto Fiorentino, Italy}\\[5mm]
   Gabriele~Veneziano
   \footnote{Email: gabriele.veneziano@cern.ch}
   \\
   {\sl\small Coll\`ege de France, 11 place M. Berthelot, 75005 Paris, France}\\
   {\sl\small Theory Division, CERN, CH-1211 Geneva 23, Switzerland}\\
     {\sl\small Dipartimento di Fisica, Universit\`a di Roma La Sapienza, Rome, Italy}
   \\[5mm]
}
\date{}

\begin{document}

\maketitle

\begin{abstract}
  We derive the limiting form of graviton radiation in gravitational scattering
  at transplanckian energies ($E \gg M_P$) and small deflection angles.  We show
  that --- owing to the graviton's spin 2 --- such limiting form unifies the
  soft- and Regge- regimes of emission, by covering a broad angular range, from
  forward fragmentation to deeply central region. The single-exchange emission
  amplitudes have a nice expression in terms of the transformation phases of
  helicity amplitudes under rotations. As a result, the multiple-exchange
  emission amplitudes can be resummed via an impact parameter $b$-space
  factorization theorem that takes into account all coherence effects. We then
  see the emergence of an energy spectrum of the emitted radiation which, being
  tuned on $\hbar/R \sim M_P^2/E \ll M_P$, is reminiscent of Hawking's
  radiation. Such a spectrum is much softer than the one na\"ively expected for
  increasing input energies and neatly solves a potential energy crisis.
  Furthermore, by including rescattering corrections in the (quantum) factorization
  formula, we are able to recover the classical limit and to find the
  corresponding quantum corrections.  Perspectives for the extrapolation of such
  limiting radiation towards the classical collapse regime (where $b$ is of the
  order of the gravitational radius $R$) are also discussed.
\end{abstract}

\newpage

\tableofcontents

\section{Introduction\label{s:intro}}

The thought-experiment of transplanckian-energy gravitational scattering was
investigated, since the
eighties~\cite{tHooft:1987rb,Muzinich:1987in,ACV87,ACV88,Verlinde:1991iu,ACV90,ACV93},
as a probe of quantum-gravity theories, mostly in connection with the problem of
a possible loss of quantum coherence in a process leading classically to
gravitational collapse. In an $S$-matrix framework such a loss would be
associated with the breakdown of unitarity at sufficiently small impact
parameters.

In the scattering regime of large energies ($\sqrt{s}\gg M_P$) but small
deflection angles (i.e., in a regime far away from that of collapse), several
authors
proposed~\cite{tHooft:1987rb,Muzinich:1987in,ACV87,ACV88,Verlinde:1991iu}, on
various grounds, an approximate semiclassical description, whose $S$-matrix
exponentiates, at fixed impact parameter, an eikonal function of order
$\alpha_G\equiv Gs/\hbar\gg 1$, which is simply related to graviton exchanges at
large impact parameters $b\gg R\equiv 2G\sqrt{s}$. Such description has its
classical counterpart in the scattering of two Aichelburg-Sexl (AS) shock
waves~\cite{Aichelburg:1970dh}.

Starting from that leading eikonal approximation, the strategy followed
in~\cite{ACV90,ACV93} consisted in a systematic study of subleading corrections
to the eikonal phase, scattering angle, and
time-delays~\cite{CC14,Camanho:2014apa,D'Appollonio:2015gpa} in terms of the
expansion parameter $R^2/b^2$ (and $\ls^2/b^2$ if working within string
theory). These corrections can be resummed, in principle, by solving a classical
field theory and one can thus study the critical region $b\sim R$ where
gravitational collapse is expected.

This program was carried out, neglecting string corrections and after a drastic
truncation of the classical field theory due to Lipatov~\cite{Li82}, in
~\cite{ACV07} (see also~\cite{Marchesini:2008yh,VW1,VW2}). It was noted there
that below some critical impact parameter value $b_c\sim R$ (in good agreement
with the expected classical critical
value~\cite{Eardley:2002re,Kohlprath:2002yh,Yoshino:2002tx,Giddings:2004xy}),
the $S$-matrix --- evaluated by taking UV-safe (regular), but possibly complex,
solutions of the field equations --- shows a unitarity deficit. This was
confirmed, at the quantum level, by a tunneling interpretation of such
restricted solutions~\cite{Ciafaloni:2008dg,Ciafaloni:2009in,Ciafaloni:2011de}.
The above results suggest that the lost information could possibly be recovered
only through use of UV-sensitive solutions which, by definition, cannot be
studied by the effective-action approach of ~\cite{ACV07} and remain to be
investigated on the basis of the underlying (string-) theory itself. It is also
possible, of course, that the apparent loss of unitarity is caused instead by
the drastic truncation made in~\cite{ACV07} of Lipatov's effective field
theory~\cite{Li82}.

On the other hand, the parallel investigation of gravitational radiation
associated with transplanckian scattering brought a worrisome surprise: even if
such radiation is pretty soft --$\bk{q}\simeq\hbar/b$ being its typical
transverse momentum--- its rapidity density $\sim\alpha_G$ is so large as to
possibly endanger energy conservation~\cite{Giudice:2001ce,Rychkov_pc}, at least
in the early na\"ive extrapolations of the available rapidity
phase-space~\cite{ACV07,VW2}.  Energy conservation can be enforced by
hand~\cite{CiVe_un}, the result being that the flat low-energy spectrum (predicted
by known zero-frequency-limit theorems~\cite{Sm77}) extends up to a cutoff at
$\omega \sim b^2/R^3$. But that would mean that a fraction $O(1)$ of
the initial energy is emitted in gravitational radiation already at scattering
angles $O(\alpha_G^{-1/2}) \ll 1$, something rather hard to accept.

This unexpected result prompted the study of the purely classical problem of
gravitational bremsstrahlung in ultrarelativistic, small angle gravitational
scattering, a subject pioneered in the seventies by Peter D'Eath and
collaborators~\cite{D'Eath:1976ri,D'Eath:1992hb} and by Kovacs and
Thorne~\cite{Kovacs:1977uw,Kovacs:1978eu}. Those papers, however, were rather
inconclusive about the ultrarelativistic limit (the method of
refs.~\cite{Kovacs:1977uw,Kovacs:1978eu}, for instance, does not apply to
scattering angles larger than $m/E \equiv \gamma^{-1}$, and thus, in particular,
to our problem). Nonetheless, two groups of authors~\cite{GrVe14,Spirin:2015wwa}
managed to discuss directly the massless limit of the classical bremsstrahlung
problem showing the absence of an energy crisis and the emergence of a
characteristic frequency scale of order $R^{-1}$ beyond which the emitted-energy
spectrum is no longer flat (within the approximations used in~\cite{GrVe14} the
spectrum decreases like $\omega^{-1}$ till the approximation breaks down at
$\omega \sim b^2/R^3$). These classical results called for a more
careful investigation of the quantum problem.

And indeed the good surprise was that --- after a careful account of matrix
elements, phases, and coherence effects --- the limiting form of such radiation
for $\alpha_G \gg 1$ takes a simple and elegant expression and has the unique
feature of unifying two well-known limits of emission amplitudes: the soft- and
the Regge-limit. As a consequence, besides reducing in a substantial way the
total emitted-energy fraction, the spectrum drifts towards characteristic
energies of order $\hbar/R \sim M_P^2/E \ll M_P$, much smaller than those expected
from the na\"ive Regge behaviour, and reminiscent of Hawking's
radiation~\cite{Hawking:1974sw} (see also~\cite{Hawking:2015qqa}) from a black
hole of mass $E$. That nice surprise, that we wish to illustrate
here in full detail, has been
presented recently in a short note~\cite{CCV15}. 

We should note incidentally that, in a different but related investigation of
transplanckian graviton production integrated over impact parameter, a similarly
surprising feature was found (even more surprisingly by a tree-level
calculation) in~\cite{Dvali:2014ila}, the typical energy of the emitted
gravitons being again of order $\hbar/R$, with a very large multiplicity of
order $s/M_P^2$ i.e. of a black hole entropy for $M \sim \sqrt{s}$ .

The above list of surprises points in the direction of a more structural role of
the gravitational radius in the radiation problem, rather than in the scattering
amplitude calculation itself, so that approaching the collapse region at quantum
level may be actually easier and more informative if made from the point of view
of the radiation associated with the scattering process.

One may wonder what's the deep reason for all that. Here we will show that our
unified limiting form of radiation --- at the first subleading level in the
parameter $R^2/b^2$ --- is due to the dual role of the graviton spin two: on the
one hand it determines, by multi-graviton exchanges, the leading AS metric
associated with the colliding particles as well as its radiative components at
first subleading level; on the other hand, it also determines the transformation
properties of the emission amplitudes for definite helicity final states.
These, in turn, are closely connected to the emission currents themselves.

For the above reasons --- after a brief introduction to eikonal scattering in
sec.~\ref{s:tes} --- we emphasize~(sec.~\ref{s:1gex}) the physical matrix
elements of the relevant emission currents whose phases --- due to the absence
of collinear singularities in gravity --- play a crucial role in both the soft
and the Regge regimes. The unified form of graviton emission is then determined
--- at the single-exchange level --- by matching the soft and Regge behaviours in
all relevant angular regions, from nearly forward fragmentation to deeply
central emission. The resulting expressions are just the Fourier transforms of
two different components of the radiative metric tensor, which, however, yield
identical results because of a transversality condition.

The next step in the construction of the emission amplitudes is to resum the
contributions of all the graviton exchanges that occur during eikonal
scattering. This is done in sec.~\ref{s:bfact}, by establishing a
$b$-factorization theorem for each single-exchange contribution, and by summing
them up with the appropriate phases due to the dependence of the helicity
amplitudes on the incidence direction. The outcome, already presented
in~\cite{CCV15}, has a classical limit that resembles (but slightly differs
from) the one of~\cite{GrVe14}. An important new result of this work is that, by
also taking into account rescattering of the emitted gravitons all over the
eikonal evolution, the classical limit of~\cite{GrVe14} is fully recovered
together with some (or perhaps all) quantum corrections to it.  This resummation
yields a coherent average over incidence directions, up to the Einstein
deflection angle $\Theta_s(b)=2R/b$, providing important (de)coherence effects
which tend to suppress frequencies of order $\omega > R^{-1}$.  The above
procedure is finally generalized to multiple emissions by constructing the
appropriate (unitary) coherent-state operator.

The spectrum is then described and analyzed in sec.~\ref{s:spec}, both in
frequency and in angular distribution. This is done, in this paper, by taking
into account the incidence angle dependence only. Including rescattering
effects, both at the classical and quantum level, is deferred to a later work.
The ensuing perspectives for the development of the present method towards the
classical collapse region (given in sec.~\ref{s:concl}) are based on the new
features of the resummation pointed out in this paper and which are typical of
the emitted gravitational radiation associated with transplanckian scattering.
Finally, a number of detailed calculations and useful remarks are left to the
appendices.

\section{Transplanckian eikonal scattering\label{s:tes}}

Throughout this paper, as in~\cite{ACV07}, we will restrict our attention to
collisions in 4-dimen\-sio\-nal space-time and in the point-particle (or quantum
field theory) limit.  Consider first the elastic gravitational scattering
$p_1+p_2 \to p'_1 + p'_2$ of two ultrarelativistic particles, with external
momenta parametrized as
\begin{equation}\label{mompar}
  p_i = E_i (1, \Tht_i, \sqrt{1 - |\Tht_i|^2}) \, ,
\end{equation}
at center-of-mass energy $2E=\sqrt{s}\gg M_P$ and momentum transfer
$Q^\mu\equiv p_1^{\prime\mu}-p_1^\mu = p_2^\mu - p_2^{\prime\mu}$ with
transverse component $\Qt=E\Tht_s$; the 2-vectors
$\Tht_i=|\Tht_i|(\cos\phi_i,\sin\phi_i)$ describe both azimuth $\phi_i$ and
polar angles $|\Tht_i|\ll 1$ of the corresponding 3-momentum with respect to the
longitudinal $z$-axis.

This regime is characterized by a strong effective coupling
$\alpha_G\equiv Gs/\hbar \gg 1$ and was argued by several
authors~\cite{tHooft:1987rb,Muzinich:1987in,ACV88,ACV90} to be described by an
all-order leading approximation which has a semiclassical effective metric
interpretation.  The leading result for the $S$-matrix $S(b,E)$ in
impact-parameter $b\equiv J/E$ space has the eikonal form
\begin{equation}\label{eikform}
  S(b,E) = \exp[2\ui\delta_0(b,E)]\;, \qquad
  \delta_0(b,E) = \alpha_G \log\frac{L}{b}\;,
\end{equation}
$L$ being a factorized --- and thus irrelevant --- IR cutoff.

Corrections to the leading form~(\ref{eikform}) involve additional powers of
the Newton constant $G$ in two dimensionless combinations
\begin{equation}\label{adimComb}
  \frac{\hbar G}{b^2}=\frac{\lp^2}{b^2} \;, \qquad
  \frac{4G^2 s}{b^2}=\frac{R^2}{b^2} \sim\alpha_G\frac{\lp^2}{b^2}
  \gg \frac{\lp^2}{b^2} \,,
\end{equation}
$\lp\equiv\sqrt{\hbar G}$ being the Planck length.
Since $\alpha_G \gg1$ we can neglect completely the first kind of corrections.
Furthermore, we can consider the latter within a perturbative framework since
the impact parameter $b$ is much larger than the gravitational radius
$R\equiv 2G\sqrt{s}$.

\begin{figure}[ht]
  \centering
  \includegraphics[width=0.6\linewidth]{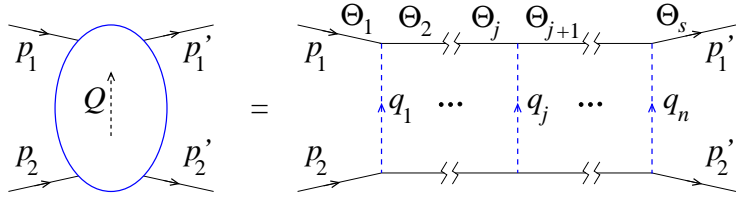}
  \caption{\it The scattering amplitude of two transplanckian particles (solid
    lines) in the eikonal approximation. Dashed lines represent (reggeized)
    graviton exchanges. The fast particles propagate on-shell throughout the
    whole eikonal chain. The angles $\Tht_j \simeq \sum_{i=1}^{j-1} \tht_i$
    denote the direction of particle 1 w.r.t.\ the $z$-axis along the scattering
    process.}
  \label{f:eikChain}
\end{figure}

In order to understand the scattering features implied by~(\ref{eikform}) we can
compute the $\Qt$-space amplitude
\begin{align}
 \frac{1}{s} \amp_\eik(s,\Qt^2)
  &= 4 \int\dif^2\bt\;\esp{-\frac{\ui\bt\cdot\Qt}{\hbar}}
  \frac{\esp{2\ui\delta_0(b,E)}}{2\ui} \nonumber \\
  &=\frac{8\pi\alpha_G}{\Qt^2}\left(\frac{4 \hbar^2}{\Qt^2 L^2}\right)^{-\ui\alpha_G}
  \frac{\Gamma(1-\ui\alpha_G)}{\Gamma(1+\ui\alpha_G)} \;, \label{QspAmp}
\end{align}
where the expression in the last line is obtained strictly-speaking by extending
the $\bt$-integration up to small $|\bt|\lesssim R$~\cite{tHooft:1987rb}, where
corrections may be large. But it is soon realized that the $\bt$-integration
in~(\ref{QspAmp}) is dominated by the saddle-point
\begin{equation}\label{saddlepoint}
  \Qt = E\Tht_s(\bt) = - E\frac{2R}{b}\hat{\bt} = - \alpha_G
  \frac{\hbar}{b}\hat{\bt} \;,
\end{equation}
which leads to the same expression for the amplitude, apart from an irrelevant
$\Qt$-independent phase factor. The saddle-point momentum
transfer~(\ref{saddlepoint}) comes from a large number $\bk{n}\sim\alpha_G$ of
graviton exchanges (fig.~\ref{f:eikChain}), corresponding to single-hit momentum
transfers $\bk{|\qt_j|}\simeq\hbar/b$ which are small, with very small
scattering angles $|\tht_j|$ of order $\theta_m\simeq\hbar/(bE)$. The overall
scattering angle --- though small for $b\gg R$ --- is much larger than
$\theta_m$ and is $|\Tht_s|=2R/b = 2\alpha_G\theta_m$, the Einstein deflection
angle.

In other words, every single hit is effectively described by the elastic amplitude
\begin{equation}
\amp_{\el}(\Qt_j) = \frac{\kappa^2 s^2}{\Qt_j^2} = \frac{\kappa^2 s^2}{E^2 \tht_j^2},
\quad \left(\kappa^2 = \frac{8\pi G}{\hbar}\right) \; ,
\end{equation}
which is in turn directly connected to the phase shift $\delta_0$:
\begin{equation}
  \delta_0(E,|\bt|) = \frac{1}{4s} \int \frac{\dif^2 \Qt}{(2\pi)^2}
  \esp{\frac{{\ui \Qt \cdot \bt}}{\hbar}} \amp_{\el}(\Qt)
  = \alpha_G \int \frac{\dif^2 \tht_s}{2\pi \tht_s^2}
  \esp{\frac{{\ui E \tht_s \cdot \bt}}{\hbar}} \; .
\end{equation}

The relatively soft nature of transplanckian scattering just mentioned is also
--- according to~\cite{ACV88} --- the basis for its validity in the
string-gravity framework. In fact, string theory yields exponentially suppressed
amplitudes in the high-energy, fixed-angle limit~\cite{Gross:1987kza} so that
several softer hits may be preferred to a single hard one in the $b\gg R$
regime.  Furthermore, this procedure can be generalized to multi-loop
contributions in which the amplitude, for each power of $G$, is enhanced by
additional powers of $s$, due to the dominance of $s$-channel iteration in
high-energy spin-2 exchange versus the $t$-channel one (which provides at most
additional powers of $\log s$). That is the mechanism by which the $S$-matrix
exponentiates an eikonal function (or, operator) with the effective coupling
$\alpha_G\equiv Gs/\hbar$ and subleading contributions which are a power series
in $R^2/b^2$ (and/or $l_s^2/b^2$). Finally, the scattering is self-sustained by
the saddle-point~(\ref{saddlepoint}), so that string-effects themselves may be
small --- and are calculable~\cite{ACV88,ACV90} --- if $b\gg R,\ls\gg\lp$ and
even at arbitrary $b$ if $R \ll \ls$.

Both the scattering angle~(\ref{saddlepoint}) (and the
$S$-matrix~(\ref{eikform})) can be interpreted from the metric point of
view~\cite{tHooft:1987rb} as the geodesic shift (and the quantum matching
condition) of a fast particle in the Aichelburg-Sexl (AS)
metric~\cite{Aichelburg:1970dh} of the other.

More directly, the associated metric emerges from the calculation~\cite{CC14} of
the longitudinal fields coupled to the incoming particles in the eikonal series,
which turn out to be
\begin{align}
  \frac14 h^{++} = h_{--} &= 2\pi R a_0(\xt)
  \delta\left(x^- -\pi R\sgn(x^+)a_0(b)\right) \;, \nonumber \\
  a_0(\xt) &= \frac1{2\pi}\log\frac{L^2}{\xt^2}\;, \qquad
  \delta_0(b,E) = \pi \alpha_G a_0(b) \;. \label{hpp}
\end{align}
Such shock-wave expressions yield two AS metrics for the fast particles, as well
as the corresponding time delay and trajectory shifts at leading level. When $b$
decreases towards $R\gg\ls$, corrections to the eikonal and to the
effective metric involving the $R^2/b^2$ parameter have to be included, 
as well as graviton radiation, to which we now turn.

\section{Limiting form of emission from single-graviton exchange\label{s:1gex}}

\begin{figure}[ht]
  \centering
  \includegraphics[width=0.3\linewidth]{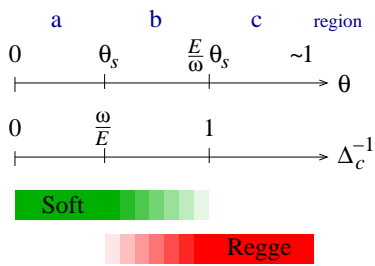}
  \caption{\it Schematics of dynamical regimes of graviton emission in
    transplanckian scattering with single-exchange (here $\hbar=1$). $\Delta_c$
    is a parameter used in sec.\ref{s:msarl}.}
  \label{f:regions}
\end{figure}

The basic emission process $p_1+p_2 \to p'_1 + p'_2 +q$ at tree level
(fig.~\ref{f:treeEmission}) of a graviton of momentum
$q^\mu:\qt=\hbar\omega\tht$ yields simple, and yet interesting, amplitudes in
various angular regimes (fig.~\ref{f:regions}) that we now consider, assuming a
relatively soft emission energy $\hbar\omega\ll E$. Note that this restriction
still allows for a huge graviton phase space, corresponding to classical
frequencies potentially much larger than the characteristic scale $R^{-1}$, due
to the large gravitational charge $\alpha_G \equiv G s / \hbar \gg 1$. We shall
consider three regimes:

\begin{figure}[ht]
  \centering
  \includegraphics[width=0.7\linewidth]{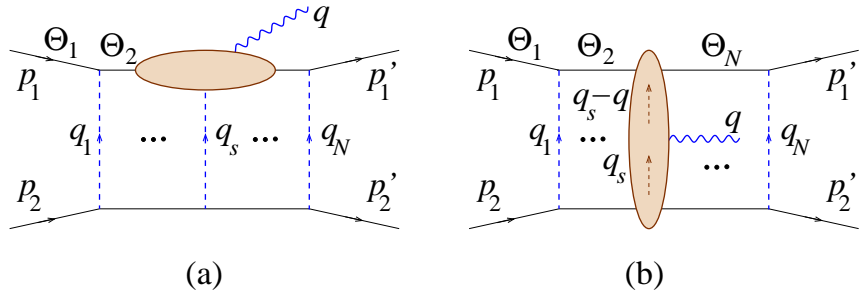}
  \caption{\it Picture and notation of generic exchange emission in (a) the soft
    and (b) the Regge limits.}
    \label{f:treeEmission}
\end{figure}
\begin{itemize}
\item[\bf a)] The regime $|\tht_s| > |\tht|$ (where $|\tht_s|=|\qt_s|/E$ is the
  single-hit scattering angle) is characterized by relatively small emission
  angles and sub-energies. If scattering is due to a single exchange at impact
  parameter $b$, then $|\tht_s| \sim\hbar/Eb\equiv\theta_m$ and $\vec{q}$ is
  nearly collinear to $\vec{p}_1$.  In that region the amplitude is well
  described by external-line insertions, but turns out to be suppressed because
  of helicity conservation zeroes.
\item[\bf b)] $|\tht| > |\tht_s| > \frac{\hbar\omega}{E} |\tht|$. In this regime
  the sub-energies reach the threshold of high-energy (Regge) behaviour, still
  remaining in the validity region of external line insertions, due to the
  condition $|\qt_s| = E|\tht_s| > \hbar\omega |\tht| =|\qt|$ which suppresses
  insertions on exchanged graviton lines.
\item[\bf c)] Finally, in the regime $|\tht_s| < \frac{\hbar\omega}{E} |\tht|$
  the soft approximation breaks down in favour of the (high-energy) H-diagram
  amplitude~\cite{ACV90} which contains internal-line insertions
  also~\cite{Li82}.
\end{itemize}

\subsection{Soft amplitudes in the Weinberg limit\label{s:sawl}}

In the soft regime (a$\cup$b), the emission amplitude $\amp_\soft^{(\lambda)}$
of a graviton with momentum $q$ and helicity (or polarization) $\lambda$ can be
expressed as the product of the elastic amplitude $\amp_\el=\kappa^2 s^2/\Qt^2$
and the external-line insertion factor
$J_W^{(\lambda)}\equiv J_W^{\mu\nu}\pol^{(\lambda)*}_{\mu\nu}$, where
$\pol^{(\lambda)}_{\mu\nu}$ is the polarization tensor of the emitted graviton
and $J_W^{\mu\nu}$ is the Weinberg current~\cite{We65} ($\eta_i=+1(-1)$ for
incoming (outgoing) lines)
\begin{equation}\label{Jw}
  J_W^{\mu\nu} = \kappa \sum_i \eta_i \frac{p_i^\mu p_i^\nu}{p_1\cdot q}
 = \kappa\left(\frac{p_1^\mu p_1^\nu}{p_1\cdot q}-\frac{p'_1{}^\mu p'_1{}^\nu}{p'_1\cdot q}
 + \frac{p_2^\mu p_2^\nu}{p_2\cdot q}-\frac{p'_2{}^\mu p'_2{}^\nu}{p'_2\cdot q}\right)
\end{equation}
and was already discussed in the planckian framework in~\cite{ACV90}.

We are interested in the projections of the Weinberg current over states of
definite positive/negative helicity, which can be conveniently defined by
\begin{equation}\label{polpm}
  \pol_{\pm}^{\mu\nu} = \frac{1}{\sqrt{2}} (\pol^{\mu\nu}_{TT}
  \pm \ui \pol^{\mu\nu}_{LT}) = \frac{1}{2} (\pol^\mu_T \pm \ui \pol^\mu_L)
  (\pol^\nu_T \pm \ui \pol^\nu_L) \; ,
\end{equation}
\begin{equation}\label{polLT}
\begin{split}
  & \pol^{\mu\nu}_{TT} = \frac{1}{\sqrt{2}} (\pol^\mu_T \pol^\nu_T - \pol^\mu_L \pol^\nu_L), \quad
  \pol^{\mu\nu}_{LT} = \frac{1}{\sqrt{2}} (\pol^\mu_L \pol^\nu_T + \pol^\mu_T \pol^\nu_L)\\
  & \pol^\mu_T = (0, -\e_{ij} \frac{q_j}{|\qt|}, 0), \quad \pol^\mu_L
  = (\frac{q^3}{|\qt|}, \textbf{0} , \frac{q^0}{|\qt|}) \mp \frac{q^\mu}{|\qt|} \; ,
\end{split}
\end{equation}
with $\e_{12} = 1$ and the $-$ and $+$ signs in $\pol^\mu_L$ corresponding to a
graviton emission in the forward and backward hemisphere respectively.

By referring, for definiteness, to the forward hemisphere, 
we define the momentum transfers $\qt_{1(2)} \equiv \pt_{1(2)} - \pt_{1(2)}'$,
$\qt = \qt_1 + \qt_2$, and the scattering angle $\qt_2 \equiv E \tht_s$, and
restrict ourselves to the forward region $|\tht|, |\tht_s| \ll1$. Giving for
ease of notation the results for a single helicity, a delicate but
straightforward calculation (app.~\ref{a:hlw}) leads to the following explicit
result in the c.m. frame with $\pt_1 = 0$:
\begin{equation}\label{Jw+-}
  J_{W-}(q^3>0;\tht,\tht_s) = \frac{J_W}{\sqrt{2}}  = \kappa\frac{E}{\hbar \omega} \;
  \left(\esp{2\ui(\phi_{\tht-\tht_s}-\phi_\tht)} - 1\right) \; ,
\end{equation}
leading to a factorized soft emission amplitude
\begin{equation}\label{fsa}
  \amp_{\soft}(\tht_s; E,\omega,\tht)
  = \amp_{\el}(E,\Qt) J_W\Big(\frac{E}{\hbar \omega},\tht,\tht_s\Big)
  = \kappa^3 s^2 \frac{1}{E \hbar \omega\tht_s^2}
  \left(\esp{2\ui(\phi_{\tht-\tht_s}-\phi_\tht)} - 1\right) \;,
\end{equation}
where $|\Qt|$ can be unambiguously identified with $|\qt_2|$ in the a) and b)
regions where eq.~(\ref{Jw+-}) is justified.

The simple expression~(\ref{Jw+-}) shows a $1/\omega$ dependence, but no
singularities at either $\tht=0$ or $\tht=\tht_s$ as we might have expected from
the $p_i\cdot q$ denominators occurring in~(\ref{Jw}). This is due to the
helicity conservation zeros in the physical projections of the tensor numerators
in~(\ref{Jw}). Therefore, there is no collinear enhancement of the amplitude in
region a) with respect to region b), while we expect sizeable corrections to it
in region c), where internal insertions are important. The helicity phase
transfer in eq.~\eqref{fsa} has a suggestive interpretation, made manifest
by introducing a ``$z$-representation'' (proven in app.~\ref{a:zrep})
\begin{equation}\label{zrep}
  \esp{2\ui \phi_{\tht}} - \esp{2\ui \phi_{\tht'}}
  = -2 \int \frac{\dif^2\zt}{{2\pi z^*}^2} \left( \esp{\ui A \zt \cdot \tht}
    - \esp{\ui A \zt \cdot \tht'} \right) \;,
 \qquad
 \begin{cases}
   z &= x + \ui y \\ \zt &= (x,y)
 \end{cases}
\end{equation}
as an integral between initial and final directions in the transverse $z$-plane
of the complex component of the Riemann tensor~\cite{GrVe14} in the AS metric of
the incident particles.

For our analysis we will need to work both in momentum and in impact parameter
space.  We define $\bt$-space amplitudes, following the normalization
convention%
\footnote{The customary helicity amplitudes with phase-space
  $\dif^3\vq/(\hbar^3 2\omega)$ are given by $\tfa(\bt,\qt)$
  ($\tfa(\bt,-\qt)^*$) for helicity $-\ (+)$ respectively.}  of~\cite{ACV07}
and~\cite{CCV15}, as:
\begin{equation}\label{defMb}
  \tfa(\bt) \equiv \frac1{(2\pi)^{3/2}}\int\frac{\dif^2 \qt_2}{(2\pi)^2}\;
  \esp{\ui\qt_2\cdot\bt}\frac1{4s}\amp(\qt_2) \;,
\end{equation}
so that, in the soft case, we have:
\begin{equation}\label{Msoft}
\begin{split}
  \tfa_\soft(\bt;E,\omega,\tht) & = \sqrt{\alpha_G}\frac{R}{\pi}\frac{E}{\hbar \omega} \int
  \frac{\dif^2 \tht_s}{2\pi\tht_s^2} \esp{\ui \frac{E}{\hbar} \tht_s \cdot \bt}
  \frac{1}{2} \left(\esp{2\ui(\phi_{\tht-\tht_s}-\phi_\tht)} - 1\right) \; .
\end{split}
\end{equation}

This definition is generalized to the backward hemisphere (jet 2) by setting
$E\tht_s=-\qt_1$ and $\qt_2=E\tht_s+\qt$, and by using the corresponding current
projections (app.~\ref{a:hlw})
\begin{equation}\label{jw2}
  J_{W-}(-q^3;-\tht,\tht_s) = J_{W-}(q^3;\tht,\tht_s)^* =J_{W+}(q^3;\tht,\tht_s)
\end{equation}
to obtain the helicity symmetry relation between backward and forward jets
\begin{equation}\label{Mb2}
  M_-(\bt;-q^3,-\qt) = M_+(\bt;q^3,\qt) \esp{-\ui\bt\cdot\qt} \;,
\end{equation}
where
\begin{align}
  M_+ &= \sqrt{\alpha_G}\frac{R}{\pi}\frac{E}{\hbar\omega}\int
    \frac{\dif^2\tht_s}{2\pi\tht_s^2}\;\esp{\ui\frac{E}{\hbar}\tht_s\cdot\bt}
    \frac12\left(\esp{-2\ui(\phi_{\tht-\tht_s}-\phi_\tht)}-1\right) \nonumber\\
    &= M_-(\bt;q^3,-\qt)^* \;.
\end{align}
Note the translational parameter $\esp{-\ui\bt\cdot\qt}$ (recalling that
particle 2 is located at $\xt=\bt$).

Let's examine the behaviour of eqs.~\eqref{fsa},\eqref{Msoft} in the various
regimes (we set $\hbar=1$ for simplicity in most of sec.~\ref{s:1gex}, except
when needed for physical understanding).  It is useful to write eq.~(\ref{fsa})
in complex notation ($\theta\equiv|\tht|\esp{\ui\phi_\tht}$, etc.) as
\begin{equation}\label{Jw+}
  \amp_\soft = \frac{\kappa^3 s^2}{E\omega}
  \left(\frac{\theta}{\theta_s}-\frac{\theta^*}{\theta_s^*}\right)
  \frac1{\theta(\theta-\theta_s)^*}\; ,
\end{equation}
which implies the approximate behaviour in the a) and b) regions
\begin{equation}\label{J+app}
  \frac{\amp_\soft}{\kappa^3 s^2} \simeq
  \begin{cases}
    \displaystyle \frac{2}{E \omega}\;\frac12\left(
      \esp{2\ui(\phi_{\tht_s}-\phi_\tht)}-1\right)\frac1{|\theta_s|^2}
    &\qquad (\theta\ll\theta_s \iff \text{region a})\\[3mm]
    \displaystyle \frac{2}{E \omega}\;\ui\sin(\phi_\tht-\phi_{\tht_s})
    \frac1{|\theta||\theta_s|} &\qquad 
    (\theta\gg\theta_s\iff \text{region b}) \;.
  \end{cases}
\end{equation}

The first behaviour is typical of the IR amplitude, showing no singularities in
the collinear ($\theta \to 0$) limit, and will be relevant for our final result
also.

The second behaviour in~\eqref{J+app}, after a simple integration in $\theta_s$
--- to which only the $\cos \phi_{\tht_s} \sin \phi_\tht$ term contributes ---
yields the result
\begin{align}\label{basicsoft}
  \tfa_\soft(\bt;E,\omega,\tht) & \Rightarrow \alpha_G\frac{R}{\pi}
  \frac{\sin\phi_\tht}{\omega b|\theta|} J_0(b\omega|\theta|)
  \qquad \left(\frac{E}{\omega}\theta_m \gg |\theta| \gg \theta_m\right)
\end{align}
(where for simplicity, we choose the $x$-axis in the transverse plane to be
aligned with $\bt$, that is $\phi_\tht \equiv \phi_\tht - \phi_\bt$), which
provides the most important term in the b) region. We note that its maximum at
$\phi_\tht = \pi/2$ is a reminder of the collinear zeroes.

As for region c), we already noticed that the soft evaluation breaks down there
in favour of the high-energy amplitude, and in sec.~\ref{s:msarl} we shall
substantially improve our $\bt$-space amplitude in all regions by matching the
soft and high-energy evaluations explicitly. The corresponding estimate, though
yielding subleading corrections in region b), will considerably change the
$b|\qt|>1$ behaviour of eq.~(\ref{basicsoft}).

\subsection{Amplitude transformation\label{s:at}}

In order to compute the emission in the general case we need to establish an
important point regarding the representation of the soft and Regge (see
sec.~\ref{s:msarl}) single-exchange amplitudes.  As already noted, the
expression~\eqref{fsa} is valid as it stands only if the initial direction (of
the momentum $\vp_1$) of the emitting particle is along the $z$-axis, or forms
with it a small angle $|\Tht_i|\ll\theta_m$, the single-hit large angle
threshold. But as shown in sec.~\ref{s:tes}, in the eikonal evolution the fast
particles scatter on average by the angle
$|\tht_s(\bt)|=2R/b=(Gs)\theta_m\gg\theta_m$, and thus we need to compute the
amplitudes in the case where the emission takes place with a generic incidence
angle $\Tht_i$, possibly much larger than $\theta_m$.

Because of Lorentz invariance, we expect the $\Tht_i$ dependence to occur
through rotation scalars (which, in the small-angle kinematics, involve the
differences $\Tht_i-\Tht_j$, the latter being angular 2-vectors of the fast
particles), and a specific transformation phase also. The latter is in turn
dependent on the definition of the helicity states $\ket{\lambda,q,\cdots}$
which is not uniquely determined. Since $\lambda$ is a Lorentz invariant for the
massless graviton, such transformation phase is only allowed by the ambiguity in
relating the $q$-states to the $z$-axis state, due to the residual rotational
invariance around the latter.  For instance, Jacob and Wick
(JW)~\cite{Jacob:1959at} relate $q$ to $z$ by a standard rotation around the
axis perpendicular to the $\bk{q,z}$ plane. Since such definition is fully
'body-fixed' (that is, independent of external observables) we expect the JW
amplitudes to be invariant for small rotations of the $z$-axis around an axis
perpendicular to it, because in such a limit the rotations involved will
commute.%
\footnote{Our amplitude is invariant under O(1) rotations around the z-axis,
  which can be separately considered.}
On the other hand, our helicity states are defined in terms of the physical
polarizations in eqs.~\eqref{polpm} and~\eqref{polLT}, which are dependent not
only on $q^\mu$, but also on the $z$-axis, which occurs as external variable in
the $T$ projection. We show in app.~\ref{a:heltras} that the ensuing relation to
JW amplitudes is a simple multiplication by $\exp(\ui\lambda\phi_\tht)$, where
the azimuthal variable is generally $\ord{1}$. Since the $\Tht_i$ rotation acts
on $\tht$ as the translation $\tht-\Tht_i$ for small polar angles, we expect the
transformation phase to be nontrivial and given by
$\exp[\ui\lambda(\phi_\tht-\phi_{\tht-\Tht_i})]$, as fully proved in
app.~\ref{a:heltras} and explicitly checked in app.~\ref{a:hlw}.

Therefore, in the forward region $\Tht_i,\Tht_f,\tht\ll 1$ --- $\Tht_f$ being
the outgoing direction of the (intermediate) fast particle --- the
momentum-space helicity amplitudes transform as (app.~\ref{a:heltras})
\begin{equation}\label{ampTrasf}
  \amp^{(\Tht_i)}(\Tht_f,\tht) = \esp{\ui \lambda(\phi_\tht-\phi_{\tht-\Tht_i})}
  \amp^{(\bs{0})}(\Tht_f-\Tht_i,\tht-\Tht_i) \;,
\end{equation}
where $\lambda$ is the helicity of the emitted graviton, $\lambda=-2$ in our case.
In $\bt$ space, i.e., by Fourier transforming w.r.t.\ $\Qt=E(\Tht_f-\Tht_i)$,
one finds
\begin{equation}\label{tfaTrasf}
  \tfa^{(\Tht_i)}(\bt,\tht) = \esp{\ui \lambda(\phi_\tht-\phi_{\tht-\Tht_i})}
  \tfa^{(\bs{0})}(\bt,\tht-\Tht_i) \;.
\end{equation}

\begin{figure}[t]
  \centering
  \includegraphics[width=0.35\textwidth]{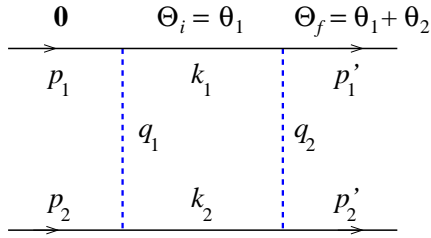}
  \caption{\it Double-exchange process. The angles describing the direction of the
    right-moving fast particle during the scattering process are shown on top of
    the upper line.}
  \label{f:doubleExchange}
\end{figure}

The expression~\eqref{ampTrasf} is easily argued for as a consistency requirement
for the insertion of the Weinberg current on the double-exchange process
(fig.~\ref{f:doubleExchange}) in the soft limit.
In fact, by the identity of Weinberg contributions
$J_p-J_{p'}=(J_p-J_k)+(J_k-J_{p'})$ we must have
\begin{equation}\label{idWc}
  J^{(\pm)}_W(\Tht_i,\Tht_f) = J^{(\pm)}_W(\bs{0},\Tht_f) -
  J^{(\pm)}_W(\bs{0},\Tht_i) \;,
\end{equation}
which agrees with the direct calculation of app.~\ref{a:hlw} as well,
from which it follows that (note that $\tht_s \simeq \Tht_f - \Tht_i$ for the
soft amplitude)
\begin{align}
  \amp_\soft^{(\Tht_i)}
  &= \kappa^3 s^2 \frac{2}{E\omega\,\tht_{s}^2}\,
  \frac12\left(
     \esp{2\ui(\phi_{\tht-\Tht_f}-\phi_\tht)}
    -\esp{2\ui(\phi_{\tht-\Tht_i}-\phi_\tht)} \right)\nonumber\\
 &= \kappa^3 s^2 \frac{2}{E\omega} \,
  \frac{\esp{2\ui(\phi_{\tht-\Tht_i}-\phi_\tht)}}{\tht_{s}^2}\, \frac12
  \left(\esp{2\ui(\phi_{\tht-\Tht_f}-\phi_{\tht-\Tht_i})} - 1 \right)
 \;. \label{MSi}
\end{align}
exactly as predicted by eq.~\eqref{ampTrasf} with $\lambda=-2$. Eq.~\eqref{MSi}
keeps the suggestive interpretation of helicity charge transfer from initial to
final state in the general case.

\subsection{Matching of soft amplitude with the Regge-limit\label{s:msarl}}

As soon as the rapidity interval $2Y_b=2\log(Eb/\hbar)$ between $p'_1$ and
$p'_2$, and the relative rapidity $Y_b-y$ between $p'_1$ and $q$ become large,
high-energy emission in the Regge limit becomes relevant, as predicted by the
Lipatov vertex~\cite{Li82} and the H-diagram~\cite{ACV90} (see
fig.~\ref{f:metricFluct}.a). More precisely, for scattering due to single
graviton exchange, the Regge limit is relevant in the region where the graviton
is emitted at a relatively large angle,
$1\gg|\theta|\gg\theta_m\equiv\hbar/(Eb)$ (as already noted, $|\theta_s| \sim
\theta_m$ in the single hit case). This large-angle region comprises regions b)
and c) discussed at the beginning of sec.~\ref{s:1gex} and is particularly
relevant for region c) in which internal line insertions are important.

\begin{figure}[t]
  \centering
  \includegraphics[width=0.26\textwidth]{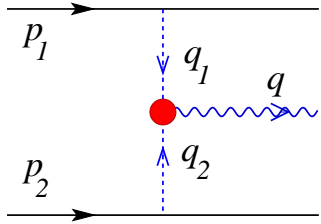}
  \caption{\it Diagrammatic picture for the emission of a graviton from two
    fast-particle scattering in the Regge limit. The blob in the middle
    represents the Lipatov vertex, i.e., the effective reggeon-reggeon-graviton
    coupling.}
  \label{f:Hamplitude}
\end{figure}

Using the same kinematic notation as in sec.~\ref{s:sawl}, the Lipatov current
is given by~\cite{Li82} (see also~\cite{ACV90}):
\begin{equation}\label{lipatov}
\begin{split}
  J_L^{\mu\nu} = & \frac{\kappa}{2}(J^\mu J^\nu - |\qt_{1\perp}|^2
  |\qt_{2\perp}|^2 j^\mu j^\nu) \quad \text{with}\\
  j^\mu \equiv & \frac{p^\mu_1}{p_1 \cdot q} - \frac{p^\mu_1}{p_2 \cdot q}, \\
  J^\mu \equiv & |\qt_{1\perp}|^2\frac{p^\mu_1}{p_1 \cdot q} -
  |\qt_{2\perp}|^2\frac{p^\mu_2}{p_2 \cdot q} + q_1^\mu - q_2^\mu - |\qt_\perp|^2 j^\mu \; ,
\end{split}
\end{equation}
where $\qt_{\perp1}$, $\qt_{\perp2}$, $\qt_{\perp}$ denote transverse
(vectorial) components to the $\vec{p}_1$ direction (which of course coincide
with $\qt_1$, $\qt_2$ and $\qt$ when $\pt_1 = \Tht_i = 0$) and the corresponding
graviton emission amplitude (considering again a single helicity for
definiteness) is
\begin{equation}
  \amp_\regge = \frac{\kappa^2 s^2}{|\qt_{1\perp}|^2 |\qt_{2\perp}|^2}
  J_L^{\mu\nu} \pol_{\mu \nu}^{(-)} \; ;
\end{equation}
note that $-|\qt_{1(2)\perp}|^2$ corresponds to the virtuality $q_{1(2)}^\mu
q_{1(2)\mu}$ in the Regge kinematics.

More quantitatively, in the c.m. frame with zero incidence angle
($\pt_1 = E \Tht_i = 0$) and in the forward region $|\tht|,|\tht_s| \ll 1$, the
amplitude takes the form~\cite{ACV07} (see also app.~\ref{a:hlw})
\begin{equation}\label{ampRegge}
  \amp_\regge(\tht;E,\omega,\tht_s) = \kappa^3 s^2
  \frac{1-\esp{-2\ui(\phi_{\qt_2}-\phi_{\qt - \qt_2})}}{\qt^2}
\end{equation}
(remember that $\qt = \omega \tht$ and $\qt_2 = E \tht_s$). The corresponding
amplitude in $\bt$ space, according to the definition~(\ref{defMb}), is given by
\begin{align}
  \tfa_\regge(\bt;E,\omega,\tht) & = \sqrt{\alpha_G}\frac{R}{\pi} \int
  \frac{\dif^2 q_2}{2 \pi |\qt|^2} \; \esp{\ui \qt_2 \cdot \bt} \frac{1}{2}
  \left( 1 - \esp{2\ui (\phi_{\qt_2-\qt}-\phi_{\qt_2})} \right) \nonumber\\
  & \equiv \sqrt{G}s\frac{R}{2} \tilde{h}(\bt, \qt)
  \equiv \sqrt{G}s\frac{R}{2} \bt^2 \int \dif^2 \zt \;
  \esp{\ui |\bt| \omega \zt \cdot \tht} h(\bt,|\bt|\zt) \; ,\label{Mregge}
\end{align}
where $\tilde{h}(\bt, \qt)$ admits the integral representation
($b\equiv b^1+\ui b^2\in\C$)
\begin{equation}
\tilde{h}(\bt,\qt) = \frac{2}{\pi} \int_0^\infty\dif\eta\;\eta\esp{-\eta} \left[
   \frac1{b\omega\theta^*(b\omega\theta^*-2\ui\eta)}+
   \frac{\esp{\ui\omega(b^*\theta+b\theta^*)/2}}{b^*\omega\theta(b^*\omega\theta+2\ui\eta)}\right] \; ,
\label{Mreggeeta}
\end{equation}
and turns out to be equal to the Fourier transform w.r.t.\ $\qt$ of the
$H$-diagram field. The latter's expression in the space of the transverse coordinate
$\xt=(x^1,x^2)$ of the emitted graviton is
\begin{equation}\label{hbx}
  h(\bt,\xt) = \frac{1-\esp{2\ui(\phi_\xt-\phi_{\xt-\bt})}}{2\pi^2 b^2}
  = \frac{x b^* - x^* b}{2\pi^2 |b|^2 x^* (x-b)} \;,
  \qquad (x\equiv x^1+\ui x^2\in\C) \;.
\end{equation}
(eqs.~\eqref{Mreggeeta} and~\eqref{hbx} are proven in app.~\ref{a:hf}).

As is the case for the soft current~\eqref{MSi}, \eqref{Mregge} is valid as it
stands only if the initial $p_1$ direction is along the $z$-axis. However, for a
generic $p_1$ direction, the amplitude in the Regge limit transforms in the same
way as the soft one in eq.~\eqref{tfaTrasf}, that is
\begin{equation}\label{Mreggetr}
  \tfa_\regge^{(\Tht_i)}(\bt;E,\omega,\tht)
  = \esp{2\ui(\phi_{\tht-\Tht_i}-\phi_\tht)}
  \tfa_\regge(\bt;E,\omega,\tht-\Tht_i)
\end{equation}
(where $E \Tht_i$ is the transverse part of the 4-momentum $p_1$), as directly proven
in app.~\ref{a:hlw}.

To connect the small-angle (soft) and large-angle (Regge) regimes of the one
graviton emission amplitude, it is convenient to rewrite eq.~(\ref{Mregge}) in
terms of the (complex) variables $\theta=q/\omega$ and $\theta_s=q_2/E$:
\begin{equation}\label{Mregge1}
  \left.\tfa_\regge(\bt;E,\omega,\tht)\right|
  = \sqrt{\alpha_G}\frac{R}{\pi}\,\frac{E}{\omega} \int
  \frac{\dif^2\tht_s}{2\pi|\theta|^2} \; \esp{\ui E\bt\cdot\tht_s} \frac12\,
  \frac{\theta\theta_s^*-\theta^*\theta_s}{\theta_s
    \left(\theta_s-\frac{\omega}{E}\theta\right)^*} \;.
\end{equation}
This expression differs from eq.~(\ref{Msoft}) by the replacement
\begin{equation}\label{Mrepl}
  \frac1{|\theta_s|^2} \frac{\theta\theta_s^*-\theta^*\theta_s}{\theta
      \left(\theta-\theta_s\right)^*} \to
  \frac1{|\theta|^2}   \frac{\theta\theta_s^*-\theta^*\theta_s}{\theta_s
    \left(\theta_s-\frac{\omega}{E}\theta\right)^*}
\end{equation}
in the integrand of eq.~(\ref{Mregge1}). By inspection, we see the important point
that Regge and soft evaluations agree in the region b), in which the condition
$|\theta|\gg|\theta_s|\gg\frac{\omega}{E}|\theta|$ insures that we are in the
``large-angle'' regime in the l.h.s.\ with negligible internal insertions in
the r.h.s., while eq.~(\ref{Mregge1}) remains the only acceptable expression in
region c), where $|\theta_s|<\frac{\omega}{E}|\theta|$.

Therefore, in order to get a reliable emission amplitude holding in all regions
($\mathrm{a}\cup\mathrm{b}\cup\mathrm{c}$), we have to match the soft with the
Regge evaluations. We start from the Fourier transform in eq.~(\ref{Msoft}) and
we then add the difference of Regge and soft evaluations of eq.~(\ref{Mrepl}) in
region c) and in part of region b), the border being parametrized by the cutoff
$\Delta_c>1$ (see fig.~\ref{f:regions}). Such difference has the form
\begin{align}
  \Delta\tfa &\equiv
  \left[\tfa_\regge-\tfa_\soft\right]_{\mathrm{c}\,\cup\,\text{(part of b)}} \nonumber \\
  &= \sqrt{\alpha_G}\frac{R}{\pi}\,\frac{E}{\omega}\frac12
  \int_0^{\Delta_c\frac{\omega}{E}|\theta|}
  \frac{\dif^2\tht_s}{2\pi|\theta_s|^2}\;
  \frac{\theta\theta_s^*-\theta^*\theta_s}{|\theta|^2}
  \left[\frac1{1-\frac{\omega}{E}
      \frac{\theta^*}{\theta_s^*}} - \frac1{1-\frac{\theta_s^*}{\theta^*}}
    \right] \esp{\ui E \bt\cdot\tht_s} \;,\label{cdiff}
\end{align}
where we require
$\left|\frac{E\theta_s}{\omega\theta}\right|=\ord{1}$, with
\begin{equation}\label{Deltac}
  \left|\frac{E\theta_s}{\omega\theta}\right| \equiv
  \left|\frac{\tilde{\theta}_s}{\theta}\right| < \Delta_c \;, \qquad
  \left|\frac{\theta_s}{\theta}\right| < \frac{\omega}{E}\Delta_c \ll 1
\end{equation}
so that we get the expression
\begin{equation}\label{rdiff}
  \Delta\tfa \simeq \sqrt{\alpha_G}\frac{R}{\pi}\,\frac12\int_0^{|\theta|\Delta_c}
  \frac{\dif^2\tilde{\tht}_s}{2\pi}\;
  \frac{\theta\tts^*-\theta^*\tts}{|\theta|^2 |\tts|^2} \left(
  \frac1{1-\frac{\theta^*}{\tts^*}}-1\right) \esp{\ui\omega\bt\cdot\bs{\tts}}\;.
\end{equation}
If we then choose $1\ll\Delta_c\ll E/\omega$ the result~(\ref{rdiff}) is weakly
cutoff dependent and, in the $\Delta_c\to\infty$ limit, is
formally equal to the negative of the Fourier transform of the soft amplitude on
the whole phase space, rescaled at $E=\omega$ or, in other words,
\begin{equation}\label{DMc1}
  \Delta\tfa \xrightarrow{\Delta_c\gg1} -\sqrt{\alpha_G}\frac{R}{\pi}
  \int\frac{\dif^2\tilde\tht_s}{2\pi|\tts|^2} \;
  \esp{\ui\omega\bt\cdot\tilde{\tht}_s} \frac12
  \frac{\theta\tts^*-\theta^*\tts}{\theta(\theta-\tts)^*}
   =-\tfa_\soft(\bt;\omega,\omega,\tht)\;.
\end{equation}
By then using eq.~\eqref{Msoft} we obtain the explicit form of the matched
amplitude (with explicit $\hbar$-dependence)
\begin{align}
  \tfa_\match &\simeq \tfa_\soft(\bt;E,\omega,\theta) -
  \tfa_\soft(\bt;\hbar\omega,\omega,\theta) \nonumber \\
  &= \sqrt{\alpha_G}\frac{R}{\pi} \int\frac{\dif^2\tht_s}{2\pi\tht_s^2} \left(
    \esp{\ui \frac{E}{\hbar}\bt\cdot\tht_s} \frac{E}{\hbar\omega}
    - \esp{\ui\omega\bt\cdot\tht_s} \right) \frac{1}{2}
  \left(\esp{2\ui(\phi_{\tht-\tht_s}-\phi_{\tht})} -1\right) \nonumber \\
  &= \sqrt{\alpha_G}\frac{R}{\pi} \esp{-2\ui\phi_{\tht}} \int
  \frac{\dif^2 \zt}{2\pi {z^*}^2} \esp{\ui b\omega\zt\cdot\tht} \left(
    \frac{E}{\hbar\omega} \log\left| \hat{\bt} -\frac{\hbar\omega}{E}\zt \right|
    - \log\left| \hat{\bt} - \zt \right| \right)  \;,
  \label{Mmatch}
\end{align}
where we have used the $z$-representation of the helicity phases~\eqref{zrep},
by rescaling the $\zt$-variable in the first term.

The final result of eq.~\eqref{Mmatch} --- derived on the basis of the
  soft-insertion formulas --- is expressed in terms of the ($\omega$-dependent)
  ``soft'' field
\begin{equation}\label{PhiRdef}
  h_s(\omega,z) \equiv \frac{1}{\pi^2 {z^*}^2} \left(
    \frac{E}{\hbar\omega} \log\left| \hat{\bt} -\frac{\hbar\omega}{E}\zt \right|
    - \log\left| \hat{\bt} - \zt \right| \right)
  \equiv - \frac{\Phi_R(\omega,\zt)}{\pi^2 {z^*}^2}
\end{equation}
in which the function $\Phi_R$ turns out to be useful for the treatment of
rescattering too (sec.~\ref{s:rescattering}). Furthermore, for relatively large
angles ($\theta \gg \theta_m \sim \hbar/(E b)$), eq.~\eqref{Mmatch} involves
values of $\hbar\omega|z|/E\lesssim\theta_m/\theta$ which are uniformly small,
and the expressions~\eqref{PhiRdef} can be replaced by their $\omega\to 0$
limits
\begin{equation}\label{Phidef}
  h_s(z) = -\frac{\Phi(\zt)}{\pi^2 {z^*}^2} \;, \qquad
  \Phi(\zt) \equiv \hat{\bt}\cdot\zt + \log\left|\hat{\bt}-\zt\right| \;.
\end{equation}
The latter quantities have a classical meaning, $h_s(z)$ as a metric component
(sec.~\ref{s:rmt}) and $\Phi(\zt)$ as modulation function in the classical
treatment of ref.~\cite{GrVe14}. As a consequence, eq.~\eqref{Mmatch} takes the
simpler form
\begin{equation}
  \tfa_\match \equiv \tfa = \sqrt{\alpha_G}\frac{R}{2} \esp{-2\ui \phi_\tht} \int
  \dif^2 \zt \; \esp{\ui b \omega \zt \cdot \tht}  h_s(z)
  \label{hsoft}
\end{equation}
that will be mostly used in the following. Replacing $h_s(z)$ by its
$\omega$-dependent form is needed if we want to treat the very-small angle
region and some quantum corrections also.

From eq.~\eqref{hsoft} we can see directly how the matching works. In fact, due
to eq.~\eqref{Mmatch}, the linear term (the log term) in eq.~\eqref{Phidef} is
in correspondence with external (internal) insertions of the emission current.
In region a), where $\zt$ is pretty large, the linear term dominates and
provides directly the soft limit. In region b), the basic soft
behaviour~\eqref{basicsoft} is reproduced but, with increasing values of $b
|\qt|$, it is actually canceled by internal insertions in region c), because in
the small $|\zt|$ limit the function $\Phi(\zt)$ is of order $\sim
\ord{|\zt|^2}$. This is confirmed by the Regge representation~\eqref{Mreggeeta}
which shows, by inspection, a $1/(b|\qt|)^2$ behaviour for $b|\qt| \gg 1$.

To summarize, our matched amplitude~\eqref{hsoft} which, by construction, should
be identical to the Regge one of eq.~\eqref{Mregge} in region c), is also a nice
interpolation in (b$\cup$c) and part of a) with $|\tht|>\theta_m$.%
\footnote{The moderate-angle restriction will become unimportant when the
  resummation of sec.~\ref{s:bfact} will extend the collinear region up to
  $\Theta_s\sim R/b \gg\theta_m$.}
For this reason we shall call
eq.~\eqref{hsoft} (eq.~\eqref{Mregge}) the soft-based (Regge-based)
representation of the same unified amplitude. Their identity can be directly
proven by the equation
\begin{equation}\label{hparts}
  \int\dif^2\zt\;\left[ \esp{-2\ui \phi_\tht} h_s(z) - b^2 h(\bt, b\zt) \right]
  \esp{\ui b \omega \zt \cdot \tht} = 0 \; ,
\end{equation}
which can be explicitly checked by switching to 
$z,\,z^*$ variables and integrating by parts. Eq.~\eqref{hparts}
is in turn a direct consequence of the differential identity
\begin{equation}\label{transv}
\frac{\partial}{\partial z} h_s 
-\frac{\partial}{\partial z^*} b^2 h 
 = 0 \;.
\end{equation}
that will be related in sec.~\ref{s:rmt} to a transversality condition of the
radiative metric tensor.

Our unified soft-Regge amplitude $\tfa$ has then, for a generic $\Tht_i$, the form
\begin{align}
  \tfa^{(\Tht_i)}(\bt,\tht) &= 
  \sqrt{\alpha_G}\frac{R}{2} \esp{-2\ui(\phi_\tht-\phi_{\tht-\Tht_i})} \int
  \dif^2 \xt \; \esp{\ui \omega \xt \cdot (\tht-\Tht_i)}  h(\bt,\xt)
  \nonumber \\
  &= \sqrt{\alpha_G}\frac{R}{2} \esp{-2\ui\phi_\tht} \int \dif^2 \zt \;
  \esp{\ui b \omega \zt \cdot (\tht-\Tht_i)}  h_s(z) \; , \label{Mrot}
\end{align}
where the Regge-based (soft-based) representation is used in the first (second)
line. Eqs.~\eqref{Mmatch} and \eqref{PhiRdef} provide an improved small-angle
description and some quantum corrections.

\subsection{Radiative metric tensor\label{s:rmt}}

\begin{figure}[t]
  \centering
  \includegraphics[width=0.75\linewidth]{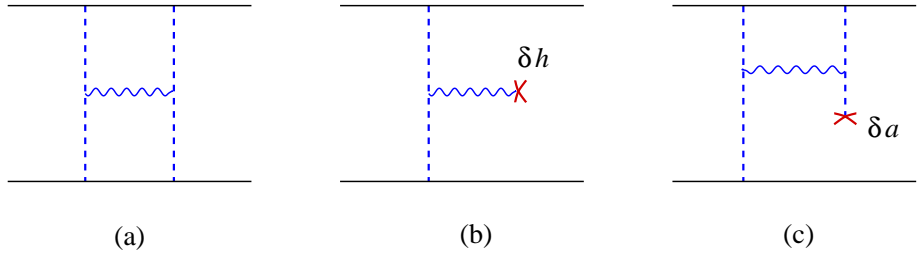}
  \caption{\it (a) The H diagram, yielding the first subleading semiclassical
    correction to the elastic amplitude. (b,c) Off-shell diagrams contributing to
    the metric fluctuations of the $h$ and the $a$ fields.}
  \label{f:metricFluct}
\end{figure}

To complete the picture of single-exchange radiation, we recall the parallel
calculation of radiative corrections to the metric fields and to the effective
action~\cite{ACV93,ACV07}. At first subleading level this amounts to calculating
the H-diagram fields $\delta h$ and $\delta a$ (fig.~\ref{f:metricFluct}.b,c)
occurring in the metric. By leaving aside time-delays~\cite{CC14} we
obtain~\cite{ACV07}
\begin{align}
  \dif s^2-\eta_{\mu\nu}\dif x^\mu \dif x^\nu &= 2\pi R \left[ a(\xt)\delta(x^-)
    \dif x^{-2}+\bar{a}(\xt)\delta(x^+)\dif x^{+2}\right] \nonumber \\
  &\quad +2(\pi R)^2\; \Re\left[ (\hat{\pol}_{\mu\nu}^{TT}
    -\ui\hat{\pol}_{\mu\nu}^{LT}) h(\xt)\Theta(x^+ x^-)\right]
  \dif x^\mu \dif x^\nu \;, \label{dsq}
\end{align}
where, starting from $a_0$ in eq.~\eqref{hpp} we expand the profile function
$a(x)$, to first order in $R^2/b^2$, and the field $h(x)$ in the form
(fig.~\ref{f:metricFluct})
\begin{subequations}\label{fieldExpn}
  \begin{align}
    a(\xt) &= a^{(0)} + \frac{R^2}{b^2}a^{(1)} + \cdots \;, & \bar{a}(\xt) &=
    a(\bt-\xt) \;, \qquad(\xt = b\zt) \\
    |\partial|^2 a^{(1)} &= \frac1{\pi}\frac1{|z|^4} 2\Phi(\zt) \;, &
    h(\xt) &= \frac{x-x^*}{2\pi^2 b x^* (x-b)} \;,
  \end{align}
\end{subequations}
and $\hat{\pol}_{\mu\nu}$ are polarization tensor operators of the form, for
instance,%
\footnote{With this prescription, the metric tensor~\eqref{dsq} satisfies the
  transversality condition $\partial^\mu h_{\mu\nu}=0$. The polarization tensors
  differ from those of~\cite{ACV07} by a factor of 1/2.}
\begin{align}
  2\hat{\pol}_{ij}^{TT} &=
  \e_{il}\e_{jm}\frac{\nabla_l\nabla_m}{|\nabla|^2}
  = \delta_{ij}-\frac{\nabla_i\nabla_j}{|\nabla|^2} \;, \\
  2\hat{\pol}_{++}^{TT} &= -\frac{\partial_+}{4\partial_-}\;, \qquad
  2\hat{\pol}_{--}^{TT} = -\frac{\partial_-}{4\partial_+} \;, \qquad
  2\hat{\pol}_{+-}^{TT} = \frac14
  \label{poltens}
\end{align}
with similar ones of the $LT$ polarization.

Such results follow from a shock-wave solution of the effective action equations
of motion which expresses all metric components in terms of the basic scalar
field $h(\xt)\equiv 4|\partial|^2\phi(\xt)$, where the explicit, single-valued
form of $\phi(\xt)$ --- not to be confused with the modulating function
$\Phi(\zt)$ --- was found in~\cite{ACV93} and is given by the single-valued
function ($x\equiv x^1+\ui x^2=bz$)
\begin{equation}\label{phimetric}
\begin{split}
  & \qquad \phi(x,x^*) = \phi(b-x^*, b-x) = \frac{1}{8\pi} \times \\
  & \left[ \log \frac{x^*}{b} \log\left(1-\frac{x}{b}\right) + \frac{x}{b}
  \log\frac{x^*}{b} + \left(1-\frac{x^*}{b}\right)\log\left(1-\frac{x}{b}\right)
  + f\left(\frac{x}{b}\right) + f\left(1-\frac{x^*}{b}\right)\right] \; ,
\end{split}
\end{equation}
where $f'(x) = \frac{x}{x-1} \log x$ is devised so as to cancel the
discontinuities at the $x=0$ and $x=b$ singularities.

From eqs.~\eqref{phimetric} and~\eqref{dsq}, we obtain, in particular, the
transverse plane metric components
\begin{equation}\label{hzz}
\begin{split}
  & h_{xx^*} = |\partial|^2 \text{Re}\, \phi = \frac{1}{8}\left(h(x) + h(b-x)\right) =
  \frac{1}{4}\text{Re}\, h(x) \;, \\
  & h_{x^*x^*}=-\partial^{*2} \text{Re}\, \phi
  = -\frac{1}{8}\left(h_s(z)+h_s(1-z)\right) = h^*_{xx} \;,
\end{split}
\end{equation}
which are closely related to the fields $h$ and $h_s$ introduced previously,
because of the derivatives
\begin{equation}\label{derivatives}
\left\{ \begin{aligned}
    4 \,\partial^{*2} \phi & = -\frac{1}{\pi^2} \frac{1}{z^{*2}} \Phi(\zt)
    = h_s(z) = \frac{\partial^*}{\partial} h(x) \\
    4 \,\partial^2 \phi & = -\frac{1}{\pi^2} \frac{1}{(1-z)^2} \Phi(\hat\bt-\zt)
    = h_s^*(1-z) = \frac{\partial}{\partial^*} h(x)
\end{aligned} \right.
\end{equation}

We note some important points. First, the $x \leftrightarrow b-x$ symmetry of
the metric is realized in the emission amplitude by exchanging jet 1 (or
$h_s(z)$) and jet 2 (or $h_s^*(1-z)$). Furthermore, the relationships of
$h_s(z)$ ($h_s^*(1-z)$) with $h(x)$ ($h^*(b-x)$) in jet 1 (jet 2), already given
in eq.~\eqref{transv} for jet 1, express the transversality conditions of the
metric components
\begin{equation}
  \partial^* h_{xx^*} + \partial h_{x^*x^*} = 0 \quad
  \left( \partial h_{xx^*} + \partial^* h_{xx} = 0 \right)
\end{equation}
and are thus rooted in the spin 2 structure of the interaction. Furthermore, in
parallel with the soft-based representation~\eqref{hsoft} for jet 1, we have by
\eqref{derivatives} the corresponding representation of the same amplitude in
jet 2
\begin{equation}\label{soft2}
  \tfa=\sqrt{\alpha_G}\frac{R}{2}\esp{2\ui\phi_\tht}\int\dif^2\zt\;
  \esp{\ui b\omega\zt\cdot\tht} h_s^*(1-z) \;.
\end{equation}
Note however that, while $h(\xt)\sim h_{xx^*}$ is a scalar, the soft fields
$h_s(z)$ ($h_s^*(1-z)$) in jet 1 (jet 2) have $J_z = 2 (-2)$ for rotations in
the transverse plane, and in fact the opposite phase is factorized in
eq.~\eqref{hsoft} (eq.~\eqref{soft2}) which are relevant for jet 1 (jet 2) and
helicity $\lambda=-2$. In all cases the resulting amplitude will come out
invariant for rotations around the $z$-axis.

Finally, by starting from the soft-based representations just mentioned at
$\Tht_i=0$ we can complete the symmetry $\xt\to\bt-\xt$ by constructing the
helicity amplitudes for both jets at $\Tht_i\neq0$, with the result
\begin{equation}\label{M12}
  \tfa_\lambda^{(-\Tht_i)}(\bt;-q^3,-\qt) = 
  \tfa_{-\lambda}^{(\Tht_i)}(\bt;q^3,\qt) \esp{-\ui\bt\cdot\qt_\perp} \;,
\end{equation}
which relates opposite helicities with opposite 3-momenta $\vq$ as function of
$\qt_\perp=\qt-\omega\Tht_i$, transverse to the $\Tht_i$
direction. The factor $\esp{-\ui\bt\cdot\qt_\perp}$ insures the translation
$\xt\to\bt-\xt$.  Eq.~\eqref{M12} can be checked in a straightforward way by
using the explicit helicity projections of app.~\ref{a:hlw}, and is anyway a
consequence of the helicity transformation properties of the soft fields in
either jet.

We then conclude that the radiative metric tensor of eq.~\eqref{dsq} (based on
the shock-wave solution of~\cite{ACV93}) is consistent with the present
single-exchange soft-Regge amplitudes and actually explains the unifying
relationships by a transversality condition of the metric tensor. We shall see
however that taking into account the helicity transformation phases is essential
for completing the calculation of graviton emission, and superimposing
single-exchange terms all along eikonal scattering.

\section{Factorization and resummation\label{s:bfact}}

\subsection{$\bs{b}$-factorization and matching\label{s:bfcrm}}

So far we have considered the radiation associated to the single-graviton
exchange contribution to the basic planckian scattering process. But we know
(sec.~\ref{s:tes}) that such high-energy process is described by the eikonal
resummation of a large number $\sim Gs/\hbar$ of single hits, so that --- for a
given impact parameter value $b$ --- the scattering angle increases from
$\theta_m=\hbar/(Eb)$ to $|\theta_s|=2(Gs/\hbar)(\hbar/Eb)=2R/b$, the Einstein
deflection angle. This fact considerably enlarges the quasi-collinear region
w.r.t.\ $\theta_s$, which might be an important source of energy loss by
radiation, so as to endanger energy-conservation~\cite{Rychkov_pc} unless
explicitly enforced.

We start noticing that, despite the enlargement of the quasi-collinear region,
the external-line insertion amplitude~(\ref{fsa}) stays unchanged, being only
dependent on the overall momentum transfer $\Qt=E\tht_s$ of the
process. Therefore, the damping of the collinear region and the large-angle
behaviour are both built in in eq.~(\ref{Jw+}) for any values of $\theta_s$, as
shown by eq.~(\ref{J+app}).

One may wonder to what extent --- or, for which $\omega$-values --- is the
external-line insertion method tenable. There are two types of internal-line
insertions: those on the fast, nearly on-shell particle lines of the eikonal
iteration, and those on the exchanged-graviton lines. We will now argue that
both kinds of internal insertions can be taken into account using the soft/Regge
matching strategy described in sec.~\ref{s:msarl}.

The first thing to notice is that fast-particle insertions are in fact already
implicitly included in the soft approximation~\eqref{fsa}. In fact, in a general
$n$-exchange contribution to eikonal scattering for each pair of propagating
lines there are two pairs of insertions, one with the mass-shell on the right
(final) and one on the left (initial), whose currents are nearly equal (to order
$\hbar\omega/E$) and opposite in sign. Thus, the purely soft emission can be
written in two equivalent ways, one as in eq.~\eqref{fsa} as a purely external
line insertion, and the other as a sum of $n$ contributions, where the insertion
--- still of the form \eqref{fsa} --- is made \emph{internally} on the fast
lines which surround the $i$-th exchanged graviton. Since in general these fast
lines will have accumulated a non-negligible transverse momentum, the $i$-th
internal insertion will be of the rotated form~\eqref{MSi}, with
$\Tht_i = \sum_{j=1}^{i-1} \tht_j$.

\begin{figure}[t]
  \centering
  \includegraphics[width=0.7\textwidth]{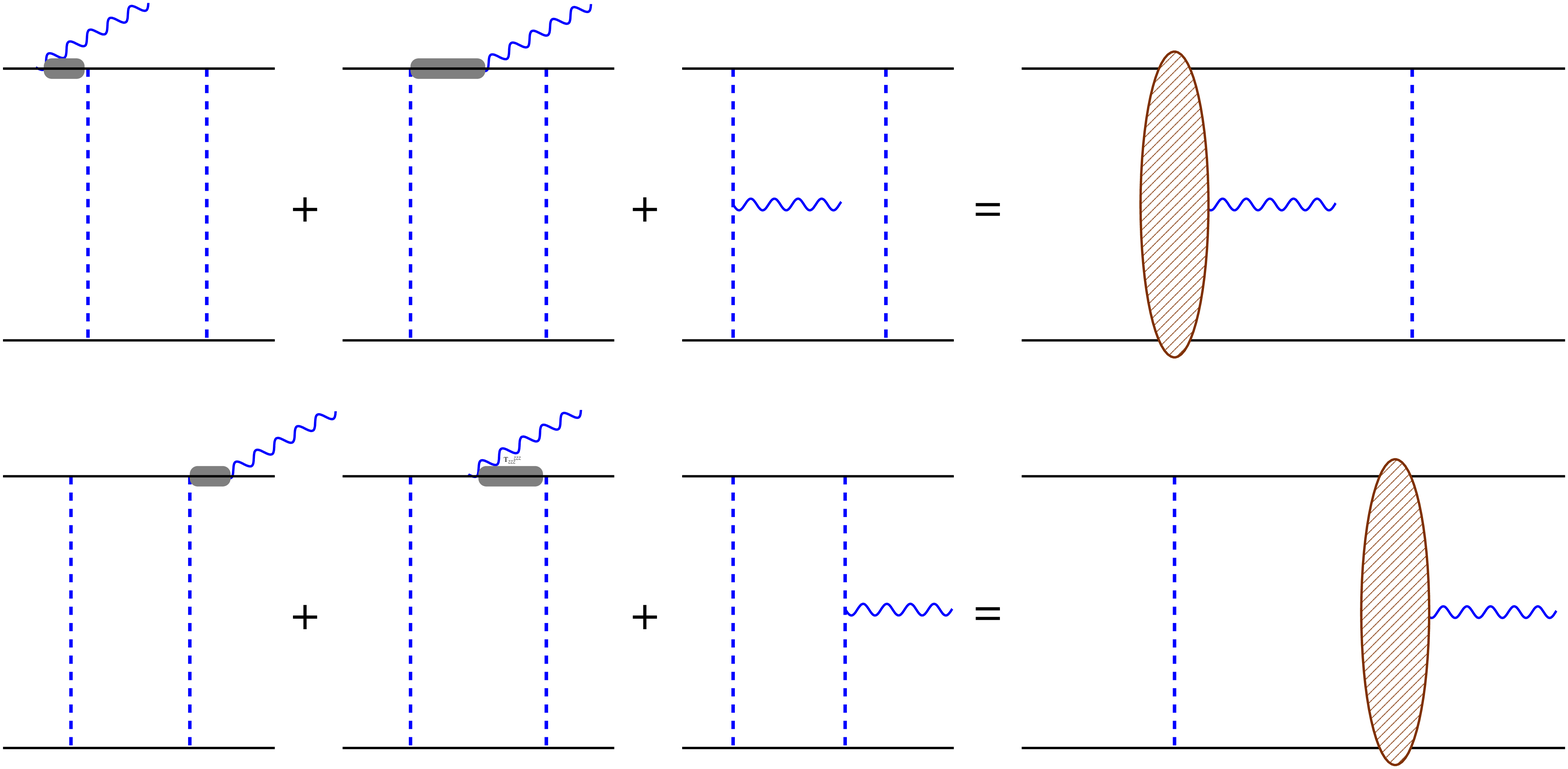}
  \caption{\it Graviton insertions for double-exchange diagrams. External line
    insertions are represented in the first column, internal line insertions on
    the fast particles are in the second column, insertions on the exchanged
    graviton in the third one. Gray shadows around the fast particles denote
    off-shell propagation. Analogous insertion diagrams from the lower line are
    understood. The sum of each row amounts to inserting a matched emission
    amplitude (hatched brown blob) in place of a graviton propagator.}
  \label{f:doubleExEmis}
\end{figure}

As for insertions on exchanged graviton lines, they are negligible if the
emitted transverse momentum $\qt=\hbar\omega\tht$ is smaller than any of the
exchanged graviton lines which, in the Regge limit, are all of order
$\bk{q_i}\sim\hbar/b$, thus leading to the condition $b|q|\ll\hbar$. The latter
is not surprising because precisely this parameter occurs in the subtraction
term of eq.~(\ref{Mmatch}) coming from the Regge estimate of region c) for a
single exchange. Fortunately this region --- which is generally multidimensional
for $n$ exchanges --- is most significant when both exchanged and emitted
momenta are of the same order $\hbar\omega\ll E$. For double exchange
(fig.~\ref{f:doubleExEmis}), for instance, this corresponds to two diagrams, and
more generally it allows to count $n$ diagrams per eikonal contribution in which
we have to compute and add $\Delta\tfa$, i.e., the difference $(\regge-\soft)$
introduced in sec.~\ref{s:msarl}. Again, this has to be adjusted to take into
account the direction of the fast legs, which is straightforward since the Regge
and soft amplitudes acquire the same transformation phase.

In the end, this means that \emph{all} internal-line insertions --- for fast
particles and exchanged particles alike --- can be accounted for by calculating
$n$ diagrams for the eikonal contribution with $n$ exchanged gravitons, where
the \emph{matched} amplitude (eqs.~\eqref{Mmatch}, \eqref{hsoft}) is
inserted in turn in correspondence to the $i$-th exchanged graviton, adjusting
for the local incidence angle $\Tht_i$ as in eq.~\eqref{MSi}.

We recall at this point the soft-based representation of the matched
amplitude~\eqref{Mmatch}, which, after adjusting for the incidence angle,
acquires the form
\begin{equation}\label{Minc}
  \tfa^{(\Tht_i)} = \sqrt{\alpha_G} \frac{R}{2} \esp{-2\ui \phi_\tht} \int
  \dif^2 \zt \; h_s(\zt) \esp{\ui b \omega \zt \cdot (\tht - \Tht_i)}\; ,
\end{equation}
and is thus simply proportional to the translated F.T. of the soft field $h_s$.

We then use this representation for each ``active'' contribution. For $n = 2$
(fig.~\ref{f:doubleExEmis}), for instance, we have $\qt_{s1} = E \tht_1$
emitting with initial angle $\Tht_1 = 0$ and $\qt_{s2} = E \tht_2$ with
$\Tht_2 = \tht_1$, and we write, accordingly,
\begin{equation}\label{active2}
\begin{split}
  & \frac{\esp{2\ui \phi_\tht} \tfa^{(2)}}{\sqrt{\alpha_G}\frac{R}{2}}
  = \frac{1}{2} \left[\int \dif^2 \zt \; h_s(z)
    \esp{\ui b \omega \zt \cdot \tht} (2\ui\alpha_G) \int \frac{\dif^2 \tht_2}{2\pi \tht_2^2} \,
    \esp{\ui \frac{E}{\hbar} \bt \cdot \tht_2} + (2\ui\alpha_G) \int
    \frac{\dif^2 \tht_1}{2\pi \tht_1^2} \;\esp{\ui \frac{E}{\hbar} \bt\cdot\tht_1}
  \right.\\
  &\times\left.\int\dif^2 \zt\; h_s(z)\esp{\ui b \omega\zt\cdot(\tht - \tht_1)}
  \right] = \frac{2\ui}{2} \int \dif^2 \zt \; h_s(z)
  \esp{\ui b \omega \zt \cdot \tht} \left[\delta_0(b)
    +\delta_0\big(|\bt - \tfrac{\hbar \omega}{E} b \zt|\big)\right] \;.
\end{split}
\end{equation}
We can see that the second ``active'' contribution, with non-zero incidence
angle $\Tht_2=\tht_1$ has a translated $\tht$-dependence, which amounts to
factorizing an eikonal with $\zt$-dependent argument. This generalized
factorization can be extended to the general case with $n>2$ exchanges, where
however the $\tht$-translation involves $\Tht_i=\sum_{j=1}^{i-1} \tht_j$, $i>2$,
yielding higher powers of the eikonal with $z$-dependent argument. In formulas,
we obtain, order by order,
\begin{equation}
\begin{split}
  &\frac{\esp{2\ui \phi_\tht} \tfa_\res}{\sqrt{\alpha_G}\frac{R}{2}}
  = \int\dif^2\zt\;h_s(z)\esp{\ui b\omega\zt\cdot \tht} \left\{ 1+\frac{2\ui}{2!}
    \Big[\delta_0(b)+\delta_0\big(|\bt-\tfrac{\hbar\omega}{E} b\zt|\big)\Big]
  \right. \\
   & \left. + \; \frac{(2\ui)^2}{3!}\Big[\delta_0^2(b)
    +\delta_0(b)\delta_0\big(|\bt - \tfrac{\hbar \omega}{E} b \zt|\big)
    +\delta_0^2\big(|\bt - \tfrac{\hbar \omega}{E} b \zt|\big) \Big]
    +\cdots \right\} \;.
\label{espRepr}
\end{split}
\end{equation}
Furthermore, the sum in square brackets is given by the expression
\begin{equation}\label{resummation}
  \big[ \ldots \big] = \frac{\esp{2\ui\delta_0(b)}-\esp{2\ui\delta_0\left(
        |\bt - \frac{\hbar \omega}{E} b \zt|\right)}}{ 2\ui\left[\delta_0(b)
      -\delta_0\big(|\bt - \tfrac{\hbar \omega}{E} b \zt|\big)\right]}
  = \esp{2\ui\delta_0(b)} \int_0^1\dif\xi\;
  \esp{-2\ui\xi\frac{Gs}{\hbar}\log\left|\hat{\bt}
      - \frac{\hbar \omega}{E}\zt \right|}\; ,
\end{equation}
so that we finally get the factorized and resummed amplitude
\begin{align}
  \frac{\tfa_\res}{\esp{2\ui \delta_0(b)}} \equiv \ampRid
  &= \sqrt{\alpha_G} \frac{R}{2} \esp{-2\ui \phi_\tht}\int \dif^2 \zt \;
  \esp{\ui b \omega \zt \cdot \tht} \int_0^1 \dif \xi\;
  \esp{-2 \ui\frac{Gs}{\hbar} \xi \log\left| \hat{\bt}
      - \frac{\hbar \omega}{E} \zt \right|} h_s(\zt)  \nonumber \\
  & \simeq \sqrt{\alpha_G} \frac{R}{2} \esp{-2\ui \phi_\tht}
  \int \dif^2 \zt \int_0^1 \dif\xi \;h_s (\zt)
  \esp{\ui \omega b \zt \cdot (\tht - \xi \Tht_s (b))} \;, \label{ResFinal}
\end{align}
where we have expanded the logarithm in the exponent and neglected higher order
terms in $\hbar / E b |\tht|$.  The latter can in principle be evaluated as
``quantum'' corrections to the basic formula of the last line.

A more symmetric expression for the resummed amplitude~\eqref{ResFinal} is
obtained in the Breit frame (also called brick-wall frame), where the initial
and final transverse momenta are equal and opposite (i.e.,
$\pm\tfrac12 E\Tht_s(\bt)$). We can reach the Breit frame by rotating the system
of $\tfrac12\Tht_s(\bt)$. According to eq.~\eqref{Minc}, this amounts to
translate $\tht\to\tht-\tfrac12\Tht_s(\bt)$, and at the end we obtain again the
expression~\eqref{ResFinal} but with $\xi$ integrated over the symmetric
interval $[-1/2,1/2]$.  In the following we shall often work in the Breit
frame.

It is important to note that the $z$-dependence in eq.~\eqref{espRepr} adds
corrections to the na\"ive factorization of $\delta_0(b)$~\cite{ACV07} which for
any given $n$ are of relative order $\hbar\omega\bk{z}/E\sim\hbar/(bE|\tht|)$
and thus may appear to be negligible in the region $|\tht| >\theta_m$. However,
this is {\em not} the case because of the counting factors of
$\bk{n}\sim Gs/\hbar$ occurring in eq.~\eqref{resummation}, which promote such
corrections to order $|\Tht_s(\bt)|/|\theta|$ making them essential for the
physics of the radiation problem at frequencies $\omega\sim R^{-1}$. The effect
of such corrections can be summarized by the introduction of the resummed field
in the Breit frame
\begin{equation}\label{hsres}
  h_s^{\res}(z) \equiv h_s(z)\frac{\sin\left(\alpha_G\log\left|\hat{\bt}
      -\tfrac{\hbar\omega}{E}\zt\right|\right)}{\alpha_G\log\left|\hat{\bt}
      -\tfrac{\hbar\omega}{E}\zt\right|}
  \simeq h_s(z) \frac{\sin\omega R\,\zt\cdot\hat{\bt}}{\omega R\,\zt\cdot\hat{\bt}}
  \qquad\left(|z|\ll\frac{E}{\hbar\omega}\right)
\end{equation}
with its resummation factor which is directly $\omega R$-dependent in the
moderate-$z$ form of the last line. Therefore, eq.~\eqref{ResFinal} can be
summarized as
\begin{equation}\label{MhRes}
  \ampRid = \sqrt{\alpha_G} \frac{R}{2} \esp{-2\ui\phi_\tht}
  \int\dif^2\zt\; h_s^\res(z) \esp{\ui b\omega \zt\cdot\tht} \;.
\end{equation}

\subsection{Rescattering corrections and classical limit\label{s:rescattering}}

We have just seen that taking into account the sizeable incidence angles in
multiple-exchange emission amplitudes provides important corrections to the
na\"ive resummation formula which involve the effective coupling
$\omega R = \frac{Gs}{\hbar} \frac{\hbar\omega}{E}$ and are thus essential for
$\omega R \simeq \ord{1}$. One may wonder, at this point, whether additional
corrections of relative order $\hbar \omega / E$ may be similarly enhanced by
multiplicity factors, thus yielding important effects as well.

We do not have a complete answer to that question. We think however that
kinematical corrections affecting incidence angles at relative order
$\hbar \omega/E$ (and occurring in the currents' projections, app.~\ref{a:hlw})
are unimportant because they actually affect the factorization procedure at
relative order $\ord{\hbar\omega/E}^2$, and are thus subleading. On the other
hand, we shall argue that \emph{dynamical} corrections due to rescattering of
the emitted graviton are to be seriously considered, even though they are
known~\cite{ACV93} to be subleading for the calculation of the scattering
amplitude of the fast particles themselves.

\begin{figure}[t]
  \centering
  \includegraphics[width=0.6\linewidth]{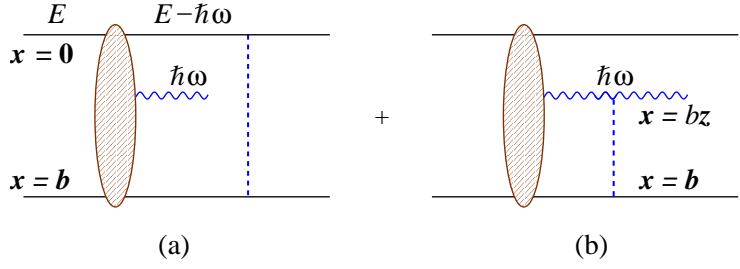}
  \caption{\it Emission diagrams with subleading corrections. In {\rm (a)} the
    eikonal scattering $\propto\delta(b)$ after graviton emission occurs with
    reduced energy $E\to E-\hbar\omega$. In {\rm (b)} the graviton at $\xt=b\zt$
    rescatters with the external particle at $\xt = b$ generating a term
    $\propto \hbar\omega\delta(|\bt-b\zt|)$.}
  \label{f:emCor}
\end{figure}

Indeed, consider for instance the contributions to the emission amplitude of the
2 graviton-exchange diagrams of fig.~\ref{f:doubleExEmis}. If the active exchange is
$\# 2$, we just have to replace $\delta_0\left(b\right)$ by
$\delta_0\left(\left| \bt - \frac{\hbar\omega}{E} b \zt \right|\right)$ because
of the nontrivial incidence angle.  But if the active exchange is $\# 1$, next
hit is a 3-body one, which involves emitted graviton interactions also,
as illustrated in fig.~\eqref{f:emCor} for an emitted graviton in jet 1 (top),
rescattering with jet 2 (bottom).  Therefore, the remaining $\delta_0$ which was
left uncorrected in eq.~\eqref{active2} should be corrected also, by replacing
it by
\begin{equation}\label{resc1}
  \delta_R(\bt,\zt) \equiv \left( 1 - \frac{\hbar\omega}{E} \right) \delta_0(b)
  + \frac{\hbar\omega}{E} \delta_0 \left(\left| \bt - b \zt \right|\right)
  = \delta_0(b) - \omega R \log |\hat\bt - \zt| \;,
\end{equation}
where we note that the fast-particle gravitational charge has been decreased by
an energy conservation effect of order $\hbar\omega/E$, while the charge of the
rescattering graviton is by itself of that order.

Since both replacements --- due to incidence angles and rescattering alike ---
hold for any one of the single hits being considered (as in eq.~\eqref{espRepr})
it follows that eq.~\eqref{resummation} should be replaced by
\begin{equation}\label{newresummation}
  \big[ \ldots \big] = \frac{\esp{2\ui \left(\delta_0(b) - \omega R \log |\hat\bt - \zt|\right)}-\esp{2\ui\delta_0\left(
        |\bt - \frac{\hbar \omega}{E} b \zt|\right)}}{ 2\ui\left[\delta_0(b) - \omega R \log |\hat\bt - \zt|
      -\delta_0\big(|\bt - \tfrac{\hbar \omega}{E} b \zt|\big)\right]}
\end{equation}
where we note the appearance, in the denominator, of the quantity $\Phi_R$
introduced in eq.~(\ref{PhiRdef}) multiplied by $2i \omega R$.  The first log in
$\Phi_R$ is due to the incidence angle while the second one is due to
rescattering. But since $-\Phi_R$ appears in the numerator if we upgrade $h_s(z)$
to $h_s(\omega,z)$ of eq.~\eqref{PhiRdef}, we simply get:
\begin{equation}\label{ResFinal2}
  \ampRid = \sqrt{\alpha_G} \frac{R}{\pi} \esp{-2\ui \phi_\tht} \int
  \frac{\dif^2 \zt}{2\pi} \; \frac{1}{{z^*}^2}
  \frac{\esp{\ui\omega b \zt \cdot\tht}}{2\ui\omega R} \left(
   \esp{-2\ui\omega R\log|\hat\bt-\zt|}-\esp{-2\ui\omega R
    \frac{E}{\hbar\omega}\log\left|\hat\bt-\frac{\hbar\omega}{E}\zt\right|}
  \right) \;,
\end{equation}
which is expressed as the algebraic sum of incidence angle and rescattering
effects.

We thus see, by inspection, that since $\Phi_R(\omega,\zt) \to \Phi(\zt)$ in the
small $\hbar\omega/E$ limit, eq.~\eqref{ResFinal2} reproduces the classical
amplitude of ref.~\cite{GrVe14} with the proper normalization according to our
conventions%
\footnote{ In order to carry out the precise comparison one should
  keep in mind that $R = 4 E_{\cite{GrVe14}}, \Phi = \frac18
  \Phi_{\cite{GrVe14}}$, and that scattering angles are defined with opposite
  sign conventions.}
and helicity $\lambda = -2$,
\begin{equation}\label{ResClas}
  \ampRid_{\mathrm{class}} = \sqrt{\alpha_G} \frac{R}{\pi} \esp{-2\ui \phi_\tht} \int
  \frac{\dif^2 \zt}{2\pi {z^*}^2} \;
  \frac{\esp{\ui \omega b \zt \cdot (\tht- \Tht_s) }}{2\ui \omega R}
  \left( \esp{-2\ui \omega R \Phi}
    -1 \right) \;,
\end{equation}
where $\Tht_s(\bt) = -\frac{2R}{b} \hat\bt$ is, as usual, the fast-particle
scattering angle.

We can also express the result (\ref{ResFinal2}) in a form similar to (\ref{ResFinal}):
\begin{equation}
 \ampRid
  = \sqrt{\alpha_G} \frac{R}{2} \esp{-2\ui \phi_\tht}\int \dif^2 \zt\;
  h_s(\omega,\zt) \esp{\ui b \omega \zt \cdot \tht} \int_0^1 \dif \xi\;
  \esp{-2 \ui\alpha_G \xi \log\left| \hat{\bt}
      - \frac{\hbar \omega}{E} \zt \right| -2\ui\omega R (1-\xi) \log|\hat\bt-\zt|}  
  \;, \label{ResFinal3}
\end{equation}
showing explicitly how, for each value of $\xi$, the incidence angle effect
depends on the evolution up to an incidence angle $\xi \Tht_s$ while
rescattering depends on the complementary interval $(1- \xi)\Tht_s$ of
incidence angles.

We can also say that the incidence angle dependence corresponds to the tilt in
the fast-particle wave-front noted in \cite{GrVe14}, so that the rescattering
counting is correctly reproduced by the simple overall subtraction in
eqs.~\eqref{ResFinal2} and \eqref{ResClas}.  Furthermore, the residual
$\hbar\omega/E$ dependence of the improved form~\eqref{ResFinal2}  produces
 quantum corrections to the classical
formula~\eqref{ResClas}. It is amazing  that the same function $\Phi_R$
yields, on the one hand, the extension of the soft field to the small-angle part
of region a) and ensures, on the other hand, the rescattering corrections at
quantum level.

In the following, we shall concentrate on the analysis of the
result~\eqref{ResFinal}, which provides what we shall call the ``geometrical''
corrections due to scattering and emission with various incidence angles all
over the eikonal evolution.  The inclusion of rescattering corrections, leading
to the classical result~\eqref{ResClas} and to its quantum
corrections~\eqref{ResFinal2} is deferred to a later work.

\subsection{The resummed amplitude and its regimes\label{s:regimes}}

The amplitude $2\ui \ampRid$~\eqref{ResFinal} is directly normalized as the
probability amplitude for the emission of a graviton in a scattering process
occurring at impact parameter $b$. Its interpretation is that of a coherent
average of the single-exchange amplitude over scattering angles
$\xi \Tht_s = -\xi \frac{2R}{b} \, \hat{\bt}$ ranging from zero to $\Tht_s$.

The final result of our calculation in the ``geometrical'' approximation and in
the Breit frame can also be expressed, more explicitly, in the form
\begin{equation}
  \ampRid = \sqrt{\alpha_G} \frac{R}{\pi} \esp{-2\ui \phi_\tht}
  \int \frac{\dif^2 \zt}{2\pi {z^*}^2} \; \esp{\ui b \qt \cdot \zt}
   \, \frac{\sin(\omega R x)}{\omega R x} \, \Phi(\zt) \;,
\label{ampSin}
\end{equation}
where $\Phi(\zt)$ was defined in eq.~\eqref{Phidef}.

In order to understand the $\omega$ dependence of eq.~\eqref{ampSin}, it is
convenient to rescale $\zt\to\omega R \tilde\zt$ so as to write
\begin{equation}\label{regrescaling}
  \ampRid = \sqrt{\alpha_G} \frac{R}{\pi} \esp{-2\ui \phi_\tht}
  \int \frac{\dif^2 \tilde\zt}{2\pi {\tilde{z}^*}{}^2} \;\
  \esp{\ui\tilde\zt\cdot \frac{2\tht}{|\Tht_s|}} \;
  \frac{\sin \tilde{x}}{\tilde{x}} \left( \frac{\tilde{x}}{\omega R}
    + \log\left| \hat{\bt} - \frac{\tilde\zt}{\omega R} \right| \right) \; .
\end{equation}

We can see that there are basically 2 regimes:
\begin{enumerate}
\item $\omega R \ll 1$. In this case $\Phi$ is dominated by the linear term,
  except if $\frac{|\tilde\zt|}{\omega R} \simeq \frac{1}{\omega|\tht| b}\ll 1$,
  in which case $\Phi\left( \frac{\tilde\zt}{\omega R} \right) \simeq
  \frac{1}{2} \left(\frac{|\tilde\zt|}{\omega R}\right)^2 \cos(2 \phi_z)$
  which is very small.  Therefore, we recover the soft limit in the form
\begin{align}
  2 \ampRid &\simeq \frac{\sqrt{\alpha_G}}{\pi\omega}\frac1{2\ui}
  \left[\esp{2\ui\big(\phi_{\tht-\frac12\Tht_s} - \phi_\tht\big)} -
    \esp{2\ui\big(\phi_{\tht +\frac12\Tht_s} - \phi_\tht\big)} \right]
  \Theta(\hbar - |\qt| b) \nonumber \\
  &= \frac{\sqrt{\alpha_G}}{\pi\omega}
  \esp{\ui\big(\phi_{\tht+\frac12\Tht_s}+\phi_{\tht-\frac12\Tht_s}-2\phi_\tht\big)}
  \left[\frac{|\Tht_s|\sin\phi_{\tht+\frac12\Tht_s,\bt}}{|\tht-\frac12\Tht_s|}\right]
  \Theta(\hbar - |\qt| b) \;, \label{softMrid}
\end{align}
where we note the close relationship of $2\ampRid$ with the soft insertion
factor in eq.~\eqref{Msoft}, evaluated --- in the Breit frame --- at scattering
angle $\Tht_s(\bt)$. The second line yields, in square brackets, the
singularity-free expression $\sin\phi_{\tht-\frac12\Tht_s,\tht+\frac12\Tht_s}$
and shows how the coupling $|\Tht_s(\bt)|=2R/b$ is recovered. Furthermore, the
cutoff $|\qt|b<\hbar$ argued on the basis of the $\Phi$-function behaviour is
consistent with the $1/(|\qt| b)^2$ behaviour of the Regge form of the
amplitude~\eqref{Mreggeeta}.
\item $\omega R \gtrsim 1$. In this regime, the amplitude starts feeling the
  decoherence factor $\sim \frac{1}{\omega R}$ due to the $\xi$-average, which
  eventually dominates the large frequency spectrum of the energy-emission
  distribution (\ref{s:cso}) and establishes the key role of $R$.  According to
  eq.~\eqref{regrescaling}, resummation effects due to the
  $\frac{\sin \tilde x}{\tilde x}$ factor are kinematically small in the region
  $|\theta_x| \gg |\Tht_s|$ because $x\sim \frac{|\Tht_s|}{|\theta_x|}\ll1$.
  Instead, in the region $|\theta_x| \lesssim |\Tht_s|$ they are important and
  tend to suppress the amplitude for $\omega R >1$. In order to see how, we
  anticipate from eq.~\eqref{enDist} the energy distribution formula
\begin{equation}\label{spectrum}
  \frac{\dif E^{GW}}{\dif \omega \, \dif \Omega} = \hbar \; |2\omega\ampRid|^2
\end{equation}
so that it is instructive to look at the combination $2\omega\ampRid$ in the
limit $\omega R \gg 1$, in which $x \sim \frac{1}{\omega R}$ is supposed to be
small, in order to avoid a higher power decrease.

Since we should have $b \omega \theta_y y \sim b \omega \theta_x x \sim\ord{1}$,
the condition $x \sim \frac{1}{\omega R}$ leads to a phase space in which
$x \ll y \sim \ord{1}$ and
$\theta_y \sim \frac{1}{b \omega} \ll \theta_x \sim |\Tht_s|=\frac{2R}{b}$. In
this region, in the Breit frame, we get the limit
\begin{equation}
  2\omega\ampRid \to \sqrt{\alpha_G} \int_{-\infty}^{+\infty}
  \frac{\dif \tilde{x}}{\pi} \; \esp{\ui \frac{b}{R} \tilde{x} \theta_x}
  \frac{\sin \tilde{x}}{\tilde{x}}
  \int_{-\infty}^{+\infty} \frac{\dif y }{2 \pi y^2} \;
  \log(1+y^2) \esp{\ui b \omega \theta_y y},
\label{bigomR}
\end{equation}
where we have set $\tilde{x} = \omega R x$ and approximated
$\Phi(x,y)\to\Phi(0,y)$.

We thus get a simple, factorized amplitude which --- by performing the remaining
integrals --- has the explicit form
\begin{equation}\label{hardMrid}
  2 \ui \omega\ampRid \to \sqrt{\alpha_G} \int_1^{\infty}
  \frac{\dif \eta}{\eta^2} \; \esp{-b \omega |\theta_y| \eta}
  \Theta\left( 1- \frac{b}{R} |\theta_x| \right)
\end{equation}
in both the forward and backward hemispheres.

The above limiting amplitude is strongly collimated around
$|\theta_y| \sim \frac{1}{b \omega} \ll \frac{1}{\omega R}$ (by which
$\phi_\tht$ si very small), for any $|\theta_x| < |\Tht_s|/2$, that is in the
$\xi$-averaging region.  Furthermore, that distribution (confirmed numerically,
see fig.~\ref{f:angle}) comes from the transverse space region $x = 0$,
which becomes dominant at large $\omega R$'s.

In fact, we can easily calculate the contribution of \eqref{bigomR} to the
integrated distribution, that is
\begin{equation}\label{eGW}
  \frac{\dif E^{GW}}{\dif \omega} = G s \, \Tht_s^2 \, \frac{4}{3} \, (1-\log2)
  \, \frac{1}{\omega R} \; ,
\end{equation}
where the coefficient comes from the $\theta_y$-integral, in agreement with the
dominant $\frac{1}{\omega R}$ behaviour of the spectrum to be discussed next.

\end{enumerate}

\subsection{Multi-graviton emission and coherent-state operator\label{s:cso}}

So far, we have considered single-graviton emission in the whole angular range.
However, the extension to many gravitons by keeping the leading terms in the
eikonal sense is pretty easy. For one emitted graviton we have factorized in
$b$-space one active (emitting) exchange out of $n$ in $n$ ways. Similarly for
two gravitons we count $n(n-1)$ pairs of active exchanges emitting one graviton
each, plus $n$ exchanges which emit two gravitons, and so on. The first ones are
independent and provide an exponential series for the single emission amplitudes
we have just resummed, the second ones yield correlated emission for a pair of
gravitons, and so on.

Resumming the independent emissions yields multiple emission amplitudes which
are factorized in terms of the single-emission ones calculated so far. Virtual
corrections are then incorporated by exponentiating both creation
($a^\dagger_\lambda(\vq)$) and destruction ($a_\lambda(\vq)$) operators of
definite helicity $\lambda$ (normalized to a wave-number $\delta$-function
commutator
$[a_\lambda(\vq),a^\dagger_{\lambda'}(\vq\,')]=\hbar^3\delta^3(\vq-\vq\,')\delta_{\lambda\lambda'}$),
as follows
\begin{equation}\label{Shat}
  \hat{S}=\esp{2\ui\delta} \exp\left\{\int\frac{\dif^3 q}{\hbar^3\sqrt{2\omega}}\;
    2\ui\left[\sum_\lambda \ampRid_\bt^{(\lambda)}(\vq) a^\dagger_\lambda(\vq)
      +\text{h.c.}\right] \right\} \;,
\end{equation}
where the helicity amplitude
$\ampRid_\bt^{(-)}(\vq)=[\ampRid_\bt^{(+)}(-\qt,q^3)]^*$ is provided by
eq.~\eqref{ResFinal} with a proper identification of variables. Since operators
associated with opposite helicities commute, the above coherent-state operator
is abelian (and thus satisfies the Block-Nordsieck theorem) but describes both
helicities, not only the infrared (IR) singular longitudinal polarization.

The $S$-matrix~\eqref{Shat} is formally unitary because of the anti-hermitian
exponent, but needs a regularization because of the IR divergence mentioned
before, due to the large distance behaviour $|h_s(z)|\sim|h(z)|\sim |z|^{-1}$ of
the relevant fields. Because of the virtual- real-emission cancellation, the
regularization parameter can be taken to be $\Delta\omega$, the experimental
frequency resolution (assumed to be much smaller than $b^{-1}$), whose role will
be further discussed in sec.~\ref{s:spec}. With that proviso, we can now provide
the normal ordered form of eq.~\eqref{Shat} when acting on the initial state,
that we identify as the graviton vacuum state $\ket{0}$, as follows:
\begin{equation}\label{normOrd}
  \esp{-2\ui\delta}\hat{S}\ket{0} = \sqrt{P_0}\prod_\lambda\exp\left\{
    2\ui\int_{\Delta\omega}\frac{\dif^3 q}{\hbar^3\sqrt{2\omega}}\;
     \ampRid_\bt^{(\lambda)}(\vq) a^\dagger_\lambda(\vq) \right\}\ket{0} \;,
\end{equation}
where
\begin{equation}\label{P0}
  P_0 = \exp\left\{-2\int_{\Delta\omega}\frac{\dif^3 q}{\hbar^3\omega}\;
    \sum_\lambda |\ampRid_\bt^{(\lambda)}(\vq)|^2\right\} 
\end{equation}
is the no-emission probability, coming from the $a,a^\dagger$ commutators.

Due to the factorized structure of eq.~\eqref{normOrd}, it is straightforward to
provide the full generating functional of inclusive distributions
\begin{equation}\label{genFun}
  \mathcal{G}[z_\lambda(\vq)] = \exp\left\{2\int_{\Delta\omega}
    \frac{\dif^3 q}{\hbar^3\omega}\;\sum_\lambda|\ampRid_\bt^{(\lambda)}(\vq)|^2
    \left[ z_\lambda(\vq)-1\right] \right\}
\end{equation}
as functional of the fugacity $z_\lambda(\vq)$.

In particular, the unpolarized energy emission distribution in the solid angle
$\Omega$ and its multiplicity density are given by
\begin{equation}\label{enDist}
  \frac{\dif E^\GW}{\dif\omega\,\dif\Omega}
  = \hbar\omega\frac{\dif N}{\dif\omega\,\dif\Omega}
  = \left. \hbar\omega^2 \sum_\lambda
    \frac{\delta\mathcal{G}}{\delta z_\lambda(\vq)} \right|_{z_\lambda=1}
  = 2\omega^2 \hbar \sum_\lambda |\ampRid_\bt^{(\lambda)}(\vq)|^2 \;.
\end{equation}
Both quantities will be discussed in the next section.

\section{The angular/frequency spectrum\label{s:spec}}

\subsection{Energy emission and multiplicity distributions\label{s:pfd}}

Starting from the soft/Regge emission amplitude in eq.~\eqref{ampSin} we obtain,
by eq.~\eqref{enDist}, the multiplicity distribution in either jet $J$ ($J=1,2$,
$z\equiv x+\ui y$, $x\equiv \zt\cdot\hat\bt$):
\begin{equation}\label{probd}
  \left.\frac{\omega\,\dif N}{\dif\omega\dif^2\tht}\right|_J = \alpha_G
    \frac{(\omega R)^2}{2\pi^2} \left[ \left|\int\frac{\dif^2\zt}{\pi{z^*}^2}\;
        \esp{\ui b\omega\zt\cdot\tht} \;
        \frac{\sin\omega Rx}{\omega Rx} \,\Phi(\zt)\right|^2 + (\tht\to-\tht)
    \right]_J \;,
\end{equation}
where the helicity sum provides the additional $\tht\to-\tht$ contribution,
equivalent to a factor of 2 after angular integration. It is convenient to look
first at the qualitative properties of the frequency dependence integrated over
angles, by distinguishing the two regimes pointed out before.

{\it i)} $\omega R\ll1$. This is the infrared singular region originally looked
at by Weinberg. The angular integration in jet 1 involves the two-dimensional
vector $\qt=\hbar\omega\tht$ and the amplitude is dominated by its leading
form~(\ref{softMrid}). If $b|\qt|/\hbar\ll\omega R$ ($|\tht|\ll\Theta_s = 2R/b$)
the amplitude is $\phi$-dependent, but is independent of $|\tht|$ because of the
cancellation of the collinear singularities due to the helicity conservation
zero, which has been extended to the whole region $|\tht|<\Theta_s$ by our
method. Therefore, the distribution acquires the form
$\sim\text{const}\frac{\dif^2\tht}{\Theta_s^2}\Theta(\Theta_s-|\tht|)$ which
effectively cuts-off the $b\qt$ integration at $b|\qt|/\hbar\geq\omega R$.

If instead $\omega R<b|\qt|/\hbar<b\omega$ we enter the intermediate angular
region $\Theta_s<|\tht|<1$ where~(\ref{hsoft}) agrees with the basic form
$\sim\frac{\sin\phi_\tht}{b|\qt|}$ of~(\ref{basicsoft}) already noticed
in~\cite{ACV90}, so that the integrated distribution is of type
\begin{equation}\label{iDist}
  \frac{\omega\,\dif N}{\dif\omega} = \alpha_G\frac{2}{\pi}\Theta_s^2\left(
    \int_{R/b}^{1}\frac{\dif^2\tht}{\pi\tht^2}\;\sin^2\phi_{b\qt}
    +\text{const}\right) = \alpha_G\frac{2}{\pi}\Theta_s^2\left(
    \log\frac{2}{\Theta_s}+\text{const}\right) \;.
\end{equation}

More precisely, by eq.~\eqref{softMrid}, we get for the energy-emission
distribution
\begin{align}
  \frac{\dif E^\GW}{\dif\omega} \equiv \hbar\omega\frac{\dif N}{\dif\omega}
  &= \frac{Gs}{\pi}\Theta_s^2\; 2\int_0^1 |\tht|\dif|\tht|
  \int_0^{2\pi} \frac{\dif\phi}{2\pi} \;
  \frac{\sin^2\phi_{\tht\bt}}{|\tht-\Tht_s|^2} \Theta(1-b\omega|\tht|)
  \nonumber \\
  & \simeq\frac{Gs}{\pi}\Theta_s^2 \left[2 \log\min\left(
     \frac{b}{R},\frac1{\omega R}\right) + \text{const}\right]\;,\label{intDist}
\end{align}
where we have changed variables $\tht+\frac12\Tht_s\to\tht$, integrated on both
jets, and used the cutoff $b\omega|\tht|<1$ due to the large $b|\qt|$
suppression.

We thus find that the typical infra-red distribution $\dif\omega/\omega$ is kept
only in the tiny region $\omega<b^{-1}$, with a rapidity plateau in the range
$|y|<Y_s=\log(2/\Theta_s)=\log(b/R)$, much restricted w.r.t.\ the full rapidity
$Y=2\log(Eb/\hbar)\sim\log s$ available in the single H-diagram evaluation. On
the other hand, the corresponding small-$\omega$ number density
$(Gs/\pi\hbar)\Theta_s^2$ agrees with that used in~\cite{ACV90} for the
calculation of the 2-loop eikonal and with the zero-frequency limit (ZFL)
of~\cite{We65,Sm77}.

{\it ii)} $1<\omega R<(Gs/\hbar)$. In this region we think it is tenable to
assume the completeness of the $\qt$ states, so that the spectrum, integrated
over $\dif^2(\omega\tht)$ of eq.~\eqref{probd} and on both jets, is obtained by
the Parseval identity in the form
\begin{equation}\label{freqd}
  \frac{\dif E^\GW}{\dif\omega} = \hbar\omega\frac{\dif N}{\dif\omega}
  = 2 Gs \frac{\Theta_s^2}{\pi} \int\frac{\dif^2\zt}{\pi |z|^4}\;\left(
    \frac{\sin\omega Rx}{\omega Rx}\right)^2 |\Phi(\zt)|^2 \;.
\end{equation}
We note the typical IR behaviour $\dif\omega/\omega$ of the number density which
is present in this formulation also, and we also note the less typical
$\tilde{x}\equiv\omega Rx$-dependence
$\left(\frac{\sin\tilde x}{\tilde x}\right)^2$ due to the coherent average over
initial directions~\eqref{hsres}, occurring in the Fourier transform of the
resummed field~(\ref{MhRes}), which essentially acts as a cutoff
$\Theta(1-\omega R|x|)$.  Its action is $\omega R$-dependent and cuts off large
values of $|x|$ for $\omega R\ll1$ and reduces the integration to small ones for
$\omega R\gg1$.

In particular, for $\omega R\gg 1$, the emitted-energy spectrum is considerably
suppressed by our treatment of the collinear region, w.r.t.\ na\"ive H-diagram
expectations. We get in fact from eq.~(\ref{freqd}) for the emitted energy
fraction
\begin{equation}\label{emitEn}
  \frac{\dif E^\GW}{\sqrt{s}\,\dif\omega} = \frac{\hbar\omega}{\sqrt{s}}\,
  \frac{\dif N}{\dif\omega} = R \Theta_s^2\frac1{\pi}\int
  \frac{\dif x\dif y}{\pi(x^2+y^2)^2} \; \left(
    \frac{\sin(\omega Rx)}{\omega Rx}\right)^2
  \left( x + \frac12 \log[(1-x)^2+y^2] \right)^2 \;.
\end{equation}

We see that the spectrum is decreasing like $1/(\omega R)^2$ for
any fixed value of $x$, in front of an integral which is linearly divergent for
$x\to 0$. This means effectively a $1/(\omega R)$ spectrum. More precisely by
integrating the averaging factor in the small-$x$ region we get, for
$\omega R\gg1$, the factor
\begin{equation}\label{sxfac}
  \int_{|x|<1}\frac{\dif x}{\pi}\; \left(\frac{\sin(\omega Rx)}{\omega Rx}
  \right)^2 \simeq \int_{|x|<1}\frac{\dif x}{\pi}\;
  \frac{\sin(2\omega Rx)}{\omega Rx} \simeq \frac1{\omega R}
\end{equation}
in front of the coefficient
\begin{equation}\label{intlog}
  \frac14 \int_{-\infty}^{+\infty}\frac{\dif y}{y^4}\;\log^2(1+y^2)
  = \frac{2}{3}\pi(1-\log2)\;,
\end{equation}
thus obtaining the asymptotic spectrum
\begin{equation}\label{asymSp}
  \frac{\dif E^\GW}{\sqrt{s}\,\dif\omega} \simeq \frac23 (1-\log2) R
  \Theta_s^2 \frac{1}{\omega R} \qquad (\omega R\gg 1) \;.
\end{equation}
Therefore, the total emitted energy fraction up to frequency $\omega_M$ is given
by
\begin{equation}\label{enFrac}
  \frac{E^\GW}{\sqrt{s}}=\Theta_s(b)^2\left[\frac23(1-\log2)\log(\omega_M R)
    +\cdots\right]
\end{equation}
in agreement with the preliminary estimate~\eqref{eGW}, and is small, of order
$\Theta_s^2$, up to a logarithm of $\omega_M$.

It is important to note that such results follow from eq.~\eqref{freqd} which is
independent of $\hbar$, and should therefore have a direct classical
interpretation. Indeed, the classical expression of~\cite{GrVe14}
--- which is here obtained by including rescattering corrections
(sec.~\ref{s:rescattering}) --- has by eq.~\eqref{ResClas} the form
\begin{equation}\label{dEclas}
  \frac{\dif E^\GW_{\mathrm{class}}}{\dif\omega}
  = 2 Gs \frac{\Theta_s^2}{\pi} \int\frac{\dif^2\zt}{\pi |z|^4}\;\left(
    \frac{\sin\omega R \Phi(\zt)}{\omega R}\right)^2  \;,
\end{equation}
which differs from~\eqref{freqd} because $\omega R\Phi(\zt)$ occurs in
exponentiated form. At small $\omega R\ll 1$ the two results are essentially
equivalent.  On the other hand, the large $\omega R$ behaviour of \eqref{dEclas}
is provided by the whole small $|z|$ region when
$\Phi\simeq-\frac12|z|^2\cos(2\phi)$, to yield the result
\begin{equation}\label{dEclar}
  \frac{\dif E^\GW_{\mathrm{class}}}{\dif\omega}
  \simeq \frac{\sqrt{s}\Theta_s^2}{2\pi\omega}
  \int_0^{2\pi}\frac{\dif\phi'}{2\pi} \int_0^\infty \frac{\dif\xi}{\xi^2} \;
  \left[1-\cos(\xi\cos\phi')\right] =
    \frac{\sqrt{s}\Theta_s^2}{2\pi\omega} \;.
\end{equation}
The latter is in agreement with the $1/\omega$ behaviour of eq.~\eqref{asymSp}
but with a slightly different coefficient $(2\pi)^{-1}\simeq 0.16$ instead of
$(2/3)(1-\log 2)\simeq 0.20$. We conclude that rescattering effects are somewhat
important at large $\omega R$, but do not change the qualitative $1/\omega R$
behaviour of the spectrum.

A related important question is whether the emitted energy
fraction~\eqref{enFrac} is limited by the quantum energy bound
$\omega_M<E/\hbar$ only, or instead should be cutoff at purely classical level.
In such a case we would expect that the $\omega R$ distribution is further
suppressed by higher order contributions to the Riemann tensor, yielding e.g.\ a
$(\omega R)^{-2}$ behaviour, or higher. An argument in favour of the classical
cutoff, advocated in ref.~\cite{GrVe14}, is detailed in app.~\ref{a:cutoff}.

\subsection{Frequency and angular dependence\label{s:azim}}

In this section we present plots of the resummed amplitude and of the
corresponding radiated energy distribution obtained by numerical evaluation. In
this way, besides verifying the asymptotic behaviours derived in
sec.~\ref{s:regimes} and~\ref{s:pfd}, we can visualize the shape of such
quantities in the transition region $\omega R\sim 1$. Furthermore, we can also
study the angular distribution of radiation and notice very peculiar features.

Our first task is to rewrite the resummed amplitude $\ampRid$ given in
eq.~\eqref{ResFinal} by means of a representation with the lowest number of
integrals, having good convergence properties. It is possible to integrate the
last line of eq.~\eqref{ResFinal} in $\dif^2\zt$ and to express the result in
terms of the special function
\begin{equation}\label{defF}
  \spf(z) \equiv \int_0^\infty \dif\eta\;\esp{-\eta}\frac{\eta}{z+\eta}
  =  1-z\esp{z}E_1(z) \;,
\end{equation}
strictly related to the exponential integral (and to the incomplete
gamma-function)\\
$E_1(z)=\Gamma(0,z)$. We are thus left with a compact
one-dimensional integral over $\xi$
\begin{equation}\label{M1int}
  \ampRid = \sqrt{\alpha_G} \frac{R}{2\pi} \esp{-2\ui\phi_\tht}
  \int_{\xi_{\min}}^{\xi_{\max}}\dif\xi\; \frac{w}{w^*}\esp{\ui w^*}\Im
  \left[\frac{\esp{\ui w}}{w}\spf(-\ui w)\right] \;, \quad
  w\equiv \omega R\left(\frac{\theta}{\Theta_s}-\xi\right) \;,
\end{equation}
where $[\xi_{\min},\xi_{\max}]$ is $[0,1]$ in the lab frame and $[-1/2,1/2]$ in
the Breit frame. The singularity at $w=0$, i.e., $\xi=\theta/\Theta_s$, is
harmless, being integrable. From the previous expression it is clear that, apart
from the prefactor $\sqrt{\alpha_G}R$, $\ampRid$ depends only on $\omega R$,
$|\theta/\Theta_s|$ and $\phi_\tht$.

\subsubsection{Energy spectrum\label{s:es}}

\begin{figure}[t]
  \includegraphics[width=0.49\linewidth]{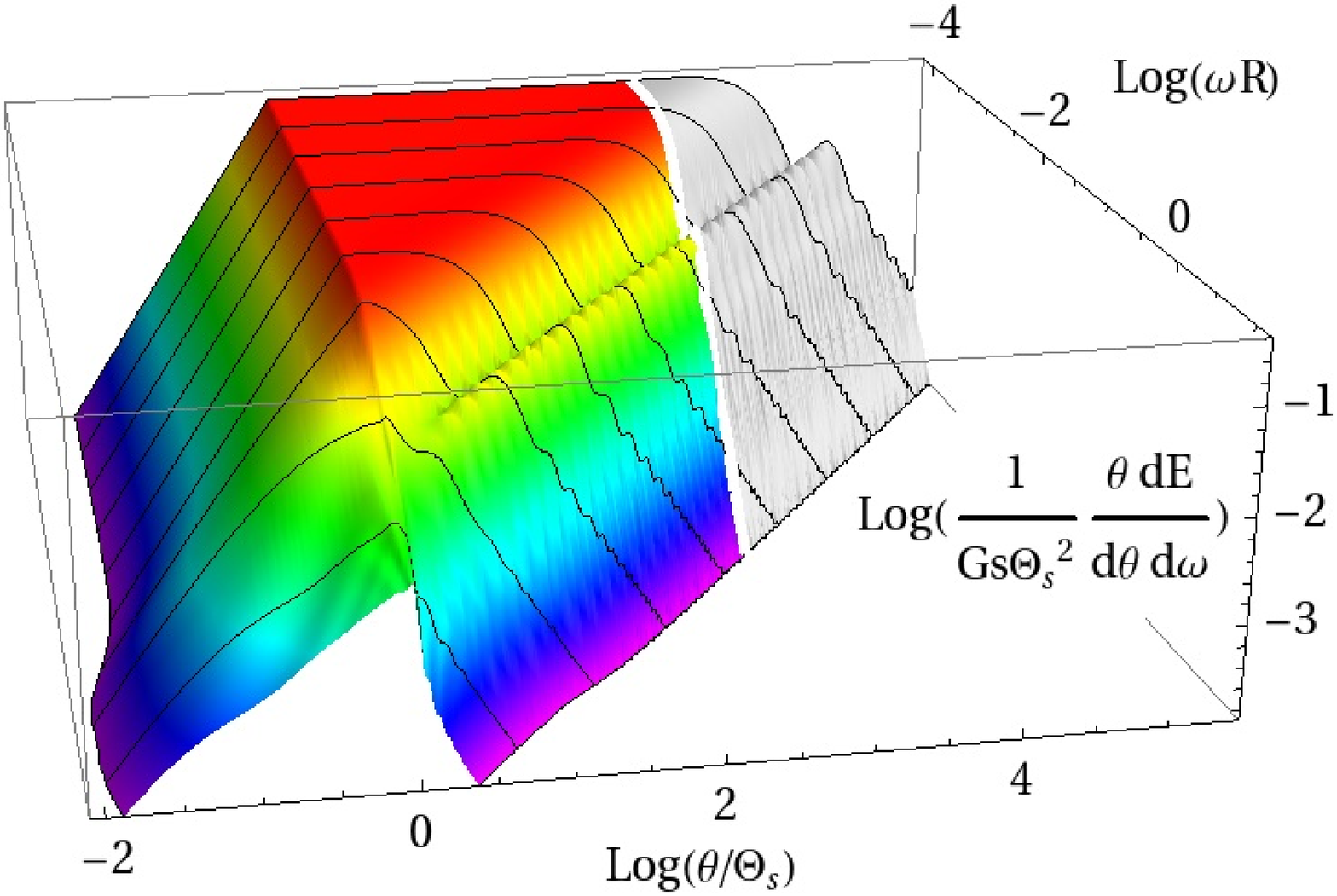}\hfill
  \includegraphics[width=0.49\linewidth]{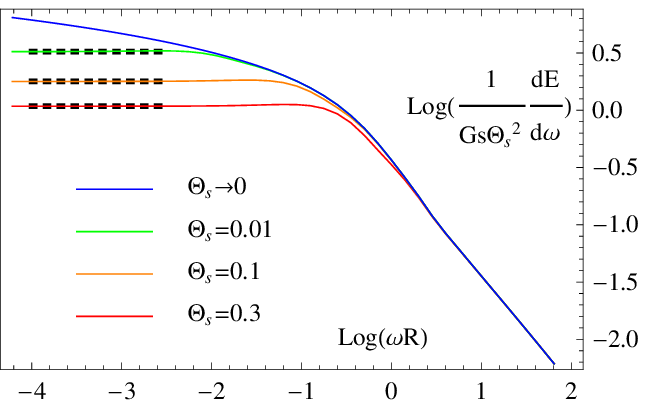}
  \caption{\it {\em a) [left]}: Azimuthally integrated spectrum vs.\ $\omega R$
    and $\theta/\Theta_s$. The shaded region on the right is excluded by the
    kinematic bound $\theta<1$ (for the choice $\Theta_s = 10^{-3}$).
    \newline
    {\em b) [right]}:  Frequency spectrum of gravitational
      radiation for various values of $\Theta_s$. For each $\Theta_s > 0$
      the ZFL value $\frac2{\pi} \log(1.65/\Theta_s)$ is obtained (dashed
      lines).}
  \label{f:resumSpec}
\end{figure}

Let us start by displaying the main features of the gravitational wave spectrum
of eq.\ (\ref{enDist}) in the ``geometrical'' approximation of
eqs.~\eqref{ampSin} and~\eqref{freqd}. In fig.~\ref{f:resumSpec}.a we plot (in
log scale) the differential emitted energy w.r.t. $\omega R$ and
$|\theta/\Theta_s|$ (i.e., after integration over the azimuthal angle $\phi$).

Fig.~\ref{f:resumSpec}.a shows very clearly that the spectrum is dominated by a
flat plateau (where kinematically accessible) whose shape can be easily
explained as follows. The spectrum falls on the left ($\theta < \Theta_s$)
because of phase space and of the absence of collinear singularities. It also
falls on the right when $\omega R = b q \Theta_s/\theta > \Theta_s/\theta$,
since then $b q >1$ (see eq.~\eqref{softMrid}). The last limitation (shaded
region on the right) is due to the trivial kinematic bound $\theta<1$.  As a
result, for fixed $\omega R < 1$ the length of plateau in $\log \theta$ is
$\log(1/\omega R)$ while it disappears completely for $\omega R >1$.

This is the reason why the spectrum in $\omega$ shown in fig.~\ref{f:resumSpec}.b
(obtained by a further integration over the polar angle $|\theta|$ and summing
the contributions of the two hemispheres) shows two very distinct regimes:

{\it(i)} $\omega R\ll 1$. In this regime the amplitude is well approximated by
eq.~\eqref{intDist}.  We see that the really infrared regime holds only in the
tiny region $\omega<1/b\iff\omega R<\Theta_s$, with a rapidity plateau up to
$|y|<Y_s\equiv\log(b/R)$ (in fig.~\ref{f:resumSpec}.a this are the deepest
horizontal sections of the plateau which are limited on the right by the shaded
region). Such rapidity plateau is much smaller than $Y=\log(Eb/\hbar)$, the
rapidity range available in the single H-diagram emission. Correspondingly, the
energy spectrum is flat, as one can see on the leftmost part of
fig.~\ref{f:resumSpec}.b for the lines with non vanishing $\Theta_s$. On the
other hand, here the small-$\omega$ number density in rapidity,
$(Gs/\pi)\Theta_s^2$, agrees with the one used in~\cite{ACV90} and with the
zero-frequency limit (ZFL) of~\cite{We65,Sm77}. For larger values of $\omega R$,
as already noticed, the length of the horizontal sections of the plateau
decreases, therefore we observe a logarithmic decrease of the energy spectrum
for $\omega R\lesssim 1$.

{\it(ii)} $1<\omega R< \omega_M R$. That is the most interesting region of the
spectrum, which in fig.~\ref{f:resumSpec}.b exhibits the large $\omega R$ decrease
$\sim 1/\omega R$, in perfect agreement with eq.~\eqref{asymSp}. This feature
originates from graviton emission all along the eikonal chain, summarized in the
resummation factor \eqref{hsres}, which contains the effective coupling $\omega
R$ and yields the decoherence effect for large $\omega R$ values which is exhibited
in fig.~\ref{f:resumSpec}.b.

It is important to note that curves for various values of $\Theta_s$ (and thus of
$b$) yield different ZFL's, as expected, but then coalesce in a common curve
for $\omega \gtrsim 1/b$, the blue curve in fig.~\ref{f:resumSpec}.b, which is
therefore universal and corresponds to eq.~\eqref{freqd}.

\begin{figure}[t]
  \centering
  \includegraphics[width=0.5\linewidth]{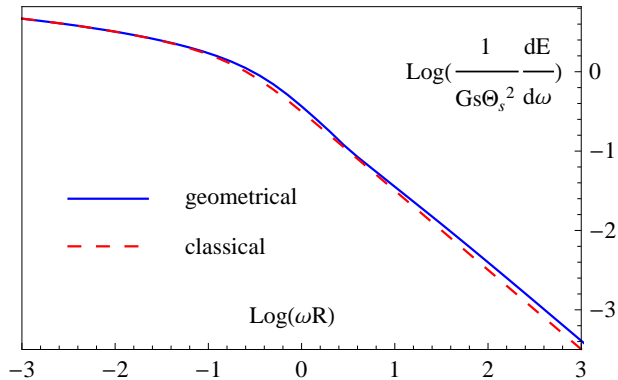}\hfill
  \caption{\it Comparison of the universal limit of our energy spectrum in the
    ``geometrical'' approximation (solid blue) with the classical result of
    ref.~\cite{GrVe14} (dashed red).}
  \label{f:quantClass}
\end{figure}

It is interesting to compare (fig.~\ref{f:quantClass}) the blue curve of
fig.~\ref{f:resumSpec}.b of our ``geometrical'' approach with the classical
counterpart of ref.~\cite{GrVe14} in eq.~\eqref{dEclas}. We can see that the
agreement is pretty good, even if the difference starts being important at large
$\omega R$ values, suggesting that rescattering corrections (not included in
eq.~\eqref{freqd}) and perhaps also quantum effects should be better evaluated
in this region. For instance, the upper limit $\omega_M$ quoted here, which
occurs in the total emitted-energy fraction \eqref{enFrac}, is provided in any
case by phase space, i.e., by the quantum frequency $E/\hbar$. But it is likely
that, as advocated in ref.~\cite{GrVe14} and illustrated in app.~\ref{a:cutoff},
the classical theory will provide by itself a physical cutoff, of order
$\omega_M\sim R^{-1}\Theta_s^{-2}$.

\subsubsection{Angular distributions\label{s:ad}}

\begin{figure}[t]
  \centering
  \includegraphics[width=0.37\linewidth]{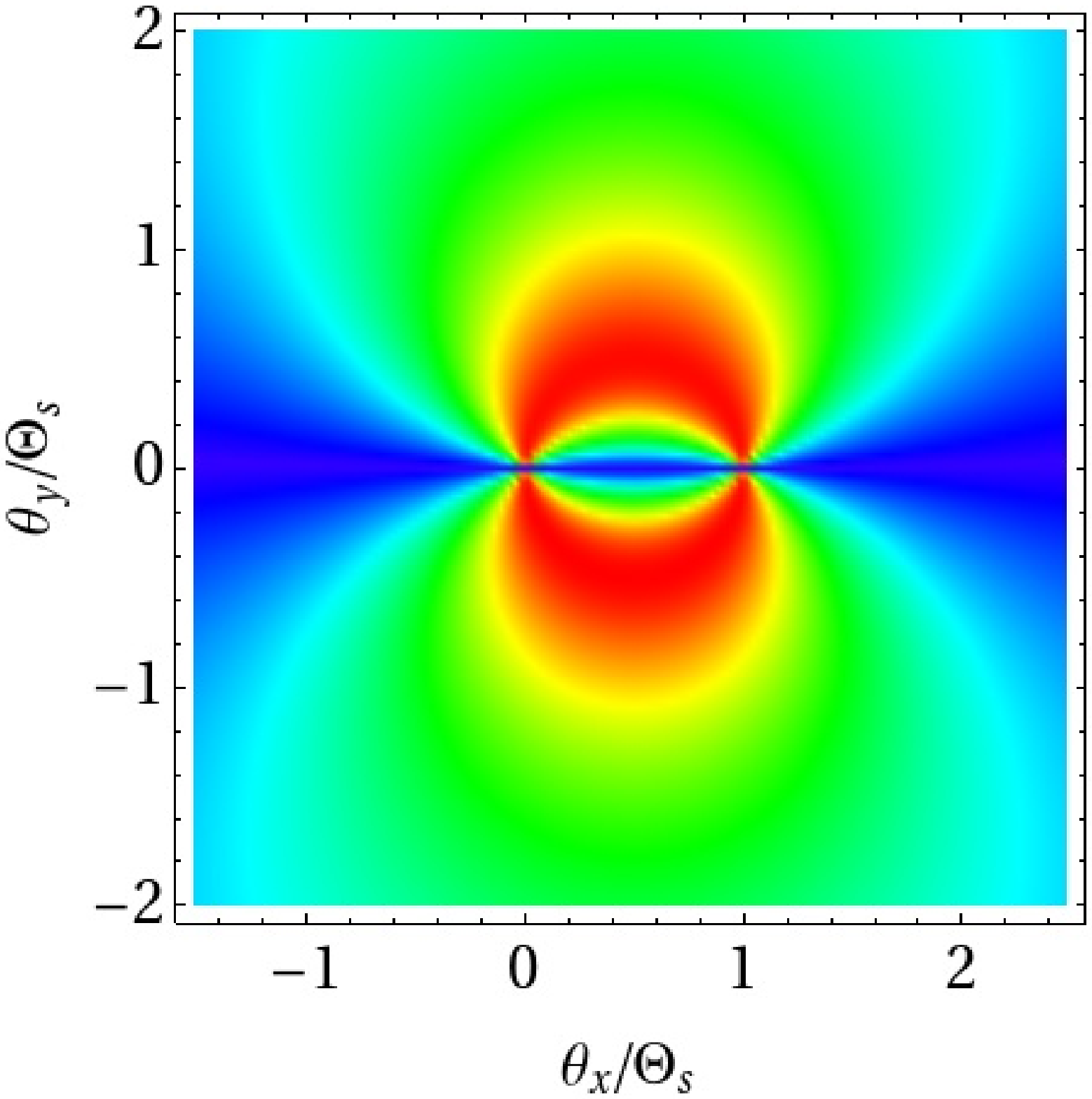}\hspace{0.12\linewidth}
  \includegraphics[width=0.37\linewidth]{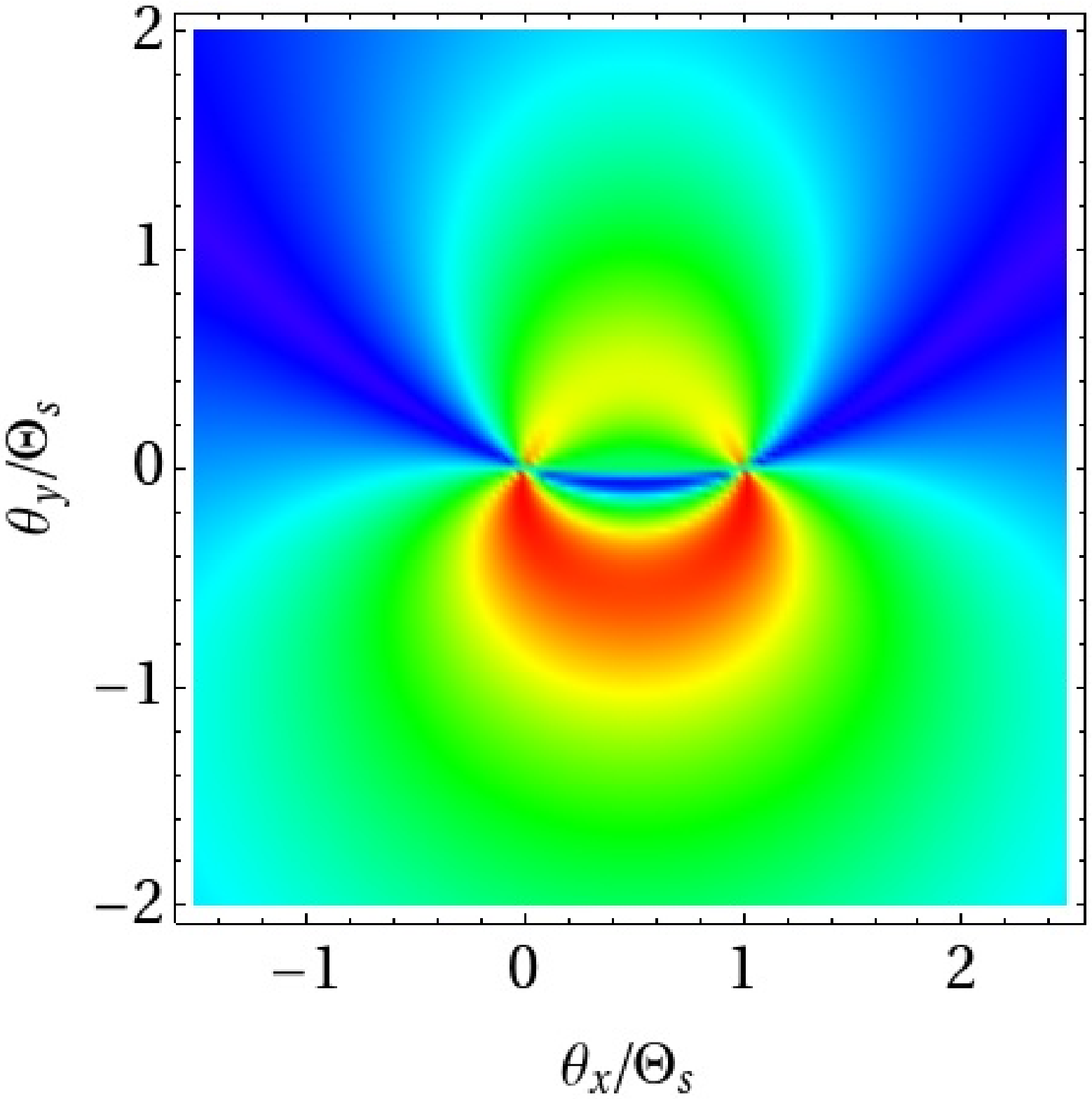} \\
  \includegraphics[width=0.37\linewidth]{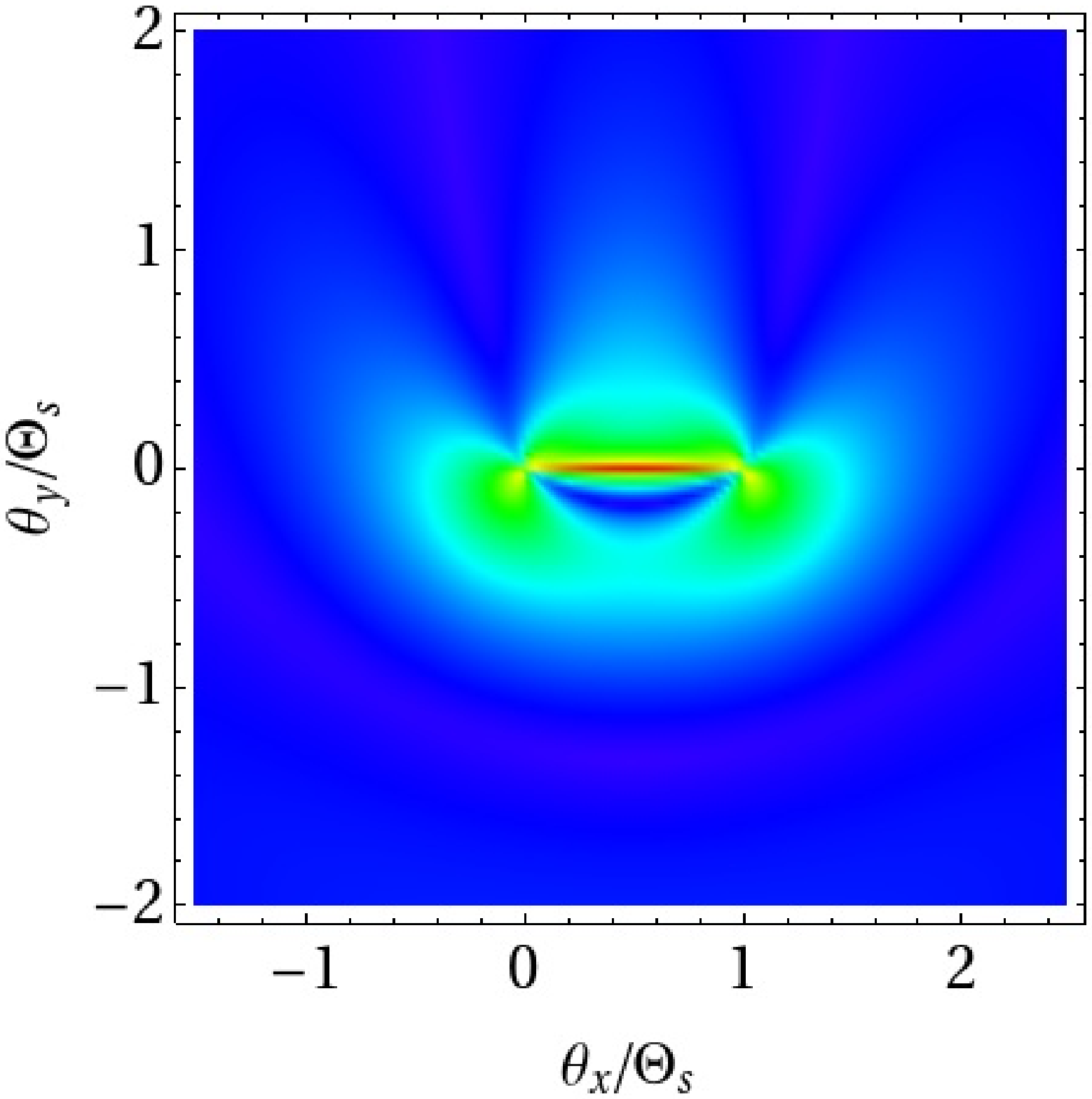}\hspace{0.12\linewidth}
  \includegraphics[width=0.37\linewidth]{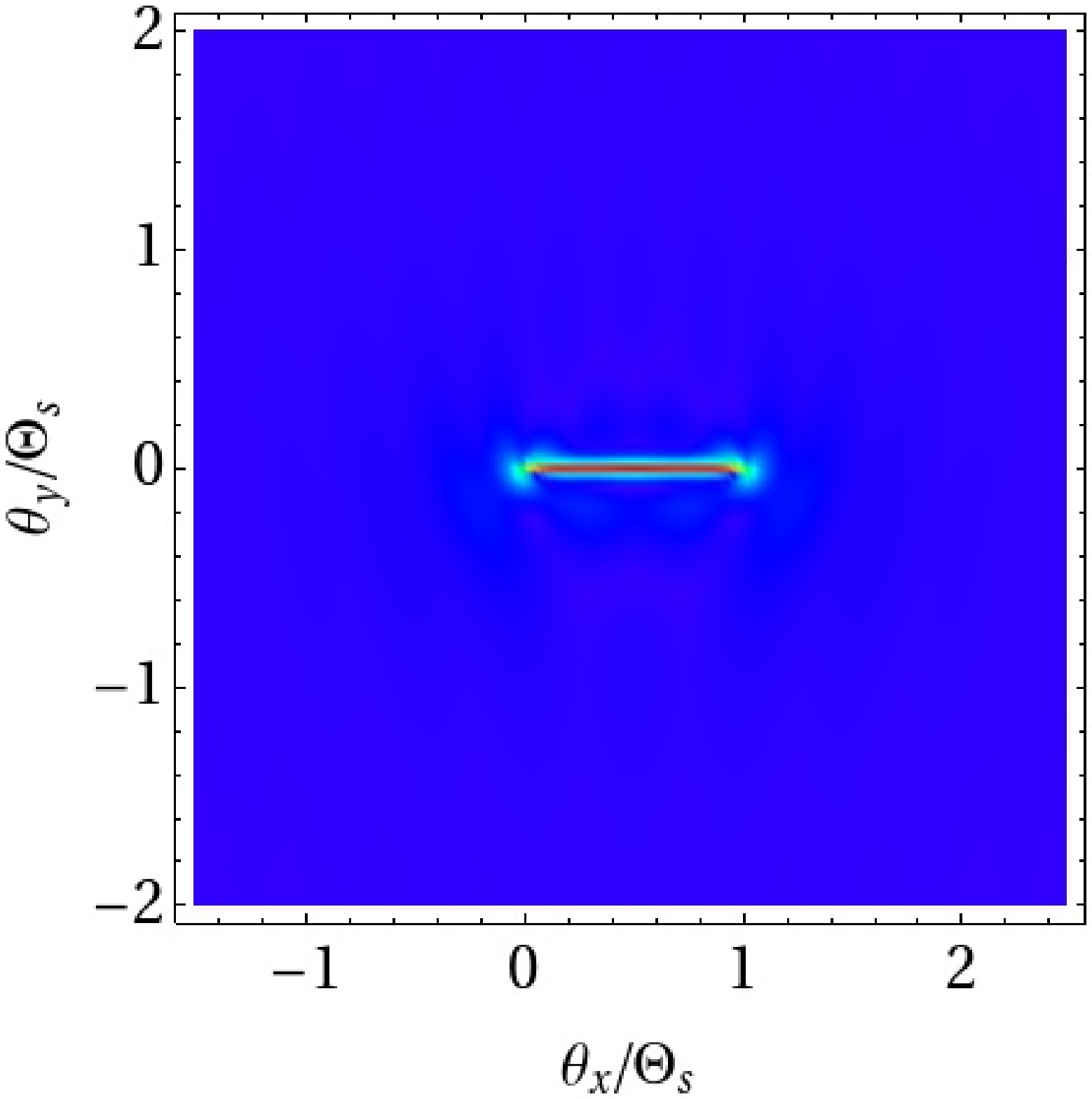} \\
  \caption{\it Modulus of the resummed amplitude $|\ampRid|$ vs. transverse
    direction $\tht$, normalized to unit $\Theta_s$, for 4 values of $\omega R$.
    From left to right, top to bottom: $\omega R =$ 0.001, 0.125, 1, 8. The
    increase of $|\ampRid|$ from 0 to its maximum (for each plot) is represented
    by colours ranging from blue to red.}
  \label{f:angle}
\end{figure}

In order to have a picture of the full angular distribution of the radiation at
various frequencies, in fig.~\ref{f:angle} we plot $|\ampRid|$ in the
$\bk{\theta_x/\Theta_s,\theta_y/\Theta_s}$-plane for some values of $\omega R$;
the maximum values of $|\ampRid|$ are attained in the red regions, while
vanishing values of the amplitude are shown in blue.

We see that for $\omega R\ll 1$ the emission is symmetric w.r.t.\ the symmetry
axes of the process, and is spread in a rather wide region around the particles'
directions, in particular at $|\phi_\tht|\simeq\pi/2$, in agreement with the
$\sin\phi_\qt$ dependence of eq.~\eqref{softMrid}. Moving to larger values of
$\omega R\sim 1$ the helicity amplitude shows an asymmetry w.r.t. the $x$-axis.
The symmetry is restored by the symmetrical behaviour of the amplitude with
opposite helicity. We note also a progressive shrinkage of the emission in the
region close to the particles' directions. At large values of $\omega R\gg1$ the
effective support of the amplitude is just a thin strip around the interval
$\theta\in[0,\Theta_s]$ whose width decreases as $1/(\omega R)$, also in this
case in agreement with the analytic estimate~\eqref{hardMrid}.

\begin{figure}[t]
  \centering
  \includegraphics[width=0.5\textwidth]{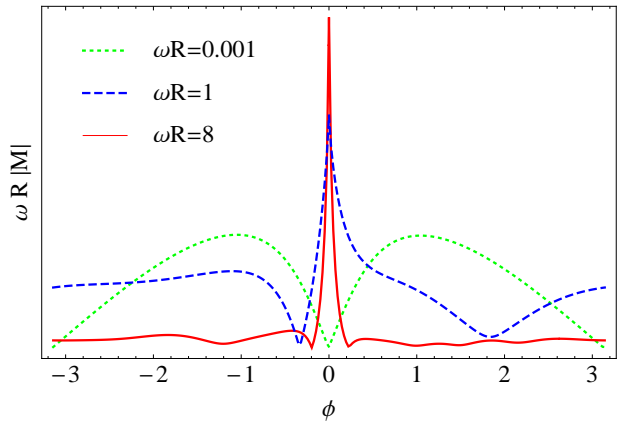}
  \caption{\it Azimuthal dependence of the rescaled resummed amplitude
    $\omega R|\ampRid|$ (in arbitrary units) for $|\theta|=\frac12|\Theta_s|$
    and for three values of $\omega R$.}
  \label{f:azim}
\end{figure}

A more quantitative graphical representation of the azimuthal dependence of the
amplitude is shown in fig.~\ref{f:azim}, where we fix the polar angle
$|\theta|=\frac12|\Theta_s|$ and plot, for various values of $\omega R$, the
amplitude versus $\phi_\tht$, rescaled by $\omega R$. At small $\omega R\ll1$
(dotted green curve) we see the expected $|\sin\phi_{\qt}|$ behaviour of
eq.~\eqref{softMrid} (it becomes exactly $|\sin\phi_\tht|$ for $|\theta|$ much
larger or much smaller than $|\Theta_s|$). At intermediate $\omega R\sim 1$
(dashed blue curve) the asymmetry in $\phi_\tht$ is evident, and the enhancement
around $\phi_\tht\simeq 0$ starts taking place. At large $\omega R \gg 1$ (solid
red curve) the amplitude shows a narrow peak at $\phi_\tht=0$, whose width and
height are inversely proportional to $\omega R$. At finite $\phi_\tht\neq 0$,
the amplitude is more and more suppressed with increasing $\omega R$, according
to eq.~\eqref{hardMrid}.

On one hand, the presence of such narrow peak (which becomes of constant height
after multiplication by $\omega$, see eq.~\eqref{spectrum}) explains the
$(\omega R)^{-1}$ decrease of the radiated energy. On the other hand, it
suggests that the radiation is sort of more and more confined along the
trajectories of the fast particles with increasing $R\propto\sqrt{s}$, at fixed
$b$ and $\omega$.

\subsection{Absorptive part and resummation effects\label{s:apre}}

We are now in a position to discuss the total emission multiplicity which is
related to the imaginary part of the resummed scattering amplitude
\begin{equation}\label{mult}
  \bk{N}_{\Theta_s} = \int_{\Delta\omega}\frac{\dif N}{\dif\omega}\; \dif\omega
  = 4 \Im\delta_\res(\alpha_G,\Theta_s)
\end{equation}
and is also related to the no-emission probability
$P_0=\esp{-\bk{N}_{\Theta_s}}$ of eq.~\eqref{P0}. According to eq.~\eqref{hsres}
its general expression is ($\Delta\omega=\ord{b^{-1}}$)
\begin{align}
  4\Im\delta_\res(\alpha_G,\Theta_s) &= \alpha_G\Theta_s^2(b)\frac{2}{\pi}
  \int_{\Delta\omega}^{E/\hbar}\frac{\dif\omega}{\omega}
  \frac{\dif^2\zt}{\pi|z|^4}\;\Phi^2(\zt) \left(
    \frac{\sin\alpha_G\log\left|\hat\bt-\frac{\hbar\omega}{E}\zt\right|}{
        \alpha_G\log\left|\hat\bt-\frac{\hbar\omega}{E}\zt\right|}
      \right)^2 \nonumber\\
  &\stackrel{\alpha_G\gg1 }{\simeq} \alpha_G\Theta_s^2(b)\frac{2}{\pi}
  \int_{\Delta\omega}^{E/\hbar}\frac{\dif\omega}{\omega}
  \frac{\dif^2\zt}{\pi|z|^4}\;\Phi^2(\zt) \frac{\sin^2\omega Rx}{(\omega Rx)^2}
  \;, \label{imPart}
\end{align}
where we have assumed $\Delta\omega=\ord{b^{-1}}$ in order to have a reliable
completeness of the $\qt$ states.

In the first line of~\eqref{imPart} we have made use of the general
expression~\eqref{hsres} of the resummed field valid for any $\alpha_G$, while
in the second line we have considered the transplanckian limit $\alpha_G\gg1$,
$\omega R$ fixed, which is the main interest of the present paper. Such two
forms show very clearly that the estimate~\eqref{imPart} for $\alpha_G$ moderate
to small is substantially different from the one in the transplanckian limit. In
the first regime the resummation factor is a power series in $\alpha_G$,
starting from 1 for $\alpha_G\to0$, limit in which~\eqref{imPart} yields just
the H-diagram result called $4\Im\delta_2$ in~\cite{ACV90}. As a consequence the
$\omega$ values can go up to $\omega\sim E/\hbar$ yielding a relatively large
emitted energy and showing the $\sim\log s$ dependence in rapidity used by
ACV~\cite{ACV90} to hint at the real part of the amplitude from a dispersion
relation. Furthermore, due to the logarithmic $\omega$-dependence of the
resummation factor, the large-$\omega$ phase space is modified rather slowly by
varying the $\alpha_G$ value, thus suggesting an intermediate regime where the
real part could be calculated also.

On the other hand in the transplanckian regime ($\alpha_G\gg 1$, $\omega R$
fixed) of the present paper, the second line of~\eqref{imPart} shows that values
of $\omega\gtrsim\ord{R^{-1}}$ are substantially suppressed, thus leading to the
reduced rapidity $Y_s=2\log(b/R)$ mentioned before, to the subsequent resolution
of the energy crisis and to the emergence of our Hawking-like radiation, which
represents the main result of our investigation.

More precisely, in order to take into account arbitrarily small values of
$\Delta\omega$ in the transplanckian case, we distinguish the soft and
large-frequency contributions as in sec.~\ref{s:pfd} by writing, to logarithmic
accuracy,
\begin{align}
  4\Im\delta_\res(\Theta_s) &= \frac{2\alpha_G}{\pi}\Theta_s^2 \left[
    \int_{\Delta\omega}^{1/R}\frac{\dif\omega}{\omega}\;\log\min
    \left(\frac{b}{R},\frac1{\omega R}\right) +\int\frac{\dif^2\zt}{\pi|z|^4}\;
    \Phi^2(\zt)\int_{|x|}^\infty\frac{\dif\tilde x}{\tilde x}\;
    \frac{\sin^2\tilde x}{\tilde{x}^2}\right] \nonumber \\
  &\simeq \frac{2\alpha_G}{\pi}\Theta_s^2\left(\log\frac{b}{R}
    \log\frac1{\sqrt{Rb}\Delta\omega}+Y_> \right) \;.\label{deltaRes}
\end{align}
We see that the large frequency integral ($\omega>R^{-1}$) yields just a
$R$-independent constant rapidity $Y_> \simeq 0.56$, while the soft one is
determined by the reduced rapidity $Y_s=\log(b/R)$, with the physical
consequences mentioned before. The phase space for the left part will eventually
disappear with increasing $R$. This suggests that in the extreme energy (and
large angle) limit, the total emission multiplicity will become just
proportional to $\alpha_G$ with a coefficient of which eq.~\eqref{deltaRes}
provides a provisional estimate.

\subsection{Towards large-angle resummation\label{s:tlar}}

It is of obvious interest to try to extend the radiation treatment presented
here to the extreme energy region $R \sim b$ where the scattering angle
$\Theta_s (b)$ becomes of order unity or larger, and a classical gravitational
collapse may take place. By following the path led by~\cite{ACV07} and mentioned
before, we encounter two kinds of effects: {\it (i)} those due to the evaluation
of the elastic eikonal function
$\delta \sim \alpha_G \, f(R^2/b^2, \lp^2/b^2)$, which becomes a
strong-coupling series showing perhaps some critical singularity at $b=R$, as in
the reduced-action model~\cite{ACV93}; and {\it (ii)} those arising from the
$\xi$-averaging, that is the coherent sum of $\delta$-exchange emissions at the
radiation level that we have just emphasized for $\delta = \delta_0$ as the
origin of the key role of $R$ in the energy emission spectrum.

By focusing on the second kind of effects, we may consider the first one as
simply the source of some structure in $\delta(b)$ and in the related
semiclassical trajectories~\cite{CC14} that will show up in the $\xi$-averaging
also. Therefore, the new features of elastic scattering in the strong coupling
regime for $R \sim b$ will provide new effects at the radiation level. A nice
picture of the present situation is exhibited in fig.~\ref{f:angle}.d, in which
the frequencies $\omega > 1/R$ send a last signal before being suppressed. This
supposedly essential message emphasizes the present span $2R/b$ of the
incidence angles $\xi \Tht_s$, and the impact-parameter direction $\hat b$. Both
parameters are expected to change with increasing $R$, because the semiclassical
trajectory is likely to approach a quasi-bound shape and the question is how
much that change will affect, by the $\xi$-averaging, the emerging radiation.

To provide an example, the impact parameter direction $\hat b$ is expected to
rotate by following the trajectory during time-evolution, and thus it is
possible (though not obvious) that the $\hat x$ and $\hat y$ directions will be
mixed by the $\xi$-averaging. If that is the case, the $y$ variable will be, on
the average, proportional to $x$ and, as a consequence, the modulation function
$\Phi(\zt)$ will be small and of order $|\zt|^2$, thus acquiring a cylindrical
symmetry and implying that the distribution~\eqref{emitEn} is of type
$1/(\omega R)^2$ (and not $1/(\omega R)$). That behaviour would yield a faster
suppression when approaching $R \sim b$ and would automatically provide a cutoff
in the energy fraction \eqref{enFrac}.

It is of course important to establish whether such a sizeable change of the
emerging radiation will really occur or not.  This is nontrivial, however,
because it requires a formal description of the $\xi$-averaging for higher
orders in the $\delta$-exchange emission also. It would appear, though, that
looking at the $\xi$-averaging by keeping the semiclassical trajectory
standpoint may produce some changes, but smooth ones, with no real hints of
information loss.

\begin{figure}[t]
  \centering
  \includegraphics[height=0.1\textheight]{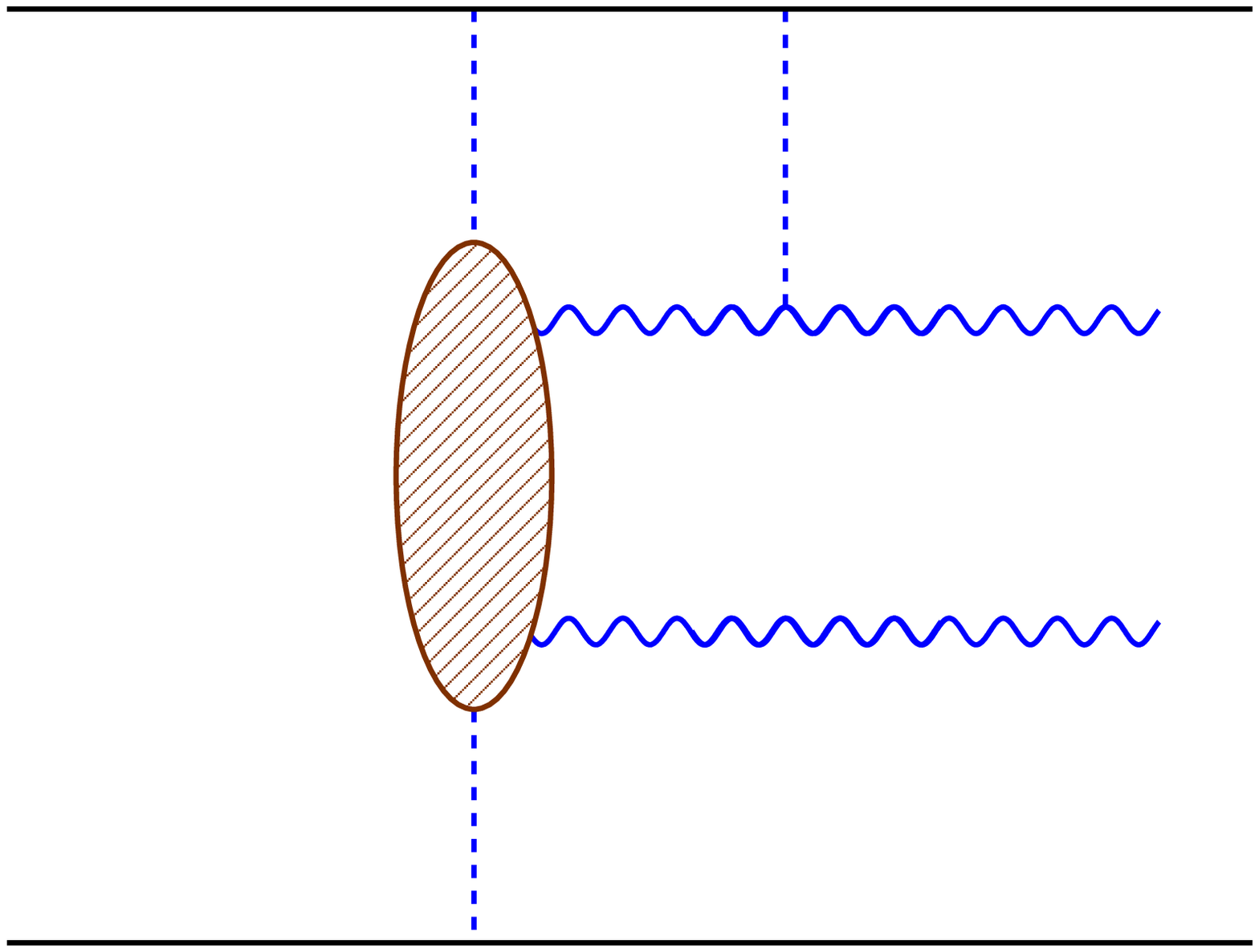}\hspace{2em}
  \includegraphics[height=0.1\textheight]{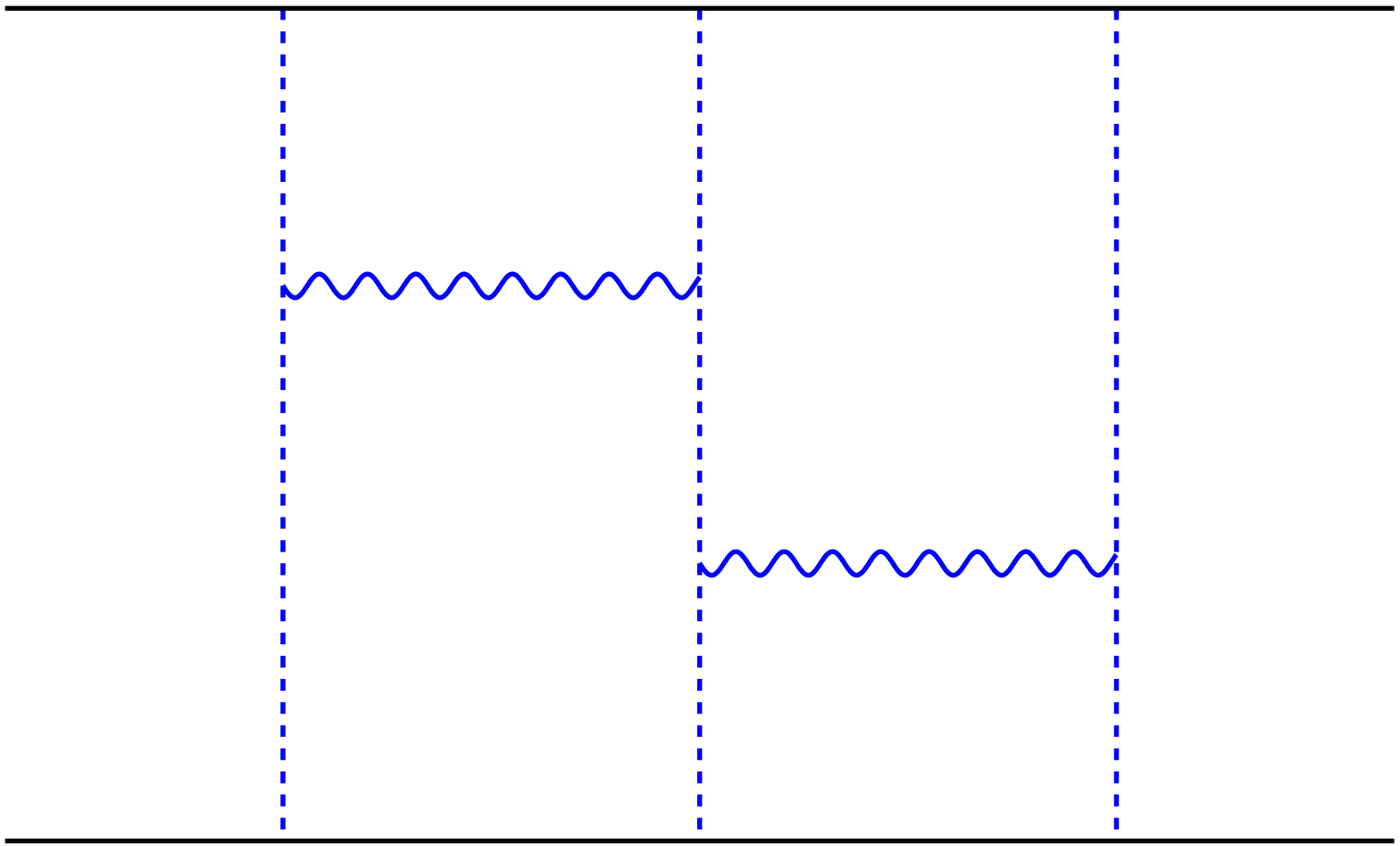}\hspace{2em}
  \includegraphics[height=0.1\textheight]{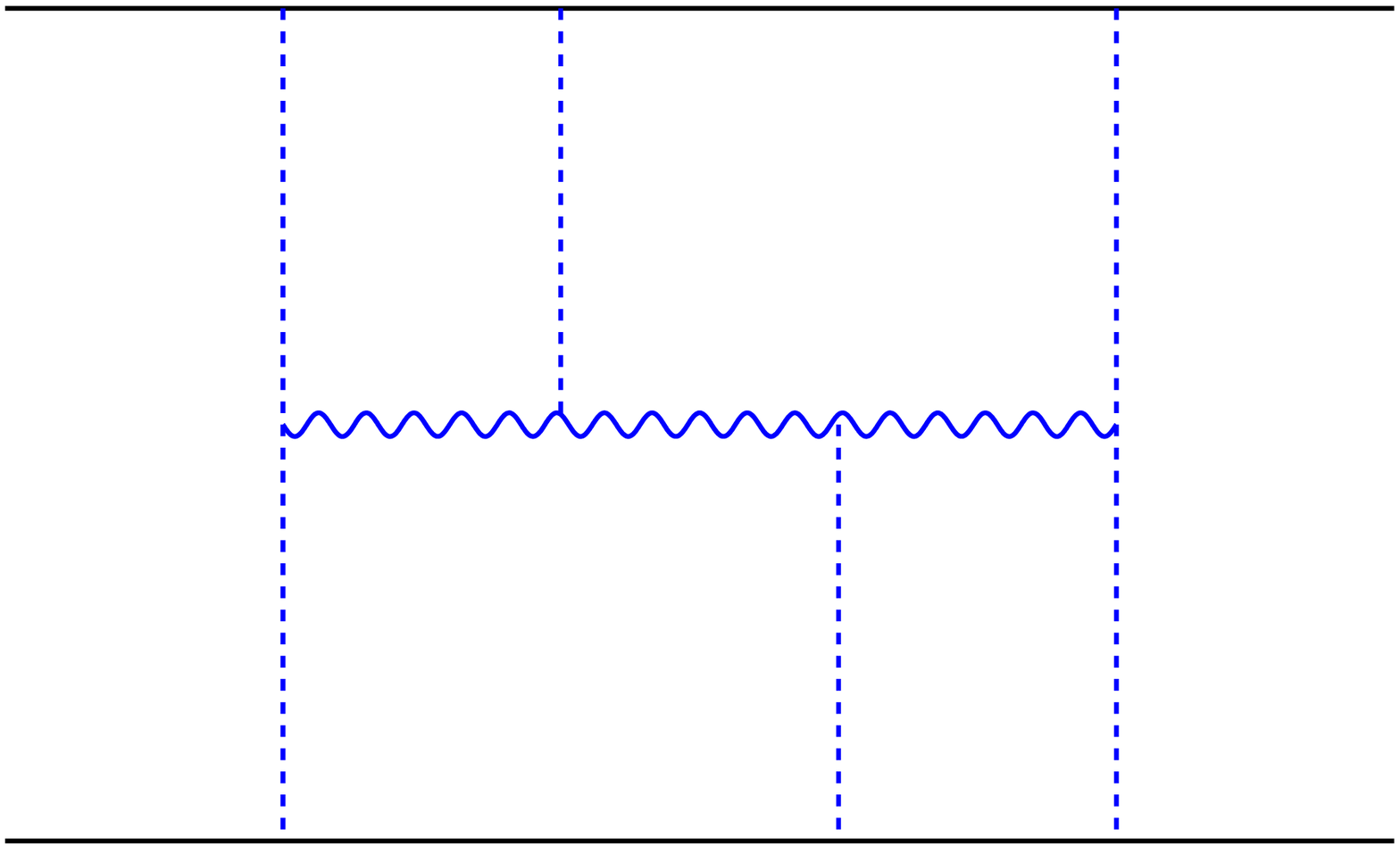}
  \caption{\it Subleading effective diagrams representing: a) two-graviton
    correlations; b) multi-H diagrams; c) rescattering corrections.}
  \label{f:multiH}
\end{figure}

If we now switch to the full quantum level, it is clear that --- besides
modifying the $\xi$-averaging by single-$\delta$ emission --- we have to modify
multiple emission also by introducing correlations in the coherent
state~\eqref{Shat} by the procedure of sec.~\ref{s:cso}. The simplest one
concerns the 2-graviton emission amplitude at order $G^3 s^2$, an example of
which is given in the diagram of fig.~\ref{f:multiH}.a, which introduces
quadratic terms $\sim a^2$, ${a^\dagger}^2$ in the exponent of~\eqref{Shat}. At
the same time such a diagram occurs in the corrections to the elastic eikonal
exemplified in fig.~\ref{f:multiH}.b,c, which contain further powers of $s$
because of $s$-channel iteration, and are thus of second order in the effective
coupling $R^2/b^2$.  The calculation of~\ref{f:multiH}.a can be devised by
following the lines of sec.~\ref{s:1gex} in the various soft and Regge regions,
so as to evaluate the correlation term.

Upgrading the present method to include the steps just described looks therefore
within reach. It may shed light on the existence of a large $\omega_M$ cutoff,
of possibly (higher-order) classical origin or quantum mechanical one. Going
much further however seems very hard, because treating both polarizations at
higher orders of the effective-action expansion is a fully two-dimensional
problem, unlike the reduced action model with one polarization in the
axisymmetric case investigated in~\cite{ACV07}. We nevertheless hope that, even
just at the next order, the present approach may provide us with some global
insight on the interplay of radiation and scattering in the strong-coupling
regime.

\section{Summary and perspectives}\label{s:concl}

We have investigated, in this paper, the peculiar features of the graviton
radiation associated to gravitational particle scattering at energies much
larger than the Planck mass. That scattering, at small deflection angles, is
described by a semiclassical $S$-matrix in $\bt$-space, which exponentiates the
eikonal function $\delta(\bt)$, of order $\alpha_G\equiv Gs/\hbar\gg1$ and
expressed as a power series in $R^2/b^2$, $R\gg\lp$ being the
gravitational radius of the system (sec.~\ref{s:tes}).

We find here that the ensuing radiation is expressed as a superposition of a
large number $\sim\alpha_G$ of single-hit emission processes, each one being
derived in a high-energy form which unifies the soft and Regge limits in the
whole angular range (sec.~\ref{s:1gex}). Combining the large emission number
($Gs/\hbar\gg1$) with the relatively small emitted energy ($\hbar\omega/E\ll1$)
produces in the emission amplitudes the effective coupling $\omega R$ which
tunes the resulting spectrum on the inverse gravitational radius
(sec.~\ref{s:bfact}). For that reason the emerging radiation is Hawking-like ---
that is with characteristic energies $\sim\hbar/R$ which decrease for increasing
input energies, even in the small-deflection angle regime in which the
$S$-matrix is explicitly unitary. In fact, as a consequence of coherence effects
in the superposition just mentioned, our unified amplitudes are found to have a
surprisingly simple interference pattern in $\omega R$, suppressing large
frequencies $\omega\gg1/R$ and reducing the radiated energy fraction to order
$\Theta_s^2$ (sec.~\ref{s:spec}).

Finally, we generalize the (quantum) factorization method of single-hit
emissions to include rescatterings of the emitted graviton and to resum them
(sec.~\ref{s:rescattering}). The ensuing emission amplitude neatly agrees with
ref.~\cite{GrVe14} in the classical limit and also includes the mentioned
quantum effects in a simple and elegant way, thus calling for further
investigation in the near future.

The ultimate goal of our thought experiment beyond the Planck scale is actually
to reach large scattering angles and the extreme-energy region $R\gtrsim b$
where a classical gravitational collapse may take place. Amati, Ciafaloni and
Veneziano proceeded a long way towards that goal from the scattering amplitude
stand point in the reduced action model~\cite{ACV07,Ciafaloni:2011de}. In such
truncated model they found that the $S$-Matrix, as functional of the UV-safe
solutions, shows an impact parameter singularity in the classical collapse
region, thus causing a unitarity deficit that they were unable to circumvent by
lack of information on the associated radiation (and on short distances).

We stress the point that, from the radiation point of view, we are better off
with the method presented here. In fact, we have just summarized two steps: the
first one yields the emission amplitudes for the single-hit process of
$\delta_0$-exchange (and corrections thereof in $\delta(b)$, sec.~\ref{s:tlar}),
the second one performs their superposition all along the eikonal deflection,
with its interference pattern. The latter may in turn feed back on higher order
corrections to the scattering amplitude itself. Therefore, by applying the
present method to an improved eikonal function, we could possibly provide the
radiation features given those of scattering and vice versa, by thus estimating
the exchange of information between them. We hope on this basis, to be able to
approach the classical collapse region in a smoother way, and to test in a more
direct way the features to be expected from a unitary evolution of the system.

\appendix
\section*{Appendices}

\section{Physical projections of the Weinberg and Lipatov currents\label{a:hlw}}

In this section we calculate the explicit projections of the Weinberg and
Lipatov currents over physical helicity states, proving in particular
eqs.~\eqref{Jw+-},~\eqref{MSi} and~\eqref{ampRegge}, the transformation
law~\eqref{ampTrasf}, and the symmetry relations~\eqref{Mb2},~\eqref{M12}.

We shall work in the gauges specified in eqs.~\eqref{polpm} and~\eqref{polLT}
which differ from the one used in~\cite{ACV90,ACV07} because the gauge vector
$\pol_L^\mu$ has the subtraction $-q^\mu / |\qt|$ ($q^\mu/|\qt|$) in the forward
(backward) emission case. Such subtraction is allowed by current conservation%
\footnote{The Lipatov current is exactly conserved, while the Weinberg current
  is conserved up to corrections of order $\ord{\omega/E}$ which we neglect
  throughout the paper.}
and is devised to suppress the longitudinal projections of external momenta in
jet 2 (jet 1) which are oppositely directed. For a generic tensor current
$J_{\mu\nu}$ we define
\begin{equation}\label{genericJ}
  J_\pm \equiv J_{\mu\nu} \pol_\mp^{\mu\nu} = \frac{1}{2} J_{\mu\nu} \pol_\mp^\mu \pol_\mp^\nu, \quad
  J_- \equiv \frac{J}{\sqrt{2}}, \quad J_+ = \frac{J^*}{\sqrt{2}}
\end{equation}
in terms of the basic complex vectors (note the nonstandard normalization
$\pol_+ \cdot \pol_- = -2$)
\begin{equation}\label{polexp1}
  \pol^\mu_{\pm} = \pol^\mu_T \pm \ui \pol^\mu_L = \left( \mp \ui \,
  \frac{|\tht|}{1+\sqrt{1-|\tht|^2}}, \mp \ui \esp{\mp \ui \phi_\tht},
  \esp{\mp \ui \phi_\tht}, \pm \ui \, \frac{|\tht|}{1+\sqrt{1-|\tht|^2}} \right) \; ,
\end{equation}
in the forward jet, where we have $q^3 = \omega \sqrt{1 - |\tht|^2} > 0$, and
\begin{equation}\label{polexp2}
  \tilde\pol^\mu_{\pm} = \pol^\mu_T \pm \ui \tilde\pol^\mu_L = \left( \pm \ui \,
  \frac{|\tht|}{1+\sqrt{1-|\tht|^2}}, \pm \ui \esp{\pm \ui \phi_\tht},
  \esp{\pm \ui \phi_\tht}, \pm \ui \, \frac{|\tht|}{1+\sqrt{1-|\tht|^2}} \right)
\end{equation}
in the backward one, where we send $q^3 \to \tilde q^3 = - q^3 < 0$, leaving the
other $q^\mu$ components unchanged.  We then see by inspection that, in the
small angle kinematics, the vectors~\eqref{polexp1} (\eqref{polexp2}) have
negligible longitudinal projection on jet 2 (jet 1) momenta, thereby making the
corresponding contributions to the currents negligible altogether, because
transverse projections are anyway of order $\ord{\theta_i^2}$ due to the lack of
collinear enhancement in the opposite jet.

\subsection{Forward hemisphere}

We need to calculate the typical scalar products
\begin{equation}\label{scalprods1}
  p_1 \cdot \pol_+ = -\ui \, E (\theta - \Theta_i) \esp{-\ui \phi_\tht},
  \quad p_1 \cdot q = \frac{E \omega}{2} |\theta - \Theta_i|^2 \; ,
\end{equation}
(restricting ourselves for cleanness of notation to the negative helicity
projection) which are given in complex notation in terms of
$\Tht_i \equiv \Tht_1$, that is the $\vec{p}_1$ incidence angle in the general
parameterization \eqref{mompar}. By using eq.~\eqref{genericJ}, the
$p_1$-contribution to the Weinberg current is then
\begin{equation}
  \frac{J^{(1)}_{W-}}{\kappa} = \frac{1}{2} \frac{(p_1 \cdot \pol_+)^2}{p_1 \cdot q}
  = -\frac{E}{\omega} \esp{2\ui (\phi_{\tht - \Tht_1} - \phi_\tht)} \; .
\end{equation}
Here we note the lack of collinear singularity due to the cancellation of the
squared numerator with the denominator in eq.~\eqref{scalprods1}.  The
contribution from $p_1'$ is analogous, with just the replacement
$\Tht_1 \to \Tht_1'$, while $p_2$ and $p_2'$ give negligible contributions in
this gauge and hemisphere, as explained before.

Therefore, introducing the coupling $\kappa$ and adding up the relevant terms we
get
\begin{align}
\label{WprefinalApp}
  J_{W-}^{(\Tht_i)}(q^3>0,\tht,\tht_s) & = \kappa \frac{E}{\omega}
  \left( \esp{2\ui (\phi_{\tht - \Tht_i - \tht_s}-\phi_\tht)}
    - \esp{2\ui (\phi_{\tht - \Tht_i} - \phi_\tht)} \right) \\
  \label{WfinalApp}
  & \simeq \kappa \frac{E}{\omega} \esp{-2\ui (\phi_\tht - \phi_{\tht - \Tht_i})}
  \left( \esp{2\ui (\phi_{\tht - \Tht_i - \tht_s} - \phi_{\tht - \Tht_i})} - 1 \right) \, .
\end{align}
Here we have used the approximate relation
\begin{equation}
\Tht_1' = \Tht_1 + \tht_s + \ord{\omega/E}
\end{equation}
neglecting the momentum conservation corrections of order $\ord{\omega/E}$,
which is allowed in regions a) + b) where the Weinberg current is relevant and
$|\qt|/|\qt_2|$ is small.

Eqs.~\eqref{WprefinalApp} and~\eqref{WfinalApp} prove formulas~\eqref{Jw+-} (in
which we have $\Tht_i = 0$ and $\Tht_f = \tht_s$) and~\eqref{MSi}, thus
confirming the transformation law~\eqref{ampTrasf} for the Weinberg current.

For the Lipatov current (including for convenience the denominator associated to
the $q_1^\mu$, $q_2^\mu$ virtualities), the negative helicity projection is
given by
\begin{equation}\label{Jjlip}
  \frac{J_{L-}^{(\Tht_i)}(q^3>0)}{|\qt_{1\perp}|^2 |\qt_{2\perp}|^2}
  = \frac{\kappa}{4} \left( \frac{(J\cdot \pol_+)^2}{|\qt_{\perp 1}|^2
      |\qt_{\perp 2}|^2} - (j \cdot \pol_+)^2 \right)
\end{equation}
(see eq.~\eqref{lipatov}), where we have
\begin{align}
  (j \cdot \pol_+) & = \frac{p_1 \cdot \pol_+}{p_1 \cdot q}
  = - \frac{2\ui}{\omega} \frac{\esp{-\ui \phi_\tht}}{(\theta - \Theta_i)^*} \; , \\
  \nonumber
  (J \cdot \pol_+) & = \left( \qt_{\perp 1}^2 - \qt_{\perp}^2 \right)
  \frac{-2\ui \esp{-\ui \phi_\tht}}{\omega(\theta - \Theta_i)^*} - 2 q_2 \cdot \pol_+ \\
  & = \left( \qt_{\perp 1}^2 - \qt_{\perp}^2 \right)
  \frac{-2\ui \esp{-\ui \phi_\tht}}{\omega(\theta - \Theta_i)^*} - 2\ui \esp{-\ui \phi_\tht} q_{2} \;.
\label{Jlip}
\end{align}
Here we have used current conservation to replace
$q_1^\mu - q_2^\mu \to -2 q_2^\mu$ and $\qt_{\perp1}$, $\qt_{\perp2}$ and
$\qt_{\perp}$ denote transverse (vectorial) components with respect to the
$\vec{p}_1$ direction, (and $q_{\perp1}$, $q_{\perp2}$ and $q_{\perp}$, the
corresponding complex versions), which are related to $\qt_1$, $\qt_2$ and $\qt$
by a rotation of angle $|\Tht_i|$ of the reference frame. Note in particular
that since we are considering a forward emission, $q_2^\mu$ has practically no
longitudinal component, while $q_1^\mu$ has; taking also into account that
$q_\mu = q_{1\mu} + q_{2 \mu}$, this implies
\begin{equation}\label{q2perp}
  \qt_{\perp2} \simeq \qt_{2}, \quad \qt_{\perp1} - \qt_{1} \simeq
  \qt_{\perp} - \qt \simeq - \omega \Tht_i.
\end{equation}

Now, rewriting $q_{\perp1} = q_\perp - q_{\perp2}$, using
$q_\perp = \omega(\theta - \Theta_i)$ and taking into account
eq.~\eqref{q2perp}, we can rewrite eq.~\eqref{Jlip} in the form
\begin{equation}\label{lip2}
  (J \cdot \pol_+) = \frac{-2\ui \esp{-\ui \phi_\tht}}{q_\perp^*}
  \left( \qt_2^2 - 2 \qt_{\perp} \cdot \qt_2 + q_\perp^* q_2 \right) =
\frac{2\ui \esp{-\ui \phi_\tht}}{\omega(\theta - \Theta_i)^*} q_2^* q_{\perp1}
\end{equation}
By substituting expression~\eqref{lip2} in eq.~\eqref{Jjlip} we finally get
\begin{equation}\label{LfinalApp}
  J_{L-}^{(\Tht_i)}(q^3>0,\qt,\qt_2) =
  \frac{\kappa}{|\qt_{\perp}|^2} \esp{2\ui(\phi_{\tht - \Tht_i} - \phi_\tht)}
  \left[ 1 - \esp{-2\ui(\phi_{\qt_2} - \phi_{\qt_{\perp} - \qt_2})} \right] \; ,
\end{equation}
which proves (by setting $\Tht_i = 0$) eq.~\eqref{ampRegge} of the text and the
transformation law to general $\Tht_i$~\eqref{ampTrasf} for the Lipatov current.
We note that the would-be singularities at $\qt_{\perp1}, \qt_{\perp2} = 0$ have
been canceled due to eq.~\eqref{lip2} and replaced by the phase difference in
eq.~\eqref{LfinalApp}, which also reduces the $\qt_\perp = 0$ singularity to a
linear integrable (in two dimensions) one.

\subsection{Backward hemisphere}

In this case the jet 2 is characterized by an incidence angle
$\tilde\Tht_i \equiv \Tht_2$ (which in the center-of-mass frame $\vp_1 + \vp_2 = 0$
is simply provided by $\tilde\Tht_i = - \Tht_1 = - \Tht_i$) and by the fact that
the emitted $q^\mu$ has a negative $\hat z$ component,
$\tilde q^3 = - \omega \sqrt{1 - |\tht|^2} < 0$.

By using the backward helicity vectors~\eqref{polexp2} we obtain
\begin{equation}\label{scalprods2}
p_2 \cdot \tilde\pol^+ = \ui \, E (\theta^* + \Theta_i^*) \esp{\ui \phi_\tht}, \quad
p_2 \cdot \tilde q = \frac{E \omega}{2} |\theta + \Theta_i|^2 \; ,
\end{equation}
and, as a consequence
\begin{equation}
  \frac{J^{(2)}_{W-}}{\kappa} = \frac{1}{2} \frac{(p_2 \cdot \tilde\pol_+)^2}{p_1 \cdot q}
  = -\frac{E}{\omega} \esp{-2\ui (\phi_{\tht - \Tht_2} - \phi_\tht)} \;.
\end{equation}
Since jet 1 is switched off, the Weinberg current is simply
\begin{equation}\label{Wbackprefinal}
J_{W-}^{(-\Tht_i)}(-q^3<0,\tht,\tht_s)  = \kappa \frac{E}{\omega} \esp{2\ui \phi_\tht}
\left( \esp{-2\ui \phi_{\tht + \tht_s + \Tht_i}} - \esp{-2\ui \phi_{\tht + \Tht_i}} \right) \; ,
\end{equation}
where we used the relation
\begin{equation}
\Tht_2' = - \Tht_i - \tht_s + \ord{\frac{\omega}{E}}
\end{equation}
and neglected, as in the forward case, the $\ord{\omega/E}$ corrections in the
a) + b) regions.

We note that eq.~\eqref{Wbackprefinal} can be recast in a form transforming like
the complex conjugate (or opposite helicity) of~\eqref{WfinalApp} by the
replacement $\tht \to - \tht$, to yield
\begin{equation}\label{Wbackfinal}
\begin{split}
  J_{W-}^{(-\Tht_i)}(-q^3<0,-\tht,\tht_s) & = \kappa \frac{E}{\omega}
  \esp{2\ui (\phi_\tht - \phi_{\tht - \Tht_i})}
  \left( \esp{-2\ui (\phi_{\tht - \Tht_i - \tht_s} - \phi_{\tht - \Tht_i})} - 1 \right) \\
  & = J_{W+}^{(\Tht_i)}(q^3>0,\tht,\tht_s) \; ,
\end{split}
\end{equation}
thus proving eq.~\eqref{jw2} and yielding the basis for the
relationship~\eqref{Mb2}, in which we also use the F.T. with respect to
$\qt_2 = E \tht_s + \qt$.

All that is left is obtaining the helicity projections of the Lipatov current in
the backward hemisphere.  Using eqs.~\eqref{lipatov} and~\eqref{genericJ} we
obtain
\begin{equation}\label{Jjlip2}
  \frac{J_{L-}^{(-\Tht_i)}(-q^3<0)}{|\qt_1|^2 |\tilde\qt_{2\perp}|^2}
  = \frac{\kappa}{4} \left( \frac{(J\cdot \tilde\pol_+)^2}{|\qt_1|^2
      |\tilde\qt_{\perp 2}|^2} - (j \cdot \tilde\pol_+)^2 \right) \; ,
\end{equation}
where we used the fact that $q_1^\mu$ has negligible longitudinal components for
an emission in jet 2 and defined (in complex notation)
\begin{equation}
  \tilde q_\perp - q \simeq \tilde q_{\perp2} - q_2
  \simeq - \omega \tilde\Theta_i = \omega \Theta_i \; .
\end{equation}
The vector currents projections are
\begin{align}
  (j \cdot \tilde\pol_+) & = \frac{p_1 \cdot \tilde\pol_+}{p_1 \cdot \tilde q}
  = - \frac{2\ui}{\omega} \frac{\esp{\ui \phi_\tht}}{(\theta + \Theta_i)} \; , \\
  \nonumber
  (J \cdot \tilde\pol_+) & = \left( \tilde\qt_{\perp 2}^2
    - \tilde\qt_{\perp}^2 \right) (j \cdot \tilde\pol_+) + 2 q_1 \cdot \tilde\pol_+
  = \frac{2\ui \esp{\ui \phi_\tht}}{\tilde\qt_\perp} \tilde q_{\perp2}^* q_{1} \; ,
\end{align}
where we performed the algebra along the lines explained before. We thus obtain
\begin{equation}\label{Lbackfinal}
  J_{L-}^{(-\Tht_i)}(-q^3<0,\qt,\qt_1) =
  \frac{\kappa}{|\tilde\qt_{\perp}|^2} \esp{-2\ui(\phi_{\tht + \Tht_i} - \phi_\tht)}
  \left[ 1 - \esp{2\ui(\phi_{\qt_1} - \phi_{\tilde\qt_{\perp} - \qt_1})} \right] \; .
\end{equation}
At $\Tht_i = 0$ the above result agrees with that in the forward jet by the
trivial replacement $\qt_1= \qt - \qt_2$, meaning that the Regge limit yields
the same form of the amplitude in either jet. On the other hand, at nonzero
$\Tht_i$ --- and by the replacement $\qt \to - \qt, \qt_1 \to - \qt_2$ --- it
yields the helicity relation
\begin{equation}
  J_{L-}^{(-\Tht_i)}(-q^3<0,-\qt,-\qt_2) = J_{L+}^{(\Tht_i)}(q^3>0,\qt,\qt_2) \; ;
\end{equation}
By finally using the transformation
$\tilde \qt_{\perp2} \to - \qt_\perp + \qt_2$ in the F.T. under the same
replacement, we prove eq.~\eqref{M12} of the text.

\subsection{$\zt$-representation\label{a:zrep}}

A generic phase difference of the form
$\esp{2\ui \phi_{\tht}} - \esp{2\ui \phi_{\tht'}}$, with
$\tht$, $\tht'$ generic 2-vectors, of the kind that appears in the physical
projections of both the Weinberg~\eqref{Msoft} and the Regge~\eqref{Mregge}
amplitudes, can be conveniently rewritten in integral form:
\begin{equation}\label{zrepA}
  \esp{2\ui \phi_{\tht}} - \esp{2\ui \phi_{\tht'}}
  = -2 \int \frac{\dif^2\zt}{{2\pi z^*}^2} \left( \esp{\ui A \zt \cdot \tht}
    - \esp{\ui A \zt \cdot \tht'} \right) \;,
\end{equation}
where $A$ is an arbitrary scale (the integration measure $\dif^2\zt / {z^*}^2$ in
the integral is scale invariant).  This is easily verified by performing first
the azimuthal, then the radial integration in $\dif^2\zt$. We get (setting for
simplicity $A = 1$):
\begin{equation}
\begin{split}
  \int \frac{\dif^2\zt}{2\pi{z^*}^2} \left( \esp{\ui \zt \cdot \tht}
    - \esp{\ui \zt \cdot \tht'} \right)
  &= - \int_0^\infty \frac{\dif |z|}{|z|} \left( \esp{2\ui \phi_\tht}
    J_2(|z| |\tht|) - \esp{2\ui \phi_\tht} J_2(|z| |\tht'|) \right) \\
  &= -\frac{1}{2}\left(\esp{2\ui\phi_\tht} - \esp{2\ui\phi_{\tht'}}\right)\;,
\end{split}
\end{equation}
where in the last step we used the scale invariance of the integration measure,
the standard integration formula $\int \frac{J_2(x)}{x} = - \frac{J_1(x)}{x}$
and the fact that $J_1(x) \sim x/2$ near $x = 0$. Eq.~\eqref{zrepA} is thus
proven.

\section{Transformation of helicity amplitudes\label{a:heltras}}

In this section we derive the transformation formula for amplitudes of definite
helicity when the momentum of the incoming particle undergoes a rotation. We are
mainly interested in small emission- and deflection-angles. Therefore in
sec.~\ref{s:sar} we derive some simple properties of small rotations, that will
be then applied in sec.~\ref{s:ampt} in order to obtain the transformation
formulas.

\subsection{Small-angle rotations\label{s:sar}}

In the eikonal approximation, the typical polar angles of the particles are
small.  This means that the 3-momentum of a particle in the forward region can
be written as
\begin{equation}\label{smallq}
  \vq=(\qt,q_z)\equiv\omega(\tht,\sqrt{1-\tht^2})\simeq\omega(\tht,1) \;,
\end{equation}
where $\omega=|\vq|$ and $\tht\equiv\qt/|\vq|$ is the intersection of $\vq$ with
the tangent plane to the unit-sphere at $\hat{z}=(0,0,1)$, so that $|\tht|$ can be
interpreted as polar angle when $|\tht|\ll 1$.
Note that such parametrization spans only half of the phase space. The backward
hemisphere is described by an analogous transverse vector spanning the tangent
plane at $-\hat{z}=(0,0,-1)$.

Any unit-vector $\hat{q}$ can be obtained by applying to $\hat{z}$ a rotation
around an axis on the $\bk{x,y}$-plane:
\begin{equation}\label{rotz}
  \hat{q} = R(\vec{\alpha})\hat{z} = \exp\{-\ui(L_x\alpha_x+L_y\alpha_y)\}\hat{z} \;,
\end{equation}
where the matrices $(L_k)_{mn}=-\ui\e_{kmn}:k=1,2,3$ are the generators
of rotations and $\vec{\alpha}=(\alpha_x,\alpha_y,0)$ denotes the rotation
vector. For small rotation angles $|\vec{\alpha}| \ll 1$, the exponentials can be
expanded to first order and one finds
\begin{equation}
  (\theta_x,\theta_y,1)=\left[1-\ui(L_x\alpha_x+L_y\alpha_y)\right]\hat{z}
  +\ord{\vec\alpha^2}\quad\text{with}\quad
  \begin{cases}
    \alpha_x = \theta_y\\ \alpha_y=-\theta_x
  \end{cases}
\end{equation}
In practice, small-angle rotations act as abelian translations on the
transverse components of unit vectors:
\begin{equation}\label{compRot}
  R(\vec\alpha_1) R(\vec\alpha_2)\hat{z} \simeq R(\vec\alpha_1) (\tht_2,1)
  \simeq(\tht_1+\tht_2,1)\simeq R(\vec\alpha_1+\vec\alpha_2)\hat{z} \;,
  \qquad (|\vec\alpha_j|\ll1)
\end{equation}
up to quadratic terms in $\vec\alpha_{1,2}$.

\subsection{Amplitude transformation\label{s:ampt}}

We are looking for a relation that connects a generic helicity amplitude with
arbitrary incoming momentum $\vp_i$ ($\Tht_i\neq0$) to another amplitude having
incoming momentum along the $\hat{z}$-axis ($\Tht_i=0$). Our procedure is based
on two main observations:
\begin{itemize}
\item[1)] helicity amplitudes are invariant under rigid rotations of {\em all}
  momenta {\em and} polarization vectors; by applying a suitable rotation we can
  bring the incoming momentum $\vp_i$ onto the $\hat{z}$-axis;
\item[2)] the rotated polarization vectors differ from the reference ones
  (eqs.~\eqref{polLT}) by a further rotation around the emitted graviton
  momentum; this is the rotation providing the helicity phase factor of the
  amplitude transformation.
\end{itemize}
Let us discuss the two points in more detail.

\paragraph{1)} A helicity amplitude $A^{(\lambda)}$ is defined by contraction of
a tensor amplitude $A^{\mu\nu}$ with some polarization tensor
$\pol^{(\lambda)}_{\mu\nu}$, e.g.,
$A^{(\lambda)}(p,q,\pol^*)=A^{\mu\nu}(p,q)\pol^{(\lambda)*}_{\mu\nu}$. The
notation shows that $A^{(\lambda)}$ explicitly depends on the emitted momentum
$q$, on the polarization vector $\pol^*$ and on additional momenta of the
process denoted by $p$.  Of course, $A^{(\lambda)}$ is invariant under Lorentz
transformations of $p,q$ and $\pol$, and in particular under any spatial
rotation $R$: $A^{(\lambda)}(Rp,Rq,R\pol^*)=A^{(\lambda)}(p,q,\pol^*)$.

In order to specify the polarization tensor $\pol^{(\lambda)}$ (and the helicity
amplitude), one has to uniquely define a pair of polarization vectors
$\pol_k:k=1,2$ which are orthogonal to the momentum $q$ of the emitted
radiation. Usually such vectors are chosen without time components, of unit
length and orthogonal to each other, in such a way that
$\{\vq,\vec\pol_1,\vec\pol_2\}$ forms a right-handed basis of 3-space. The
remaining degree of freedom is a rotation of the pair
$\{\vec\pol_1,\vec\pol_2\}$ around the $\vq$ axis. In this paper,
$\pol_1=\pol_T$ and $\pol_2=\pol_L+$ gauge-term $\propto q$.

We completely specify the polarizations by requiring $\vec\pol_T$ to be
orthogonal to a given momentum $\vp$: $\vec\pol_T\cdot\vp=0$.  According to this
recipe, we can write (omitting the 3D arrows) $\pol_k=\pol_k(p,q)$. Such
procedure is frame independent, therefore we have
\begin{equation}\label{rotPol}
  \pol_k(Rp,Rq) = R\,\pol_k(p,q) \;,
\end{equation}
meaning that if we simultaneously rotate $p$ and $q$, the polarization vectors
$\pol_k$ necessarily get rotated by the same matrix $R$. The same holds for
$\pol_k^*$.

In this paper we chose $p_1=E\hat{z}$ as reference vector orthogonal to
$\pol_T$, and defined accordingly
\begin{equation}\label{ampInc}
  \amp^{(\Tht_i)}(\Tht_f,\tht) \equiv A^{(\lambda)}\big(p_i,p_f,q,\pol^*(p_1,q)\big) \;,
\end{equation}
$p_i$ and $p_f$ being the incident and final momenta of the fast particle.  We
can relate this amplitude to a zero-incidence-angle one by applying a rotation
$R$ to the arguments of $A^{(\lambda)}$, in such a way that
$p'_i \equiv R p_i = p_1$:
\begin{align}\label{ampR1}
  \amp^{(\Tht_i)}(\Tht_f,\tht)
  &= A^{(\lambda)}\big(R p_i,R p_f,R q,R \pol^*(p_1,q)\big) \nonumber \\
  &= A^{(\lambda)}\big(p_1,R p_f,R q, \pol^*(R p_1,Rq)\big)
 \equiv A^{(\lambda)}\big(p_1,p'_f,q', \pol^*(p'_1,q')\big)
\end{align}
where a prime means ``rotated by $R$'' and we have used eq.~\eqref{rotPol} in
the second step.

In the case of small polar angles ($\tht,\Tht_i,\Tht_f\ll1$) we can write
$q=\omega(\tht,1)$, $p_i=E(\Tht_i,1)$ etc., and the previous rotation $R$
amounts just to a translation of $-\Tht_i$ on the angular variables:
$q'=\omega(\tht-\Tht_i,1)$, $p'_1=E(-\Tht_i,1)$ and so on.

\paragraph{2)} The last quantity in eq.~\eqref{ampR1} would be just
$\amp^{(\bs{0})}(\Tht_f-\Tht_i,\tht-\Tht_i)$, if only $p_1$ (and not $p'_1$)
would appear ``inside'' the polarization $\pol^*$. But $\pol(p'_1,q')$, being
orthogonal to $q'$, is obtained by applying to $\pol(p_1,q')$ a rotation around
$q'$:
\begin{equation}\label{polRot}
  \pol^*(p'_1,q') = R_{q'}(\alpha)\pol^*(p_1,q')
  = \esp{\ui\lambda\alpha}\pol^*(p_1,q')
\end{equation}
where $\alpha$ is a suitable (possibly large) rotation angle, and we have
exploited the fact that a polarization tensor of helicity $\lambda$ acquires a
phase factor under rotations. To sum up:
\begin{equation}\label{rotAmp}
  \amp^{(\Tht_i)}(\Tht_f,\tht) =
  \esp{\ui\lambda\alpha}\amp^{(\bs0)}(\Tht_f-\Tht_i,\tht-\Tht_i) \;.
\end{equation}

\begin{figure}[ht]
  \centering
  \includegraphics[width=0.9\linewidth]{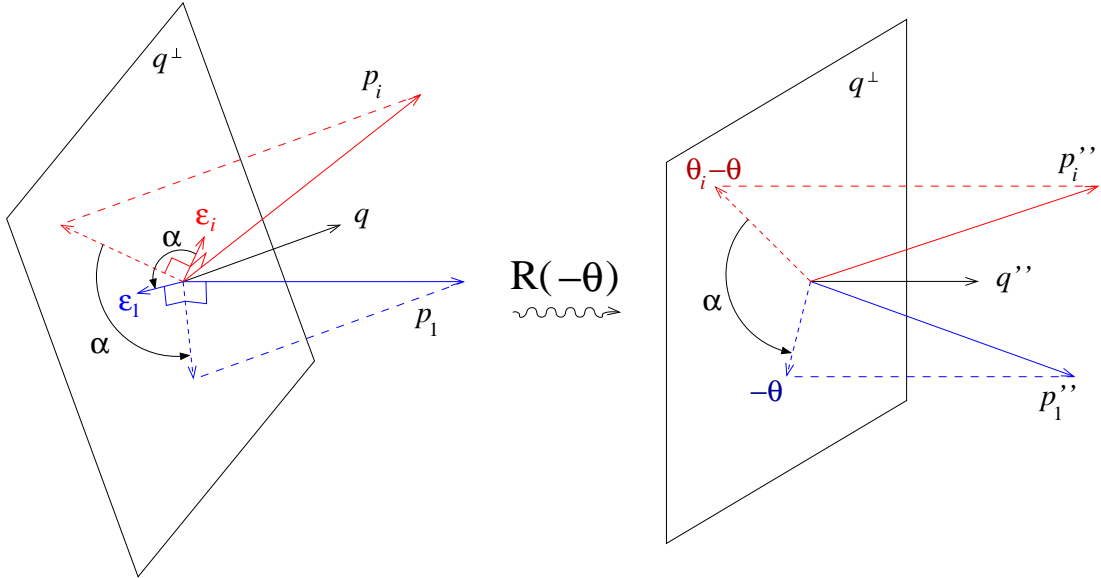}
  \caption{\it Spatial representation of the graviton momentum $q$ (black) and of
    the two \{reference-momentum, $T$-polarization\} pairs
    $\{p_1,\pol_T(q,p_1)\}$ (blue) and $\{p_i,\pol_T(q,p_i)\}$ (red) in the
    original frame (left) and in the rotated frame (right) where $q\leadsto q''$
    lies on the $z$ axis.}
  \label{f:helrot}
\end{figure}

It remains to compute the rotation angle $\alpha$. This is easily derived by
rotating the original system by the (small) angle $-\tht$
in such a way that $q\leadsto q''$ is put on the $z$ axis so that the
$\bk{q''}^\perp$ plane becomes the transverse plane, as depicted in
fig.~\ref{f:helrot}. Explicitly:
\begin{align*}
  q   &= \omega(\tht,1) \quad\, \leadsto q'' = \omega(\bs{0},1) \\
  p_i &= E(\Tht_i,1)    \;\, \leadsto p''_i = E(\Tht_i-\tht,1) \\
  p_1 &= E(-\bs0,1)     \;    \leadsto p''_1 = E(-\tht,1)
\end{align*}

It is now evident that the azimuth of $p''_1$ in the $\bk{q''}^\perp$ plane is
$\varphi_{p''_1}=\phi_{-\tht}=\phi_\tht+\pi$ and that of $p''_i$ is
$\varphi_{p''_i}=\phi_{\Tht_i-\tht}=\phi_{\tht-\Tht_i}+\pi$, and their
difference is just the sought rotation angle
\begin{equation}\label{alpha}
  \alpha=\varphi_{p''_1}-\varphi_{p''_i} = \phi_\tht-\phi_{\tht-\Tht_i} \;,
\end{equation}
because the $\pol_T$'s are orthogonal to the projections of $p_i$ and $p_1$ on
the $\bk{q}^\perp$ plane.

The same reasoning applies in the case of backward emission, e.g., with
$\vq=\omega(\tilde\tht,-1)$ directed in the opposite direction w.r.t.\
fig.~\ref{f:helrot} . In this case the rotation angle needed to align $q$ along
$-\hat{z}$ is just $\tht=-\tilde\tht$. Furthermore, the azimuthal angle $\alpha$
between the polarization vectors must be counted in the opposite direction,
because the thumb of the right hand now points towards the negative
$z$-direction. In conclusion
\begin{equation}\label{-alpha}
  \tilde\alpha = - (\phi_{-\tilde\tht}-\phi_{-\tilde\tht-\Tht_i})
  = - (\phi_{\tilde\tht}-\phi_{\tilde\tht+\Tht_i}) \;.
\end{equation}

\subsection{Relation with Jacob-Wick conventions\label{a:jw}}

The relation of our helicity amplitudes with those defined by Jacob and Wick
(JW)~\cite{Jacob:1959at} can be understood by comparing in the two frameworks
the choice of the polarization vectors for a generic graviton 3-momentum $q$
with polar angle $\theta$ and azimuth $\phi$.

Let $q = R(\omega\hat{z})$, where, according to JW conventions, $R$ is the
rotation matrix of angle $\theta$ and axis along $\hat{z}\times q$ (thus
belonging to the transverse plane). Such matrix $R$ is conveniently written in
terms of the usual Euler angles $(\alpha,\beta,\gamma)=(\phi,\theta,-\phi)$, so
that it can be represented as the product of 3 rotations along the $y$ and $z$
axis:
\begin{equation}\label{euler}
  R = R_{\phi,\theta,-\phi} = R_z(\phi) R_y(\theta) R_z(-\phi) \;.
\end{equation}
JW define a reference helicity state when the particle (here the graviton) has
momentum $\omega\hat{z}$ along the positive $z$-axis. This means that they
implicitly fix a pair of polarization vectors orthogonal to $\omega\hat{z}$ ---
i.e., $\pol_1=\hat{y}$ and $\pol_2=-\hat{x}$ --- so as to build the right-handed
orthogonal basis $\{\omega\hat{z},\pol_1,\pol_2\}$ .  The transformation of the
helicity state in eq.~(6) of~\cite{Jacob:1959at} correspond to rotate the
graviton momentum {\em and} the polarization vectors with the matrix $R$ of
eq.~\eqref{euler}, in such a way that the right-handed basis adapted to $q$ is
$\{q,R\pol_1,R\pol_2\}$.

On the other hand, our convention~\eqref{polLT} of the polarization vectors
requires $\pol_T$ to be orthogonal to both $q$ and $\hat{z}$, and it is easy to
see that
\begin{equation}\label{polTrot}
  \pol_T = R_{\phi,\theta,0}\pol_1 = R_z(\phi) R_y(\theta) \pol_1 \;,
\end{equation}
where the rotation matrix $R_{\phi,\theta,0}$ differs from that of JW by the
vanishing of the last Euler angle $\phi$, which doesn't affect the action
$\omega\hat{z}\to q$, but changes the orientation of the polarization vectors in
the $\bk{q}^\perp$ plane.

In practice, our right-handed basis $\{q,\pol_T,\pol_L\}$ is obtained by
applying to the JW reference basis $\{\omega\hat{z},\pol_1,\pol_2\}$ the
rotation $R_{\phi,\theta,0}$ of eq.~\eqref{polTrot}. As a consequence, our
polarization vectors are rotated by an angle $+\phi$ around $q$ w.r.t.\ those of
JW. It follows that
$2^{-1/2}\pol_{\pm} = R_q(\phi)\pol_\pm^{\jw} = \esp{\mp\ui\phi}\pol_\pm^{\jw}$
while for the helicity amplitudes (involving contractions with
$\pol_{\pm}^{\mu\nu*}$) we have
\begin{equation}\label{relAmp}
  M_\lambda(q) = \esp{\ui\lambda\phi}M_\lambda^{\jw}(q) \;.
\end{equation}

Let us now rederive the amplitude transformation phase of eqs.~(\ref{rotAmp},
\ref{alpha}). In the JW conventions, the amplitudes are invariant under
rotations bringing $\omega\hat{z}\leftrightarrow q$, provided the rotation axis
is in the transverse plane. In the case of small emission angles $\tht\ll 1$ ---
adopting now the 2D angular notations --- such rotations are translations in the
transverse components of forward momenta like $q$, $p_i$, $p_f$, as explained in
the previous subsections~\ref{s:sar},\ref{s:ampt}.  Therefore, starting with the
amplitude $M_\lambda^{\jw(\Tht_i)}(\Tht_f,\tht)$ and applying first a small
(JW-like) rotation $R(-\tht)$ bringing $q\to\omega\hat{z}$ and then another
small (JW-like) rotation $R(\Tht_i-\tht)$ we find that
\begin{equation}\label{jwamp}
  M_\lambda^{\jw(\Tht_i)}(\Tht_f,\tht) =
  M_\lambda^{\jw(\Tht_i-\tht)}(\Tht_f-\tht,\bs0) =
  M_\lambda^{\jw(\bs0)}(\Tht_f-\Tht_i,\tht-\Tht_i) \;,
\end{equation}
up to terms $\ord{\tht,\Tht_i}^2$ in the arguments of $M^{\jw}$.
By then recalling eq.~\eqref{relAmp} we immediately obtain
\begin{align*}\label{helPhase}
  M_\lambda^{(\Tht_i)}(\Tht_f,\tht)
  &\stackrel{\eqref{relAmp}}{=}
    \esp{\ui\lambda\phi_\tht}M_\lambda^{\jw(\Tht_i)}(\Tht_f,\tht) \\
  &\stackrel{\eqref{jwamp}}{=}
    \esp{\ui\lambda\phi_\tht}M_\lambda^{\jw(\bs0)}(\Tht_f-\Tht_i,\tht-\Tht_i)\\
  &\stackrel{\eqref{relAmp}}{=}
    \esp{\ui\lambda\phi_\tht}\esp{-\ui\lambda\phi_{\tht-\Tht_i}}
    M_\lambda^{(\bs0)}(\Tht_f-\Tht_i,\tht-\Tht_i) \;.
\end{align*}

\section{The $\bs{h}$ field\label{a:hf}}

\subsection{Calculation of the $\bs{h}$ field in coordinate space}

According to the analysis of~\cite{ACV07}, the two real components $h_{TT}$ and
$h_{LT}$ of the radiation field are conveniently collected into a single
complex-valued field $h\equiv h_{TT}+\ui h_{LT}$ that admits the integral
representation (see eq.~(2.14) of~\cite{ACV07})
\begin{equation}\label{hIntRep}
  h(\bt,\xt) = 2 \int\frac{\dif^2\qt_1}{(2\pi)^2}\frac{\dif^2\qt_2}{(2\pi)^2}
  \frac{1-\esp{2\ui\phi_{12}}}{(\qt_1+\qt_2)^2}\;
  \esp{-\ui[\qt_1\cdot\xt+\qt_2\cdot(\xt-\bt)]} \;,
\end{equation}
where $\phi_{ij}\equiv\phi_i-\phi_j$.

\begin{figure}[ht]
  \centering
  \includegraphics[width=0.3\textwidth]{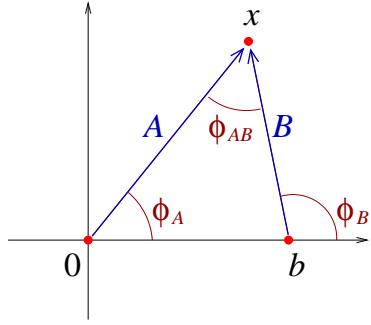}
  \caption{\it The variables of the transverse plane introduced in the
    computation of $h(\xt)$. Here the impact parameter $\bt$ defines the origin
    of azimuthal angles, i.e., the real axis of the corresponding complex
    plane.}
  \label{f:coordTrasverse}
\end{figure}

By denoting with $A\equiv|\xt|$ and $B\equiv|\xt-\bt|$ the moduli of the
external vectors (see fig.~\ref{f:coordTrasverse}) appearing in the last
exponent, and by explicitly writing out the various azimuthal angles, we rewrite
$h$ in the form
\begin{equation}\label{hAzimAng}
  h = \frac1{2\pi^2}\int_0^\infty \dif q_1\; q_1 \int_0^\infty \dif q_2\; q_2
  \int_0^{2\pi} \frac{\dif\phi_1}{2\pi} \int_0^{2\pi}
  \frac{\dif\phi_2}{2\pi} \; \frac{(1-\esp{2\ui\phi_{12}})
    \esp{-\ui(q_1 A \cos\phi_{A1} + q_2 B \cos\phi_{2B})}
  }{(q_1+q_2\esp{\ui\phi_{12}})(q_1+q_2\esp{-\ui\phi_{12}})}
  \;,
\end{equation}
where $q_i\equiv|\qt_i|$ and $\phi_A$ ($\phi_B$) is the azimuthal angle of
the 2D vector $\xt$ ($\xt-\bt$). Since
$\phi_{AB}=\phi_{A1}+\phi_{12}+\phi_{2B}$, the integrations
over $\phi_1$ and $\phi_2$ actually provide a double convolution, which can
be diagonalized by a Fourier transform. In practice, by computing the partial
waves w.r.t.\ the angle $\phi_{AB}$, we obtain
\begin{align}\label{hm1}
  h_m(A,B) &\equiv \int_0^{2\pi}\frac{\dif\phi_{AB}}{2\pi} \;
  \esp{\ui m \phi_{AB}} h(\bt,\xt) \\
  &= \frac{(-1)^m}{2\pi^2} \int\dif q_1\;q_1 J_m(q_1 A)
  \int\dif q_2\;q_2 J_m(q_2 B) \nonumber \\
  &\quad \times \int\frac{\dif\phi_{12}}{2\pi} \;
  \frac{\esp{\ui m \phi_{12}}\;(1-\esp{2\ui\phi_{12}})}{
    (q_1+q_2\esp{\ui\phi_{12}})(q_1+q_2\esp{-\ui\phi_{12}})} \;, \nonumber
\end{align}
where we used the relation
\begin{equation}\label{BesselJm}
  \int_0^{2\pi}\frac{\dif\phi}{2\pi}\; \esp{\ui m \phi} \esp{-\ui x \cos\phi}
  = \ui^{-m} J_m(x) \qquad (m\in\Z) \;.
\end{equation}
The last (azimuthal) integral in eq.~(\ref{hm1}) is easily computed by
transforming it in a contour integral over the unitary circle in the complex
plane of the variable $z\equiv\esp{\ui\phi_{12}}$. For $m\geq 0$ we have
\begin{equation}\label{phi12int}
  \Phi_m(q_1,q_2) \equiv \int\frac{\dif z}{2\pi\ui}\;
  \frac{z^m (1-z^2)}{(q_1+q_2 z)(q_1 z+q_2)}
  = (-1)^m \left(\frac{q_<}{q_>}\right)^m \frac1{q_>^2} \qquad(m\geq 0)
\end{equation}
since only the pole at $z=-q_</q_>$ is enclosed by the contour.

For $m=-1$ the additional pole at $z = 0$ provides a contribution that exactly
cancels the one at $z=-q_</q_>$: $\Phi_{-1}(q_1,q_2) = 0$.

The azimuthal integral for $m\leq -2$, after the change of variable $z\to 1/z$,
keeps its original structure, determining the (anti)symmetry property
$\Phi_m = -\Phi_{-m-2}$. Note also that $\Phi_m$ is symmetric in the exchange
$q_1 \leftrightarrow q_2$.

Let us then proceed with the computation of $h_m(A,B)$ for $m\geq 0$. We have
\begin{align}
  2\pi^2 h_m(A,B) &= \int_0^\infty\dif q_1 \dif q_2 \; J_m(q_1 A) J_m(q_2 B)
  \left(\frac{q_<}{q_>}\right)^{m+1} \nonumber \\
 &= \int_0^\infty\dif q_1 \int_0^{q_1} \dif q_2 \; J_m(q_1 A) J_m(q_2 B)
  \left(\frac{q_2}{q_1}\right)^{m+1} + \{A \leftrightarrow B\} \;. \label{intJJ}
\end{align}
By expressing the $q_2$ variable in terms of the ratio $\rho\equiv q_2/q_1$, the
$q_1$ integration reduces to the orthogonality relation for Bessel functions:
\begin{align}
  2\pi^2 h_m(A,B) &= \int_0^1\dif \rho \; \rho^{m+1} \int_0^\infty \dif q_1 \; q_1
  J_m(q_1 A) J_m(q_1 \rho B) + \{A \leftrightarrow B\} \nonumber \\
  &= \frac1{AB}\left(\frac{A}{B}\right)^{m+1} \Theta(B-A) + \{A \leftrightarrow B\}
    \;.\label{hm}
\end{align}

The $h$-field can now be obtained by summing the Fourier series
\begin{align}\label{fourierSum}
  2\pi^2 h(\xt) &= \sum_{m=-\infty}^{\infty} h_m(A,B) \esp{-\ui m \phi_{AB}} 
  = \sum_{m=0}^\infty [h_m(A,B)-h_{-m-2}(A,B)] \esp{-\ui m \phi_{AB}} \\
  &= \sum_{m=0}^\infty \frac{\esp{-\ui m \phi_{AB}}-\esp{\ui(m+2)\phi_{AB}}}{AB}
  \left[\left(\frac{A}{B}\right)^{m+1}\Theta(B-A)
    +\left(\frac{B}{A}\right)^{m+1}\Theta(A-B)\right] \nonumber \\ \nonumber
 &= \frac1{\cA^*\cB}\left[
   \Theta(B-A)\!\left(\frac{\cA^*}{\cB^*-\cA^*}-\frac{\cA}{\cB-\cA}\right) +
   \Theta(A-B)\!\left(\frac{\cB}{\cA-\cB}-\frac{\cB^*}{\cA^*-\cB^*}\right) \right]
\end{align}
where we introduced the complex numbers $\cA\equiv A\esp{\ui\phi_A}$ and
$\cB\equiv B\esp{\ui\phi_B}$. It turns out that the square brackets in the
last equation are equal, and we finally obtain
\begin{equation}\label{hx}
 h(\xt) = \frac1{2\pi^2} \frac{\cA^*\cB-\cA\cB^*}{\cA^*\cB |\cA-\cB|^2}
 = \frac1{2\pi^2} \frac{x-x^*}{b x^*(x-b)} = \frac{1-\esp{2\ui\phi_{AB}}}{2\pi^2 b^2}
\end{equation}

The components $h_{TT}$ and $h_{LT}$ correspond to the real and imaginary part
of $h$, respectively, and read ($\phi_A=\phi_{xb}$)
\begin{subequations}\label{hij}
  \begin{align}
    h_{TT}(\xt;\bt) &= \frac{1-\cos(2\phi_{AB})}{2\pi^2 b^2} =
    \frac{\sin^2\phi_{AB}}{\pi^2 b^2} = \frac{\sin^2\phi_{xb}}{\pi^2
      |\xt-\bt|^2} \\
    h_{LT}(\xt;\bt) &= \frac{-\sin(2\phi_{AB})}{2\pi^2 b^2} =
    \frac{\sin\phi_{xb}}{\pi^2
      |\xt-\bt|^2}\left(\frac{|\xt|}{b}-\cos\phi_{xb}\right) \;.
  \end{align}
\end{subequations}
Some remarks are in order:
\begin{itemize}
\item The final form confirms the UV-safe solution of the differential equation
  (2.15) of~\cite{ACV07}.
\item The simple expression of the solution in the r.h.s.\ of eq.~(\ref{hx}) has
  the same form of the phase factors in the integral
  representation~(\ref{hIntRep}) coming from H-diagram vertices, evaluated at
  the angle $\phi_{AB}=\phi_{x,x-b}$. In particular, the $h_{TT}$ component has
  a geometrical significance, embodied in the relation
  \begin{equation}\label{geomArea}
    \frac{\sin\phi_{AB}}{b} = \frac{\xt\wedge(\xt-\bt)}{b|\xt||\xt-\bt|}
    = \frac{2\;\text{Area}}{|\xt||\xt-\bt|b} \;.
  \end{equation}
\end{itemize}

\subsection{Calculation of the $\bs{h}$ field in momentum space}

The 2-dimensional Fourier transform of the complex field $h(\bt,\xt)$ w.r.t.\ the
transverse variable $\xt$ is given by
\begin{equation}\label{htildeDef}
  \tilde{h}(\bt,\qt) \equiv \int\dif^2\xt\; \esp{\ui\qt\cdot\xt} h(\bt,\xt)
  = \frac1{2 \pi^2 \qt^2}\int\dif^2\qt_2\;
  \esp{\ui\qt_2\cdot\bt} \left[1-\esp{2\ui\phi_{12}}\right] \;.
\end{equation}
In fact, by replacing $h(\bt,\xt)$ with the integral
representation~\eqref{hIntRep}, the integration in $\xt$ just provides a delta
function $\delta^2(\qt-\qt_1-\qt_2)$ which is then used to perform the
integration in $\qt_1 = \qt-\qt_2$, according to eqs.~(2.11-12) of~\cite{ACV07}.

By introducing the complex variables
\begin{equation}\label{qComp}
  q\equiv q_x + \ui q_y \;, \qquad q_2\equiv q_{2x}+\ui q_{2y}
\end{equation}
the angular factors can be written in rational form:
\begin{align}
  \esp{2\ui\phi_1} = \frac{q_1}{q_1^*} = \frac{q-q_2}{(q-q_2)^*} \;, \qquad
  \esp{-2\ui\phi_2} = \frac{q_2^*}{q_2}
 \nonumber \\
 1-\esp{2\ui\phi_{12}} = 1-\frac{q-q_2}{(q-q_2)^*}\frac{q_2^*}{q_2}
  = \frac{q_2 q^* - q_2^* q}{q_2(q-q_2)^*}
\end{align}
Without loss of generality we can orient the impact parameter vector along the
real axis: $\bt=(b,0)$. In this way $\qt_2\cdot\bt = q_{2x}b$ and we obtain
\begin{equation}\label{htildeAB}
  2\pi^2\tilde{h} =
  \frac1{q}\int\dif^2 \qt_2 \; \frac{\esp{\ui q_{2x}b}}{q^*-q_2^*}
 - \frac1{q^*}\int\dif^2\qt_2 \;\frac{\esp{\ui q_{2x}b} q_2^*}{q_2(q^*-q_2^*)}
 \equiv \frac{I_1}{q} - \frac{I_2}{q^*} \;.
\end{equation}
The first integral is straightforward:
\begin{equation}\label{Aintegral}
  I_1 = \int_{\R^2}\dif q_{2y}\dif q_{2x}\;
  \frac{\esp{\ui q_{2x}b}}{q^*-q_{2x}+\ui q_{2y}}
 = -2\pi\ui \, \esp{\ui q^* b} \int_{-\infty}^{+\infty}\dif q_{2y} \;
  \Theta(q_{2y}-q_y) \esp{-q_{2y}b}
 = -2\pi\ui \frac{\esp{\ui q_x b}}{b} \;,
\end{equation}
where the $q_{2x}$ integral has been performed by closing the contour in the
upper complex half-plane, where the simple pole at $q_{2x}=q^*+\ui q_{2y}$ is
found provided $q_{2y}-q_y > 0$.

Also in the second integral of eq.~(\ref{htildeAB}) the $q_{2x}$ integration is
performed in the upper complex half-plane, where two simple poles can be found:
the previous one and another one at $q_{2x}=-\ui q_{2y}$ provided $q_{2y}<0$.
Explicitly
\begin{align}
 I_2 &= \int_{\R^2}\dif q_{2y}\dif q_{2x}\;
  \frac{\esp{\ui q_{2x}b} (q_{2x}-\ui q_{2y})}{
    (q_{2x}+\ui q_{2y})(q^*-q_{2x}+\ui q_{2y})} \nonumber \\
 &= 2\pi\ui \int_{-\infty}^{+\infty} \dif q_{2y} \left\{
  \Theta(-q_{2y})\frac{-2\ui q_{2y} \esp{q_{2y}b}}{q^*+2\ui q_{2y}}
 -\Theta(q_{2y}-q_y)\frac{q^* \esp{\ui q^* b} \esp{-q_{2y}b}}{q^*+2\ui q_{2y}}
 \right\} \nonumber \\
 &= 2\pi\ui\left\{ -\frac1{b}+q^*\left[
     \int_0^{\infty}\frac{\esp{-t}\dif t}{q^*b-2\ui t} - \esp{\ui q_x b}
     \int_0^{\infty}\frac{\esp{-t}\dif t}{q b+2\ui t} \right]\right\} \;,
 \label{Bintegral}
\end{align}
where we made the substitutions $t=-q_{2y}b$ and $t=(q_{2y}-q_y)b$ in the two
integrals of eq.~(\ref{Bintegral}), respectively. The latter are related to the
exponential-integral special function $E_1$ defined by
\begin{equation}\label{defE1}
  E_1(z)\equiv \int_z^{+\infty} \frac{\esp{-t}}{t}\;\dif t \;, \quad
  (|\arg(z)|<\pi)
\end{equation}
so that, by combining eqs.~(\ref{htildeAB},\ref{Aintegral},\ref{Bintegral})
we finally obtain
\begin{equation}\label{htilde}
  \tilde{h}(\qt;\bt) = \frac{\ui}{\pi b}\left(
    \frac1{q^*}-\frac{\esp{\ui q_x b}}{q} \right) +
  \frac{\esp{\frac{\ui}{2}q^*b}}{2\pi} \left[ E_1\big(\frac{\ui q^* b}{2}\big)
   + E_1\big(-\frac{\ui q b}{2}\big) \right]
\end{equation}
We notice that $\tilde{h}$ obeys a simple conjugation property:
$\tilde{h}^* = \esp{-\ui \qt\cdot\bt}\tilde{h}$ and therefore
the combination
\begin{equation}\label{realComb}
  \esp{-\frac{\ui}{2}\qt\cdot\bt} \tilde{h} = \left[
  \esp{-\frac{\ui}{2}\qt\cdot\bt} \tilde{h} \right]^* \in \R
\end{equation}
is real valued, and reads
\begin{align}
  \esp{-\frac{\ui}{2}\qt\cdot\bt} \tilde{h} &=
  \frac{\ui\esp{-\frac{\ui}{2}q_x b}}{\pi q^* b} +
  \frac{\esp{-\frac12 q_y b}}{2\pi} E_1\big(\frac{\ui q^* b}{2}\big)
  \quad + \text{c.c.} 
 = \frac{\ui}{\pi}\esp{-\frac{\ui}{2}q_x b} \left[
  \frac1{q^*b} - \int_0^\infty \frac{\esp{-t}\;\dif t}{q^*b-2\ui t}\right]
  \quad + \text{c.c.} \nonumber \\
 &=  \frac{2}{\pi}\esp{-\frac{\ui}{2}q_x b}
  \int_0^\infty \frac{\esp{-t}\;\dif t}{(q^*b-2\ui t)^2}
  \quad + \text{c.c.}
  =  \frac{2}{\pi q^* b}\esp{-\frac{\ui}{2}q_x b} 
  \int_0^\infty \frac{t\esp{-t}\;\dif t}{q^*b-2\ui t}
  \quad + \text{c.c.} \;, \label{comb}
\end{align}
where integrations by parts have been performed in the last steps.

\section{An argument for the cutoff
  $\bs{\omega_{max} \sim R^{-1} \Theta_s^{-2}}$\label{a:cutoff}}

In this appendix we repeat, in more explicit terms, the argument advocated
in~\cite{GrVe14} for an upper cutoff on the $\omega$ spectrum.  To this purpose,
we should write the frequency spectrum of the emitted energy in terms of the so called news functions
as:
\begin{equation}\label{dEdo}
  \frac{dE}{d\omega} \propto |c(\omega)|^2
\end{equation}
and the energy emitted per unit retarded time $u \sim t-r$ as:
\begin{equation}\label{dEdu}
  \frac{dE}{du} \propto |\tilde{c}(u)|^2
\end{equation}
where $c$ and $\tilde{c}$ are one-dimensional Fourier transforms of each other.
From eqs.~(\ref{dEdo},\ref{dEdu}) we have, up to numerical constants,
\begin{equation}
  c(\omega)\sim \sqrt{Gs} ~\Theta_s \log^{\frac12} (\frac{1}{\omega R})
  \;,\quad (\omega < R^{-1}) \;; \qquad
  c(\omega)  \sim  \sqrt{Gs}~ \Theta_s (\omega R)^{-\frac12}
  \;, \quad (\omega >  R^{-1}) \;,
\end{equation}
and we find
\begin{equation}
  \tilde{c}(u) \sim  \sqrt{Gs}~ \Theta_s  u^{-1} \log^{\mp\frac12}
  (\frac{u}{R}) \;, \quad (u > R) \;; \qquad
  \tilde{c}(u)  \sim   \sqrt{Gs}~ \Theta_s (u R)^{-\frac12}
  \;, \quad (u < R) \;,
\end{equation}
where the - (+) holds for the even (odd) part of $\tilde{c}(\omega)$ under
$\omega\to-\omega$.

We thus get the following power time-history for GW emission:
\begin{equation}
  \frac{\dif E^\GW}{\dif u}
  = Gs ~\Theta_s^2 u^{-2} \log^{\mp1} (u/R) \;, \quad (u > R) \;; \qquad
   \frac{\dif E^\GW}{\dif u}
  = Gs~\Theta_s^2 (uR)^{-1} \;, \quad (u < R) \;.
\end{equation}
At this point we note that the latter behaviour exceeds a generally believed
(so-called Dyson) bound on the maximal power in gravitational-wave energy
emission (see e.g. ~\cite{Cardoso:2013krh}): $P_{GW} \le c^5 G_N^{-1} \rightarrow 1$ in our
units, if $u < R \Theta_s^2$.  But this precisely corresponds to saying that for
$\omega > R^{-1} \Theta_s^{-2}$ the spectrum should soften in order for the
bound on the power to be respected at very early times after the collision.

\bibliographystyle{h-physrev5}
\bibliography{coherence}

\end{document}